\documentclass[journal]{IEEEtran}

\usepackage{graphicx}      % include this line if your document contains figures
\usepackage{float}
\usepackage{caption}
\usepackage{subcaption}
\usepackage{epsfig} % for postscript graphics files
\usepackage{mathptmx} % assumes new font selection scheme installed
\usepackage{times} % assumes new font selection scheme installed
\usepackage{amsmath} % assumes amsmath package installed
\usepackage{amssymb}  % assumes amsmath package installed
\usepackage{amsfonts}
\usepackage{xspace}
\usepackage{pifont}

\usepackage{graphicx}
\graphicspath{{Figures/}}

\usepackage{comment}
\usepackage{todonotes} %%MF \usepackage[disable]{todonotes} to disable them without removing them
\presetkeys{todonotes}{inline,fancyline}{}
\usepackage{color}

\usepackage{cite}
\usepackage[multiple]{footmisc}
\usepackage{hyperref}

\usepackage{algorithm}
\usepackage[noend]{algpseudocode}

\usepackage{tikz}
\usetikzlibrary{automata,arrows}
\usetikzlibrary{matrix}

\makeatletter

\newcommand{\tinytodo}[2][]
{\todo[caption={#2}, size=\scriptsize, #1]{\renewcommand{\baselinestretch}{0.9}\selectfont#2\par}}

\newcommand{\AR}[1]{\tinytodo[backgroundcolor=yellow]{#1}}
\newcommand{\MR}[1]{\tinytodo[backgroundcolor=green]{#1}}
\newcommand{\MF}[1]{\tinytodo[backgroundcolor=lightgray]{#1}}
\newcommand{\MFedit}[2]{{\color{blue}#2}} %% The first argument is the original text, which is replaced by the second argument

\newcommand{\Rmnum}[1]{\expandafter\@slowromancap\romannumeral #1@}

\newcommand{\tick}{\ensuremath{\mathit{tick}}\xspace}
\newcommand{\ticks}{\ensuremath{\mathit{ticks}}\xspace}

\newcommand{\NS}{\ensuremath{\mathit{NS}}\xspace}
\newcommand{\ns}{\ensuremath{\mathit{ns}}\xspace}
\newcommand{\NSP}{\ensuremath{\mathit{NSP}}\xspace}

\newcommand{\NP}{\ensuremath{\mathit{NP}}\xspace}

\newcommand{\SigmaFor}{\ensuremath{\Sigma_{\mathit{for}}}\xspace}
\newcommand{\hatSigmaFor}{\ensuremath{\hat{\Sigma}_{\mathit{for}}}\xspace}
\newcommand{\hatSigmauc}{\ensuremath{\hat{\Sigma}_{\mathit{uc}}}\xspace}
\newcommand{\hatSigmauo}{\ensuremath{\hat{\Sigma}_{\mathit{uo}}}\xspace}

\newcommand{\SigmaNS}{\ensuremath{\Sigma_{\mathit{NS}}}\xspace}
\newcommand{\deltaNS}{\ensuremath{\delta_{\mathit{NS}}}\xspace}
\newcommand{\SigmaNSP}{\ensuremath{\Sigma_{\mathit{NSP}}}\xspace}
\newcommand{\deltaNSP}{\ensuremath{\delta_{\mathit{NSP}}}\xspace}
\newcommand{\deltaNP}{\ensuremath{\delta_{\mathit{NP}}}\xspace}
\newcommand{\Sigmauc}{\ensuremath{\Sigma_{\mathit{uc}}}\xspace}

\newcommand{\BLock}{\ensuremath{\mathit{BLock}}\xspace}
\newcommand{\TLock}{\ensuremath{\mathit{TLock}}\xspace}
\newcommand{\OBS}{\ensuremath{\mathit{OBS}}\xspace}
\newcommand{\Uncon}{\ensuremath{\mathit{Uncon}}\xspace}
\newcommand{\BS}{\ensuremath{\mathit{BS}}\xspace}
\newcommand{\bs}{\ensuremath{\mathit{bs}}\xspace}
\newcommand{\BPre}{\ensuremath{\mathit{BPre}}\xspace}
\newcommand{\head}{\ensuremath{\mathit{head}}\xspace}
\newcommand{\Mmax}{\ensuremath{M_{\mathit{max}}}\xspace}
\newcommand{\Lmax}{\ensuremath{L_{\mathit{max}}}\xspace}

\makeatother
\newtheorem{defn}{Definition}
\newtheorem{property}{Property}

\newtheorem{example}{Example}
\newtheorem{theorem}{Theorem}
\newtheorem{lemma}{Lemma}
\newtheorem{corollary}{Corollary}
\newtheorem{proof}{Proof}
\newtheorem{remark}{Remark}
\newcommand{\emptyseq}{\varepsilon}

\begin{document}
\title{Networked Supervisory Control Synthesis of Timed Discrete-Event Systems}

\author{Aida~Rashidinejad~\IEEEmembership{} \and
        Michel~Reniers~\IEEEmembership{} \and
        Martin~Fabian~\IEEEmembership{}% <-this % stops a space

\thanks{This research has received funding from the European Union’s Horizon 2020 Framework Programme for Research and Innovation under grant agreement no 674875.}

\thanks{Aida Rashidinejad and Michel Reniers are with the Control Systems Technology Group, Department of Mechanical Engineering, Eindhoven University of Technology
%, P.O.Box 513, 5600 MB\ Eindhoven, The Netherlands 
(e-mail: \{a.rashidinejad, m.a.reniers\}@tue.nl).}% <-this % stops a space

\thanks{Martin Fabian is with Department of Electrical Engineering, Chalmers University of Technology, Sweden (e-mail: fabian@chalmers.se).}%
}

\maketitle

\begin{abstract}
Conventional supervisory control theory assumes full synchronization between the supervisor and the plant. This assumption is violated in a networked-based communication setting due to the presence of delays, and this may result in incorrect behavior of a supervisor obtained from conventional supervisory control theory.
%% \MF{It seems a bit unclear here, what do we mean by `´failure''? Maybe better to say something like "incorrect behavior"?}
This paper presents a technique to synthesize a networked supervisor handling communication delays.
For this purpose, first, a networked supervisory control framework is provided, where the supervisor interacts with the plant through control and observation channels, both of which introduce delays. The control channel is FIFO, but the observation channel is assumed to be non-FIFO so that the observation of events may not necessarily be received by the supervisor in the same order as they occurred in the plant.
%\MF{Should we not be more specific here about the nature of the observation channel?}
It is assumed that a global clock exists in the networked control system, and so the communication delays are represented in terms of time.
Based on the proposed framework, a networked plant automaton is achieved, which models the behavior of the plant under the effects of communication delays and disordered observations.
%\MF{Do we really need to repeat FIFO and non-FIFO here? Does it add clarity?}
Based on the networked plant, the networked supervisor is synthesized, which is guaranteed to be (timed networked) controllable, nonblocking, time-lock free, (timed networked) maximally permissive, and satisfies control requirements for the plant. 

%To represent the effects of delays introduced by the control channel, events enabled by the supervisor and disordering introduced by the observation channel 
%In our framework, we use different notations for the enablement and observation of an event by the supervisor to indicate that they do not occur simultaneously as it is executed in the plant.
\end{abstract}

\begin{IEEEkeywords}
Discrete-event systems, time delays, networked control, maximal permissiveness, nonblockingness, safety,  supervisory control, synthesis.
\end{IEEEkeywords}

\IEEEpeerreviewmaketitle

%\MR{Shouldn't there be a reference to the WODES 2020 paper on synthesis of timed automata (as a very much used model of timed behaviour)?}
%\AR{But then we need to add a paragraph describing the difference between real-time modeling and discrete time and reasoning why we chose the second ... I am not sure if this would be helpful or confusing (by giving too much information)}

\section{Introduction}
%1. Benefits and challenges of networked control of systems.
\IEEEPARstart{N}{etworked} control of systems has gained a lot of attention in recent years. By eliminating unnecessary wiring, the cost and complexity of a control system are reduced, and nodes can more easily be added to or removed from the system. More importantly, there are applications in which the system is required to be controlled over a distance such as telerobotics, space explorations, and working in hazardous environments~\cite{gupta:10}.

Networked control of systems is challenging due to network communication problems among which delays have the highest impact~\cite{Heemels:10}. In this regard, many works have appeared in the literature investigating the effects of communication delays on the performance of a control system with time-based dynamics~\cite{Antsaklis:07,gupta:10,Heemels:10}. However, there is less work considering networked control of discrete-event systems (DESs).

%2. DES, Safety and SCT.
A DES consists of a set of discrete states where state transitions depend only on the occurrence of instantaneous events. DESs are used for modeling many types of systems, e.g., manufacturing processes, traffic, and queuing systems~\cite{Cassandras:99}.
In DESs, time is typically neglected meaning that events can occur independently of time. However, there are control applications in which time is an important factor to be considered, such as minimizing the production-cycle time in a manufacturing process~\cite{Wonham:19}. To consider time in control of a DES, the concept of a timed discrete-event system (TDES) has been introduced, in which the passage of a unit of time is indicated by an event called \tick~\cite{Wonham:94}.

Supervisory control theory is the main control approach developed for DESs~\cite{Ramadge:87}. To achieve desired (safe) behavior, a supervisor observes events executed in the plant and determines which of the next possible events must be disabled. Supervisory control theory synthesizes nonblocking supervisors that ensure safety, controllability, and nonblockingness for the plant and do not unnecessarily restrict the behavior of the plant (maximal permissiveness)~\cite{Cassandras:99}.
%\MR{Is using both supremal and maximal permissiveness in the previous sentence not too much of the same? Remove supremal?} Supervisory control synthesis of TDESs has been further proposed by Brandin and Wonham~\cite{Wonham:94}.

%3. The focus of this paper.
\MFedit{ %% Suggestions for re-write
In the conventional supervisory control theory, the plant and supervisor are assumed to interact synchronously; control commands (from the supervisor) are simultaneously received by the plant, and the observations of events executed in the plant are simultaneously received by the supervisor. \MF{I am not sure how picky we should be here... normally the supervisor "enables" (or disables) events, it doesn't "send" them. With observation it makes slightly more sense to talk of an observation being "received", but really the observation is "made" simultaneously as the event is executed.}\AR{Does it help to say control commands sent by $S$ like now?}
\MR{I agree with Martin that in the conventional setting there is no receiving and sending. We should only start using these these terms once our framework is discussed.}
\MF{I think we need to do something here\dots In the conventional SCT, all events are generated by G, and S disables events by not defining them from the corresponding state, or enable them and in that case when G generates the event S synchronously follows. }
Based on the \MFedit{simultaneous}{synchronous} interaction assumption, the controlled behavior is obtained  by synchronizing the events executed in the plant and supervisor.
However, the \MFedit{simultaneous}{synchronous} interaction assumption fails in a networked supervisory control setting due to the presence of delays in communication channels between the plant and supervisor.
%\MF{It is unclear here what "assumption" it is that fails. This assumption needs to be spelled out. Just saying "synchronizing events" is not enough in the intro, I would say. I guess the word "simultaneous" needs to be included somehow.}
}

In conventional supervisory control theory~\cite{Ramadge:84,Cassandras:99}, the plant generates all events, while the supervisor can disable some of the events and observes synchronously the execution of events in the plant. Based on this synchronous interaction, a model of the controlled plant behavior can be obtained by synchronous composition of the respective models of the plant and the supervisor. However, the synchronous interaction assumption fails in a networked  supervisory control setting, due to the presence of delays in the communication channels between the plant and supervisor.
%4. Literature review on asynchronous SC and NSC.
%\AR{Is it a good idea to categorize literature based on the important aspects in which they differ and finally providing a table as a summary or is it too much to put in a journal paper!?}

There are several works in the literature investigating supervisory control of DES under communication delays.
There are three important properties that these works may focus on:

%\begin{enumerate}
1) Nonblockingness.
For many applications, it is important to guarantee that the supervised plant does not block (as an additional control requirement)~\cite{Lin:17NB,Cassandras:99,yin2015synthesis}.
%\AR{A supervised plant is nonblocking if it can always reach a desired (marked) state from any state reachable from the initial state.}

%\MF{Can we summarize what they say?}

%\item Control objective. The control (safety) requirements may be given as avoiding a set of illegal states, or avoiding illegal sub-strings (events occurring alternately)~\cite{Cassandras:99}. 
%By modeling requirements as automata, both types of requirement are allowed to be considered.
%The control objective of a few papers such as~\cite{Timed:15,Rashidinejad18} is limited to only illegal states avoidance. However, in~\cite{Balemi:92,Park:06,ParkTime:08,Lin:17,CDC:17}, the objective is to satisfy a set of control requirements modeled as automata.

2) Maximal permissiveness.
A supervisor must not restrict the plant behavior more than necessary so that the maximal admissible behavior of the plant is preserved~\cite{Cassandras:99,Wonham:19}.

3) Timed delays (delays modeled based on time). In most of the existing approaches such as in~\cite{Balemi:92,Park:06,Park:12,Lin:14,Lin:17Det,Lin:17NB,Liu:19,Rashidinejad:19}, communication delays are measured in terms of a number of consecutive event occurrences. As stated in ~\cite{Lin:17Det,Rashidinejad18,Zhao:17}, it is not proper to measure time delay only based on the number of event occurrences since events may have different execution times. %\MR{Rephrase previous sentence.}
Here, as in TDES~\cite{Wonham:19}, the event \tick is used to represent the passage of a unit of time, which is the temporal resolution for modeling purposes.
%\end{enumerate}

%\AR{I need to re-formalize the literature review somehow ... so, existing works present some conditions on control requirement ... if these conditions are satisfied the result is proved in some works to be nonblocking or maximally permissive.However, if the conditions are not satisfied, they fail to provide the supremal sublanguage of the requirement satisfying the necessary conditions as stated in \cite{Lin20}}
%\AR{I am thinking how to put this here and also in the table ...}

Supervisory control synthesis under communication delays was first investigated by Balemi~\cite{Balemi:92}. To solve the problem, Balemi defines a condition called \emph{delay insensitive language}.
A plant has a delay insensitive language whenever any control command, enabled at a state of the plant, is not invalidated by an uncontrollable event.
Under this condition, supervisory control under communication delays can be reduced to the conventional supervisory control synthesis~\cite{Balemi:92}.
In other words, if a given plant has a delay insensitive language, then the conventional supervisor is robust to the effects of delays.
The benefit of this method is that
nonblockingness and maximal permissiveness are already guaranteed by the supervisor if it exists (as they are guaranteed in the conventional supervisory control theory).
However,
the imposed condition restricts the applications for which such a supervisor exists.

%\MF{It is common to avoid writing "we", possibly except when referring to the authors of the current paper. Here "we" is used with a generic meaning "anyone". I think this should be avoided.} 
%\MR{It may be better if we can actually show that there are relevant cases where the conditions are not met.}

In~\cite{Park:06,Park:12}, a condition called \emph{delay-observability} is defined for the control requirement such that the existence of a networked supervisor depends on it.
%The delay-observability condition is satisfied if the occurrences of consecutive uncontrollable events are limited to the communication delay bound
The delay-observability condition is similar to the delay insensitivity condition generalized for a sequence of uncontrollable events so that a control command is not invalidated by a sequence of consecutive uncontrollable events.
%\MR{good addition in my opinion}
%\MR{what does that mean?}
In~\cite{Park:06,Park:12}, nonblockingness is guaranteed. However, maximal permissiveness is not guaranteed. Also, no method is proposed to obtain the supremal controllable and delay-observable sublanguage of a given control requirement~\cite{Park:12}.

%The delay observability condition is restrictive as it may not necessarily hold for many practical applications. 

%In~\cite{Lin:14,Shu:15,Zhao:17,Alves:17,Lin:17Pre,Lin:17NB,Lin:17Det}
In a more recent study, Lin introduced new observability and controllability conditions under the effects of communication delays called \emph{network controllability} and \emph{network observability}~\cite{Lin:14}.
The approach presented by Lin has been further modified in ~\cite{Lin:14,Shu:15,Zhao:17,Alves:17,Lin:17Pre,Lin:17NB,Lin:17Det}.
In all these works, the problem of supervisory control synthesis under communication delays is defined under certain conditions (network controllability and network observability or the modified versions of them).
When the conditions are not met (by the control requirement), the synthesis does not result in a (networked) supervisor~\cite{Lin:14,Shu:15,Alves:17,Zhao:17,Lin:17NB}.
As discussed in~\cite{Lin:17Det}, delayed observations and delayed control commands make it (more) challenging to ensure nonblockingness of the supervised plant (compared to the conventional non-networked setting when there is no delay). To guarantee nonblockingness, additional conditions are imposed on the control requirement in~\cite{Lin:17NB}, but maximal permissiveness is not investigated. %\AR{In~\cite{Lin:17Det}, an algorithm is provided to compute the supremal controllable and observable sub-language of the control requirement under communication delays. However, no synthesis method is provided to achieve a maximally permissive supervisor.}

In~\cite{Liu:19}, an online predictive supervisory control synthesis method is presented to deal with control delays. The supervisor is claimed to be maximally permissive. However, this is not formally proved. This is also the case in~\cite{Shu:15} as they do not formally prove the maximal permissiveness although they establish the steps to achieve it.
In~\cite{Lin:17Pre}, a predictive synthesis approach is proposed to achieve a networked supervisor which is guaranteed to be maximally permissive in case it satisfies the conditions. 
Nonblockingness is yet not investigated in~\cite{Lin:17Pre}.
None of the works following Lin's method consider simultaneously nonblockingness and maximal permissiveness.
Moreover, as discussed in a recent study by Lin, in case that the conditions are not met by the control requirement,
%\cite{Lin20} is a recent study by Lin providing an overview and outline of the networked supervisory control synthesis problem. As stated in~\cite{Lin20},
there is no method so far to compute the supremal sublanguage satisfying the conditions ~\cite{Lin20}.

%In order to find a (maximally permissive) supervisor which preserves the maximal admissible behavior of the plant, the maximal sub-language of the control requirement satisfying the controllability and observability conditions should be computed~\cite{Lin:17Det,Lin:17Pre}.

   % A method to compute the maximal controllable sub-language of a given requirement is given in~\cite{Lin:17}.
    % Note that delays are not modeled based on time.
    %\MF{I have a hard time following the text above, it is rather fragmented. I think the "also" and "further" confuse things. These terms are used as connectives, but here the things they refer to seem largely unconnected.}

%\MR{Dedicated synthesis approach?}}
In~\cite{Rashidinejad18,Rashidinejad:19}, a new synthesis algorithm is proposed in which the effects of communication delays are taken into account in the synthesis procedure instead of in extra conditions to be satisfied by the plant/control requirement.
%\AR{category 1 considers the effects of delays in extra conditions on plant/ specifications ... category 2 considers them in observability and controllability conditions so still extra conditions on requirements, right? category 3 considers thos effects in the synthesis procedure ... so no extra conditions needed to be checked ... so basically to me category one and 2 seem to be the same}
%In this sense, there exist only a few papers investigating nonblockingness of the supervised plant under communication delays~\cite{Balemi:92,Lin:17NB,ParkTime:08,Rashidinejad18}.
%If the algorithm terminates and results in a (networked) supervisor, then it is nonblocking and maximally permissive.
\cite{Rashidinejad:19} investigates supervisory control of DES in an asynchronous setting.
The asynchronous setting does not take time into account, but it is guaranteed that (if the algorithm terminates) the synthesized (asynchronous) supervisor satisfies nonblockingness. Maximal permissiveness is still an open issue in~\cite{Rashidinejad:19}. \cite{Rashidinejad18} focuses on timed delays, but it does not formally prove nonblockingness or maximal permissiveness.
 
In~\cite{Zhu19} as a more recent study, first, the control and observation channels are modeled. Then, both the plant and control requirements are transformed into a networked setting. Using these transformations, the problem of networked supervisory control synthesis is reduced to conventional supervisory control synthesis. 
Using conventional supervisory control synthesis, the resulting supervisor is controllable and nonblocking for the transformed plant and the transformed control requirements. However, it is not discussed if the supervisor satisfies these conditions for the (original) plant.
%The synthesized supervisor is said to be nonblocking. However, this is not formally proved. Moreover,  the effects of observation delay in bringing non-determinism to the plant states are not taken into account. This problem is already considered in controllability or observability conditions in~\cite{Lin:17NB} or in the synthesis algorithm~in \cite{Rashidinejad18,Rashidinejad:19}. Therefore, there is no guarantee that the synthesized supervisor provides nonblockingness. 
%%MF \MF{Here above, you basically claim that~\cite{Zhu19} lies. Do you really want to claim that? Are you 100\% sure that their claim is incorrect? \MR{As Aida tried to indicate, their claim is probably fine, but it is not the claim that you want/need. They have a result for the transformed problem and it remains to be seen that that also is a solution for the original plant.}} \AR{Please check here.} \MR{I think the text is fine now.}

Furthermore, although it is important to consider time in the presence of delays, only a few papers investigate networked supervisory control of TDES~\cite{ParkTime:08,Zhao:17,Alves:17,Rashidinejad18,Miao:19} (where communication delays are modeled based on a consistent unit of time) as it introduces new complexities and challenges.

Table \ref{tab:review} gives an overview of the existing works.
To the best of our knowledge, none of these works studies supervisory control synthesis of discrete-event systems under communication delays such that
delays are modeled based on time, and the delivered supervisor guarantees both nonblockingness and maximal permissiveness as is done in this paper.

\newcommand{\cmark}{\ding{51}}
\newcommand{\xmark}{\ding{55}}
\begin{table}[h]
    \centering
    \begin{tabular}{|c|c|c|c|c|}
    \hline
         Citation &  Timed & Nonblocking & Permissive\\
         \hline
         \cite{Balemi:92,Park:12} & \xmark & \cmark & \cmark\\
         \hline
         \cite{Lin:14,Zhu19,Liu:19,Shu:15}& \xmark & \xmark & \xmark\\ 
         \hline
         \cite{Park:06,Lin:17NB,Rashidinejad:19} & \xmark & \cmark & \xmark \\
         \hline
         \cite{Lin:17Pre} & \xmark & \xmark & \cmark \\
%         \hline \cite{Rashidinejad18} & \cmark & \xmark & \xmark \\
%         \hline
%         \cite{Rashidinejad:19} & \xmark & \cmark & \xmark \\
         \hline
         \cite{ParkTime:08,Rashidinejad18,Alves:17,Zhao:17,Miao:19} & \cmark & \xmark & \xmark\\
         \hline\hline
         This Paper & \cmark & \cmark & \cmark \\
    \hline
    \end{tabular}
    \caption{Overview of existing works.}
    \label{tab:review}
\end{table}

Our work is close to~\cite{Rashidinejad18} in terms of the networked supervisory control setting and to~\cite{Rashidinejad:19} in terms of the synthesis technique. Similar to~\cite{Rashidinejad18} and~\cite{Rashidinejad:19}, the following practical conditions are taken into account:

%\begin{enumerate}
1) A controllable event can be executed in the plant only if it is commanded (enabled) by the supervisor.

2) An uncontrollable event is not commanded (enabled) by the supervisor; it occurs spontaneously in the plant. % Same assumption as in WoDES and CASE
   % \AR{Does this condition include \emph{tick} as well? Shall we consider that if \emph{tick} is enabled at a state where there is no forcible event enabled there, it occurs in the networked supervised plant. If there is a forcible event, \emph{tick} can occur if it is also enabled by $\NS$. We have 2 options: 1. consider this condition in the operator we propose to achieve the networked supervised plant, and then prove that networked controllability always hold (like what we have in CASE), or 2. we do not consider this condition in the definition of operator and prove controllability for the synthesized networked supervisor.}
   
3) Any event, controllable or uncontrollable, executed in the plant is observable to the supervisor. % Same assumption as in WoDES and CASE

4) A control command sent by the supervisor reaches the plant after a constant
    %%MF Is "fixed>" a good word? "bounded"? "limited"? "known"? "predetermined"? "determined"? "finite"? "constant"? "known"?
amount of time delay. The command may not necessarily be %%MF immediately?
accepted by the plant, in which case it will be removed from the control channel when the next \tick occurs.
    %it will remain in the control channel.
    %\AR{@Michel, the last sentence should be: in which case it will be removed from the control channel (due to Definition \ref{dfn:list})? or better not to mention it here at all?}  %%MF Forever?
    % Same assumption as in WoDES and CASE
    %\MR{The most natural interpretation is to assume it is gone, so perhaps it does not have to be mentioned explicitly?}
    %\MF{I think this should be mentioned; it has always been a bit of an issue for me (see old comment on the next item). It must be removed, right? Else it will block consecutive commands, not? And then arises the question of \emph{when} is it removed? And how? By whom?}
%\MR{Items 4 and 7 are alike. Put together / closer?}
Also, the observation of a plant event, controllable or uncontrollable, occurs after a constant amount of time delay. % % Same assumption as in WoDES
%\MR{Is it the case that the events that arrive through the observation channel must be observed or may these be neglected by the supervisor (in a similar way as the cplant may skip the events that arrive through the control channel? Why this difference?}
%\AR{I think it is fair to compare the observation with control commands; both are some information being sent through channels. So, the observation of events is an information receives by the supervisor. Similarly, the control command is an information received by the plant (and the plant does not ignore it).
%Not accepting a control command by the plant means that it receives the information (control command) but does not execute the event relevant to the command, maybe this needs to become clearer here?}
%\MR{Is this really what happens in the composition in case the enabling event reaches the plant, but the plant does not have the corresponding event as an alternative?}

%\MR{Itmes 3 and 5 are very much alike, maybe we should put them closer together?}

5) The control channel is assumed to be FIFO, so \MFedit{consecutive}{}control commands sent by the supervisor will reach the plant in the same order as they have been sent. %%MF So previous commands not immediately accepted will block future commands?
However, the observation channel is non-FIFO, and so consecutive events that occur in the plant may be observed by the supervisor in any possible order.
For instance, if the events $a$ and $b$ occur in that order between two $\tick$s in the plant, they may be observed in the other order.
    % Same assumption as in WoDES and CASE
%    \MF{I am not sure about the word "fixed", it can mean many things. On line 239 in the LaTeX file I list a few possible meanings. The more precise we can be, the better.}
%\end{enumerate}
\MFedit{Note that \MFedit{}{also} the control channel \MFedit{can also}{could} be non-FIFO\MFedit{. A non-FIFO control channel can}{, which would} make the results more conservative.}{}
Here, we investigate the situation where only the observation channel is non-FIFO.  See Section \ref{remark:nonfifocontrolchannel} for a discussion on how the proposed solution is adapted for a non-FIFO control channel.
%However, a non-FIFO control channel is not considered here as it results in too conservative solutions.
%\MR{I consider this a bad reason not to do it.}\AR{Is it better now?}
%\MF{Above, a TDES is defined to have fixed time bounds. That makes it strange to here say that the plant is a TDES with no fixed time bounds. I think that it is the previous text that needs to change, since in the formal def of TDES (1), there is no mention of fixed time bounds, is there?}\AR{Here is the first time we give details about TDES ... maybe I already removed some previous description}
%Compared to~\cite{Rashidinejad18, Rashidinejad:19}, this paper brings the following contributions. \MR{Here you say that the contributions are given, and then continue talking about TDES old style. That is confusing.}
%\AR{The previous paragraph was saying in what terms our work is close  to our WoDES and CASE papers (similarities). Here, we may want to talk about (differences) what is improved compared to those. So, shall we start here by saying: This paper improves~\cite{Rashidinejad18,Rashidinejad:19} in the following aspects:} \AR{or any other better sentence that you suggest :), and then we can itemize the following (3 items: modeling purposes, synthesis technique, requirement generalization)}
%\MR{For me this is OK}

This paper improves~\cite{Rashidinejad18,Rashidinejad:19} in the following aspects:

%\begin{enumerate}
1) Modeling purposes. In~\cite{Wonham:19}, a TDES is generally derived from a DES by restricting the execution of each event within a lower and an upper time bound specified to the event. %%MF time bounds?
Also, a TDES should satisfy the ``activity-loop free" (ALF) assumption
%, meaning that at any state of a TDES, either a transition with an active event is enabled or the \tick event is defined. In other words,
to guarantee that the clock never stops~\cite{Wonham:19}.
%A TDES is generally displayed by a ``timed transition graph" (TTG). \MR{I have no clue what it means to display a TDES?}
Fixing time bounds for events and imposing the ALF condition restrict the applications that can be modeled as TDESs.
In this paper, the plant is already given as a TDES.
%\AR{(not a DES with a given set of time bounds for events)}. %\MR{Then, why are we calling it TDES and not TTG?}
Namely, the plant behavior is represented by an automaton, including the event \tick with no specific relationship between the occurrences of \tick and other events. 
%Namely, the plant is a TDES in which events do not necessarily occur between lower and upper time bounds specified to each event.
%\MF{To follow-up on my comment above\dots Is it not better to first define TDES without time bounds, and then restrict S(?) to be a TDES with time bounds? Now it is done the other way around, a TDES is defiend with time bounds and then teh plant is said to be a TDES without time bounds.}
%\MF{Or\dots the time bounds in the plant are always between $-\inf$ and $+\inf$?}
%\MF{This is not a full sentence, probably needs to be joined with the previous one.}
To relax the ALF condition, the concept of time-lock freeness is introduced as a property, expressing the time progress of the system.
%\MF{I do not think "check" is the right word here.}
%The plant may not necessarily be time-lock free.
Time-lock freeness, similar to nonblockingness, is guaranteed by the networked supervisor.
%Nonblockingness and time-lock freeness will be considered in the synthesis problem to be provided by the supervisor.

2) Synthesis technique. Inspired from the idea introduced in~\cite{Rashidinejad:19} to synthesize an asynchronous supervisor for DES, the synthesis method proposed in~\cite{Rashidinejad18} for networked supervisory control of TDES is improved.
For this purpose, first, the networked supervisory control (NSC)
framework is modeled. Then, a networked plant automaton is proposed, modeling the behavior of the plant in the NSC
framework. Based on the networked plant, a networked supervisor is synthesized. It is guaranteed that the networked supervisor provides nonblockingness, time-lock freeness, and maximal permissiveness.

3) Control requirement. The control requirement in~\cite{Rashidinejad18,Rashidinejad:19} is limited to the avoidance of illegal states. Here, the networked supervisory control synthesis is generalized to control requirements modeled as automata. 
%\end{enumerate}

In the following, the NSC framework is introduced in Section~\ref{section:basic NSP}. For the NSC framework, an operator is proposed to give
the networked supervised plant. Moreover, the conventional controllability and maximal permissiveness conditions are modified to timed networked controllability and timed networked maximal permissiveness conditions suitable for the NSC framework. Then, the basic networked supervisory control synthesis problem is formulated which aims to find a timed networked controllable and timed networked maximally permissive networked supervisor guaranteeing nonblockingness and time-lock freeness of the networked supervised plant. In Section~\ref{section:synthesis}, first, the networked plant is defined as an automaton representing the behavior of the plant under communication delays and disordered observations. Furthermore, a technique is presented to synthesize a networked supervisor that is a solution to the basic networked supervisory control problem. In Section~\ref{section:requirements}, the basic networked supervisory control synthesis problem is generalized to satisfy a given set of control requirements. Relevant examples are provided in each section. Finally, Section~\ref{section:conclusions} concludes the paper.
To enhance readability, all technical lemmas and proofs are given in the appendices.
%\AR{Shall we give the lemmas in the appendix as well as say this here? \MR{If a lemma has no intrinsic value or clear meaning then there is no point in placing it in the main text. So, for each lemma make this assessment and act accordingly.}}
%6. Sections.
%\AR{I am thinking about the following issues: 1. Should be events from $\Sigma_e$ assumed to be forcible? 2. Should we have an assumption on $G$ that $\forall w\in L_m(G), wt^*\in L_m(G)$ to make it reasonable that $G$ may stay at a marked state (what about activity loop free assumption in Wonham's book?)? 3. Should we consider equivalent states? (Is it needed?)}

%\MR{We may want to put the longer proofs in appendices? Discuss with Martin!}\AR{yes}

\section{Basic NSC Problem}
\label{section:basic NSP}
\subsection{Conventional Supervisory Control Synthesis of TDES}
A TDES $G$ is formally represented as a quintuple 
\begin{equation*}
G=(A, \Sigma, \delta, a_{0}, A_{m}),
\end{equation*}
where $A, \Sigma$, $\delta: A \times \Sigma \rightarrow A$, $a_{0}\in A$, and $A_{m}\subseteq A$ stand for the %%MF finite
set of states, the %%MF finte
set of events, the (partial) transition function, the initial state, and the set of marked states, respectively. 
The set of events of any TDES is assumed to contain the event $\tick \in \Sigma$. 
%\AR{The event \tick is used to represent the passage of a unit of time, which is the temporal resolution for modeling purposes~\cite{Wonham:19}.}
%\MF{I also cannot help to think that "constant time delay" and "observed [...] in any possible order" seem to contradict each other. Say the plant executes abcd, and this is observed as dcba. Is then not a delayed more than d? I know there is an assumption of zero time between the ticks, but is this really clear here?}
%\AR{shall we add a note here? sth like \tick presents the smallest measurable time unit ???}
%\AR{From Wonham's book: However, we imagine measuring time only with a global digital clock with output tickcount: R + → N, Temporal conditions will always be specified in terms of this digital clock time; real-valued time as such, and the clock function tickcount, will play no formal role in the model. The temporal resolution available for modelling purposes is thus just one unit of clock time.}
%\AR{go for temporal resolution and refer to Wonham}
The set $\Sigma_a = \Sigma \setminus \{ \tick \}$ is  called the set of active events.
The notation $\delta(a,\sigma)!$ denotes that $\delta$ is defined for state $a$ and event $\sigma$, i.e., there is a transition from state $a$ with label $\sigma$ to some state.
The transition function is generalized to words in the usual way: $\delta(a,w)=a'$ means that there is a sequence of subsequent transitions from state $a$ to the state $a'$ that together make up the word $w\in\Sigma^*$. Starting from the initial state, the set of all possible words that may occur in $G$ is called the language of $G$ and is indicated by $L(G)$; $L(G):=\{w\in\Sigma^*\mid\delta(a_0,w)!\}$.
Furthermore, for any state $a\in A$, the function $\textit{Reach}(a)$ gives the set of states reachable from the state $a$; $\textit{Reach}(a) :=\{a'\in A\mid \exists w\in\Sigma^*, \delta(a,w)=a'\}$.
States from which it is possible to reach a marked state are called nonblocking. An automaton is nonblocking when each state reachable from the initial state is nonblocking; for each $a\in \textit{Reach}(a_0)$, $\textit{Reach}(a) \cap A_m \neq \varnothing$.
$L_m(G)$ denotes the marked language of $G$; $L_m(G):=\{w\in L(G)\mid \delta(a_0,w)\in A_m\}$. 
States from which time can progress are called \emph{time-lock free} (TLF). %\MF{Sure not in *all* outgoing transitions, it is enough with one, not?}\AR{Sounds logical but I am not sure because then it is not inline with ALF condition in TDES! (it may not be important, but is it practical?)}
An automaton is TLF when each state reachable from the initial state is TLF; for each $a\in \textit{Reach}(a_0)$, there exists a $w\in \Sigma^*$ such that $\delta(a, w\,\tick)!$.
%%MF What does the underline mean=
%\begin{remark}
%\AR{Is the following enough description for the differences between automata in this paper and Wonham TDES?}
%As stated in~\cite{Wonham:19}, a TDES is a DES for which the execution of each event is restricted between lower and upper bounds.
%Moreover, a TDES should satisfy the activity loop free (ALF) assumption meaning that at any state of a TDES either a transition with an active event is enabled or at least the tick transition is defined. In other words, a TDES never stops the clock~\cite{Wonham:19}. The way of modeling TDES in~\cite{Wonham:19}, and the ALF condition restricts the applications that can be modeled as TDES. In this paper, we suppose that \emph{tick} is also an event that can be considered in the modeling from the scratch. Moreover, besides nonblockinness, we consider a time-lock freeness condition to be satisfied by a supervisor. Hence, the plant may not necessary be nonblocking or time-lock free, and these conditions will be considered in supervisory control synthesis to be satisfie by the supervisor.
%\hfill $\blacksquare$ 
%\end{remark}
%%MF The above and below have nothing to do with each other, and so should not be in the same paragraph.
%%MF In fact, I think that the finiteness of A could be mentioned where A is first introduced (see above).
\MFedit{In this paper, we frequently use 
%the concepts of \emph{subautomaton} and
\emph{natural projection} operator~\cite{Cassandras:99}.
}{}

\begin{comment}
\begin{defn}[Subautomaton~\cite{Cassandras:99}]
Given two TDES $G'=(A',\Sigma',\delta',a'_0,A'_m)$ and $G=(A,\Sigma,\delta,a_0,A_m)$, $G'$ is a subautomaon of $G$, denoted by $G'\subseteq G$, if for all $w\in L(G')$: $\delta'(a'_0,w)=\delta(a_0,w)$. In other word, $x'_0=x_0$, $X'\subseteq X$, and $L(G')\subseteq L(G)$. Moreover, $X'_m=X_m\cap X'$ ($L_m(G')\subseteq L_m(G)$).
\hfill $\blacksquare$
\end{defn}
\end{comment}

\begin{defn}[Natural Projection~\cite{Cassandras:99}]
\label{dfn:proj}
For sets of events $\Sigma$ and $\Sigma'\subseteq\Sigma$, $P_{\Sigma'}: \Sigma^* \rightarrow \Sigma'^*$ is defined as follows: for $e \in \Sigma$ and $w \in \Sigma^*$,
\begin{align*}
    P_{\Sigma'}(\epsilon)&:=\epsilon,\\
    P_{\Sigma'}(we)&:=\begin{cases}
    P_{\Sigma'}(w) e &\text{if $e\in\Sigma'$,}\\
    P_{\Sigma'}(w) &\text{if $e\in\Sigma\setminus\Sigma'$.}
    \end{cases}
\end{align*}
The definition of natural projection is extended to a language $L\subseteq\Sigma^*$; 
$P_{\Sigma'}(L):=\{w'\in\Sigma'^*\mid \exists w\in L, P_{\Sigma'}(w)=w'\}$
\cite{Cassandras:99}.
\hfill $\blacksquare$
\end{defn}

%\begin{remark}
Natural projection is an operation which is generally defined for languages. However, it is also possible to apply it on automata~\cite{Ware:08}. 
For an automaton with event set $\Sigma$, $P_{\Sigma'}$ first replaces all events not from $\Sigma'$ by the silent event $\tau$. Then, 
using a determinization algorithm (such as the one introduced in~\cite{Hopcroft:01}), %%MF No other algorithm can be used?
the resulting automaton is made deterministic.
A state of a projected automaton is then marked if it contains at least one marked state from the original automaton (see~\cite{Hopcroft:01} for more details).
Using the notation $\delta_P$ for
%{Let $\delta_P:2^A\times\Sigma^*\rightarrow  2^A$ be}
the transition function of the projected automaton, we state the following properties of this construction: (1) for any $w \in \Sigma^*$, if $\delta(a_0,w)=a_r$ then $\delta_P(A_0,P_{\Sigma'}(w)) = A_r$ where $A_0$ is the initial state of the projected automaton, and $A_r \subseteq A$ is a set with $a_r \in A_r$, (2) for any $w \in \Sigma^*$, if $\delta(a_0,w) \in A_m$, then $\delta_P(A_0,P_{\Sigma'}(w))$ is a marked state in the projected automaton.

In the rest of the paper, the plant is given as the TDES $G$ represented by the automaton $(A,\Sigma_G,\delta_G,a_0,A_m)$ with $\Sigma_G=\Sigma_a\cup\{\tick\}$ and $\Sigma_a\cap\{\tick\}=\varnothing$.
%\MR{Strictly speaking it is not said that $\tick$ is not also in $\Sigma^{a}$.}
%\AR{$G$ is changed to $(A,\Sigma_G,\delta_G,a_0,A_m)$ with $\Sigma_G=\Sigma_a\cup\{\tick\}$ regarding Martin's comment ... please let me know if you notice that $\Sigma$ or $\delta$ is still used anywhere for $G$}s
%\MF{On page 4, left, there is a remark about finite automata, but I think, in fact, using the Kleene-closure already assumes finite automata. Infinite state automata (like Petri nets) have larger repressive power than Kleene languages. So that paragraph can be removed.}
%\AR{It is not given as a remark anymore ... but I think it is better to have the following description here since we will use this later in the proof of Lemma \ref{lemma:finiteNP}}
Also, as it holds for many applications, $G$ is a finite automaton \cite{Wonham:19}. A finite automaton has a finite set of states and a finite set of events \cite{Carrol:89}.

Here, it is assumed that all events in $G$ are observable. A subset of the active events $\Sigmauc\subseteq \Sigma_a$ is uncontrollable.
$\Sigma_c=\Sigma_a\setminus\Sigmauc$ gives the set of controllable active events.
The event \tick is uncontrollable by nature. However, as in~\cite{Wonham:94}, it is assumed that \tick can be preempted by a set of forcible events $\SigmaFor\subseteq \Sigma_a$.
%When a forcible event is enabled at a particular state, \tick may be preempted there in the supervised plant. In fact, disabling \tick does not mean stopping the clock in the system; however, it means that the occurrence of an event is forced by timely preemption.
Note that forcible events can be either controllable or uncontrollable. For instance, closing a valve to prevent overflow of a tank, and the landing of a plane are controllable and uncontrollable forcible events, respectively~\cite{Wonham:19}.
%Then, $\Sigma_a_{c} := \Sigma_a\setminus\Sigma_a_{uc}$ is the set of controllable active events.
%We also use the notation $\ut{\Sigma}_c := \Sigma_c\setminus\{tick\}$ to refer to controllable active events.
%Both controllable and uncontrollable active events can be forcible.
Note that for synthesis, the status of the event \tick lies between controllable and uncontrollable depending on the presence of enabled forcible events. To clarify, when the event $\emph{tick}$ is enabled at some state $a$ and also there exists a forcible event $\sigma\in\SigmaFor$ such that $\delta_G(a,\sigma)!$, then \tick is considered as a controllable event since it can be preempted. Otherwise, $\emph{tick}$ is an uncontrollable event. In the figures, forcible events are underlined. The transitions labelled by controllable (active or \tick) events are indicated by solid lines and the transitions labelled by uncontrollable (active or \tick) events are indicated by dashed lines. %%MF\MF{This is better moved to the caption of Fig 8.}\AR{Also all other information about the figures because they are given in text?}
%\MR{in synthesis we behave in a certain way w.r.t. that event. Discuss!} \MR{My point is that the event does not behave as a controllable or uncontrollable event. It is only the synthesis procedure that notices the status and does something with it.}
%\MR{Before you can talk about a supervisor in the definition below you first need to define what a supervisor is!} 
%\AR{we will also have an informal description of a supervisor in introduction that a supervisor observes events and makes decisions ...}

If the plant $G$ is blocking, then a  supervisor $S$ needs to be synthesized to satisfy nonblockingness of the supervised plant. 
%\MF{To be picky\dots S can never make G nonblocking, S can only make the controlled system $G||S$ nonblocking.}\noindent
$S$ is also a TDES with the same event set as $G$.
Since the plant and supervisor are supposed to work synchronously in a conventional non-networked setting,
%\MF{Things get a bit blurred for me her. What does ``conventional setting'' really mean. I thought it meant the ordinary untimed SCT\dots? Maybe we say ``non-networked''?\AR{conventional/non-networked both refer to the same thing: a supervisor and a plant with synchronous interactions between them ... can be either for DES/TDES}}\noindent
the automaton representing the supervised plant behavior is obtained by applying the \emph{synchronous product} indicated by $S||G$ \cite{Cassandras:99}.
Generally, in the synchronous product of two automata, a shared event can be executed only when it is enabled in both automata, and a non-shared event can be executed if it is enabled in the corresponding automaton. Since the conventional supervisor $S$ has the same event set as $G$, each event will be executed in $S||G$ only if the supervisor enables (allows) it. $S$ is controllable if it allows all uncontrollable events that may occur in the plant. This is captured in \emph{conventional controllability for TDES}. %The formal definition of controllability is presented in Definition \ref{dfn:cont.TDES}.

\begin{comment}
\begin{defn}[Conventional Controllability for TDES (reformulated from~\cite{Wonham:94})]
\label{dfn:cont.TDES}
%\AR{Should it be exactly the same as Wonham as he defied $Elig_G$ function or my own version given below :)?}
Given a plant $G$ with uncontrollable events $\Sigmauc$ and forcible events $\SigmaFor$, a (conventional) supervisor $S$, is controllable w.r.t.\ $G$ if for all
$w\in L(S||G)$ and 
%\begin{equation*}
$u\in \begin{cases}
       \Sigmauc &\quad\text{if}\quad \exists\sigma\in\SigmaFor,~ w\sigma\in L(\mathit{S||G})  \\
       \Sigmauc\cup\{tick\} &\quad\text{otherwise,}
\end{cases}$,\\
%\end{equation*} 
whenever $wu\in L(G)$, then $wu\in L(S||G)$.
\hfill
$\blacksquare$
\end{defn}
\end{comment}

%\AR{We will use controlability of requirement $R$ w.r.t. $G$ as well. So, shall we generalize the following definition to $G$ and a random TDES not mentioning it only as a supervisor? \MF{I see no harm. In fact, this often pops up: controllability is defined for a \emph{supervisor}, is it then OK to use it as a property of the spec?}}

\begin{defn}[Conventional Controllability for TDES (reformulated from~\cite{Wonham:19})]
\label{dfn:cont.TDES}
Given a plant $G$ with uncontrollable events $\Sigmauc$ and forcible events $\SigmaFor$, a TDES %(conventional) supervisor
$S$, is controllable w.r.t.\ $G$ if for all
$w\in L(S||G)$ and $\sigma\in \Sigmauc\cup\{tick\}$, if $w\sigma\in L(G)$,
\begin{enumerate}
    \item \label{defCtrlStandard}$w\sigma\in L(S||G)$ , or
    \item \label{defCtrlForcible}$\sigma=\tick$ and $w\sigma_f\in L(S||G)$ for some $\sigma_f\in \SigmaFor$.\hfill
$\blacksquare$
\end{enumerate}

\end{defn}
Property~\eqref{defCtrlStandard} in the above definition is the standard controllability property (when there is no forcible event to preempt \tick); $S$ cannot disable uncontrollable events that $G$ may generate. However, if a forcible event is enabled, this may preempt the time event, which is captured by Property~\eqref{defCtrlForcible}.

%\MF{The formatting of the above def makes it hard to read. It looks as if there is something missing after "and". I also think there is something wrong after the $\exists$, should that part really be subscripted?}
%\AR{Is it better to put \{cases\} also in the text you mean ??}
%\MF{Yes, maybe like this. At least it is easier to parse now.}

A supervisor $S$ is called \emph{proper} for a plant $G$ whenever $S$ is controllable w.r.t.\ $G$, and the supervised plant $S||G$ is nonblocking.

\begin{defn}[Conventional Maximal Permissivenesss]
\label{dfn:MaxPer}
A proper supervisor $S$ is \emph{maximally permissive} for a plant $G$, whenever $S$ preserves the largest behavior of $G$ compared to any other proper supervisor $S'$; for any proper $S'$: $L(S'||G)\subseteq L(S||G)$.
\hfill$\blacksquare$
\end{defn}
%\AR{Shall we provide a formal definition of maximally permissive supervisor here as well?\MF{I think that this is formal enough. But, is it well-defined what $\subseteq$ means for TDES? Typically max perm is defined in terms of language inclusions, is that what is mean here? Is it clear? And BTW... why does it say "acceptable"? Is this meant to refer to nonblockng? I think ``acceptable'' can be removed there.} }

For a TDES, a proper and a maximally permissive supervisor can be synthesized by applying the synthesis algorithm proposed in~\cite{Wonham:19}.

\subsection{Motivating Examples}
%\MFedit{Example \ref{exp:ME1}, Example \ref{exp:ME2}, and Example \ref{exp:ME3}}{Examples~\ref{exp:ME1}, \ref{exp:ME2}, and \ref{exp:ME3}}
%\MF{In the descriptions of examples 1, 2, and 3, I think we need to be more explicit on the conflict between max perm and nb. For instance, in Ex 1:
%•	to be nb, S must not disable a in a1;
%•	to be max perm, S must enable a in a0.
%Currently, only one of those is made explicit. And I think it is the same for all three examples.}
%\AR{Please check if this is solved now}
This section discusses the situations where a proper and maximally permissive conventional supervisor $S$ fails in the presence of observation delay (Example~\ref{exp:ME1}), non-FIFO observation (Example  \ref{exp:ME2}), or control delay (Example \ref{exp:ME3}).

\begin{example}[Observation Delay]
\label{exp:ME1} 
Consider the plant depicted in Figure \ref{fig:ME1P}. 
To be maximally permissive, $S$ must not disable $a$ at $a_0$, and to be nonblocking, $S$ must disable $a$ at $a_2$.
Now, assume that the observation of the events executed in $G$ are not immediately received by $S$ due to observation delay. Starting from $a_0$, imagine that $u$ occurs, and $G$ goes to $a_2$. Since $S$ does not observe $u$ immediately, it supposes that $G$ is still at $a_0$ where it enables $a$. Then, $a$ will be applied at the real state where $G$ is, i.e., $a_2$, and so $G$ goes to $a_3$ which is blocking.
%\MF{Don't we need to say here that for S to be maximally permissive, it must enable a at the state a0? And then, since the observation of u is not immediate, a will remain enabled when the plant has moved to the a2 state until S observes the occurrence of u. Only then can S disable a.}\AR{edited}
%\hfill $\blacksquare$

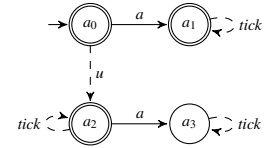
\begin{figure}[htb]
\centering
%\begin{subfigure}[b]{0.4\linewidth}
%\centering
\begin{tikzpicture}[>=stealth',,shorten >=0.8pt,auto,node distance=2.2cm,scale = 0.6, transform shape]

\node[initial,initial text={},state,accepting]           (A)                                    {$a_0$};
\node[state,accepting]         (B) [right of=A]                       {$a_1$};
\node[state,accepting]                   (C) [below of=A]                       {$a_2$};
\node[state]                   (D) [right of=C]                       {$a_3$};

\path[->] (A) edge [above]   node [align=center]  {$a$} (B)
(A) edge [right,dashed]   node [align=center]  {$u$} (C)
(C) edge [above]   node [align=center]  {$a$} (D)
(B) edge [loop right, dashed]      node [align=center]  {$\tick$} (B)
(D) edge [loop right, dashed]      node [align=center]  {$\tick$} (D)
(C) edge [loop left, dashed]      node [align=center]  {$\tick$} (C);
\end{tikzpicture}
\caption{Plant for Example \ref{exp:ME1}.}\label{fig:ME1P}
%\end{subfigure}
%\hfill
%\begin{subfigure}[b]{0.58\linewidth}
%\centering
%\begin{tikzpicture}[>=stealth',,shorten >=0.8pt,auto,node distance=2.4cm,scale = 0.5, transform shape]

%\node[initial,initial text={},state,accepting]           (A)                                    {$s_0$};
%\node[state,accepting]         (B) [right of=A]                       {$s_1$};
%\node[state,accepting]                   (C) [below of=A]                       {$s_2$};
%\node[state]                   (D) [right of=C]                       {$a_3$};

%\path[->] (A) edge [above]   node [align=center]  {$a$} (B)
%(A) edge [right,dashed]   node [align=center]  {$u$} (C)
%(B) edge [loop right,dashed]      node [align=center]  {$\tick$} (B);
%(D) edge [loop right]      node [align=center]  {$\tick$} (D)
%\end{tikzpicture}
%\caption{Conventional Supervisor}\label{fig:ME1S}
%\end{subfigure}
%\caption{Plant and Conventional Supervisor for Example \ref{exp:ME1}.}\label{fig:ME}
\end{figure}
\end{example}

\begin{example}[Non-FIFO Observation]
\label{exp:ME2}
Consider the plant $G$ depicted in Figure \ref{fig:ME2P}.
To be nonblocking, $S$ must disable $a$ at $a_3$, and to be maximally permissive, $S$ must not disable $a$ at $a_6$.
%Note that $a_3$ is not TLF, and so $S$ does not guarantee time-lock freeness.
%\MR{Is it really necessary to discuss or involve TLF here? What about making it TLF trivially. I see that you need it later. I propose not to mention the non-TLF here, but only later where it is used.}
%\MF{Strictly speaking, the above is not true. S has to enable also $a$ in some other states. The text says ``any \emph{other} controllable event'', which would mean $b$ here. I think something about ``other appropriate states'' needs to be added.}
%\MF{I have problems understanding the sentence about $S$. Is it not better to just say that with non-FIFO observation, $S$ is not guaranteed to be able to determine whether $G$ is in $a_3$ or in $a_6$, and then to say that to be nonblcoking $S$ must disable $a$ in $a_3$, but to be max perm it must enable $a$ in $a_6$.}
%\AR{In each example, I am trying to convey the message that if there is delay in control or observation, or observation is non-FIFO, then (how) $S$ fails to guarantee nonblockingness ... 
%\MF{But is that really true? It seems to me that there is a conflict between being nonblocking and being maximally permissive. Due to delays, S cannot be both. Is not this the problem?
%I think it is better to talk in terms of ``must'', like ``to be nonblocking S must disable $a$ in $a_3$'', and ``to be maximally permissive S must enable $a$ in $a_6$'', and due to delays it cannot know whether G is in $a_3$ or $a_6$.}}
Now, assume that the observation channel is non-FIFO, i.e., events may be observed in a different order as they occurred in $G$.
Starting from $a_0$, imagine that $G$ executes $\tick\,a\,b$ and goes to $a_3$.
%$S$ disables the event $a$ after the observation of the word %\MFedit{$\underline{tick}ab$}{$\tick.a.b$}
%$\tick ab$ ($G$ is at state $a_3$).
%\MF{I do not really understand the underline of \tick. Is it not better to use dots as separators?} 
%\MF{The above sentence is awkward. If there is no comm delay, then why is it necessary to point out that the control channel is FIFO? Also, if there is no comm delay, does not observation occur immediately? And I think the word "however" is used in an unconventional way. Typically "however" starts or ends a sentence.}\AR{edited}
Since the observation channel is non-FIFO, $S$ may receive the observation of $\tick\,a\,b$ as $\tick\,b\,a$ after which it does not disable $a$. However, $G$ is actually at $a_3$ and by executing $a$, it goes to $a_4$ which is blocking.
%\MF{Is a4 really a time-lock state? It enables \tick. Do I misunderstand something? It is blocking though, is that not enough?}
%\hfill 
%\hspace*{\fill}
%$\blacksquare$

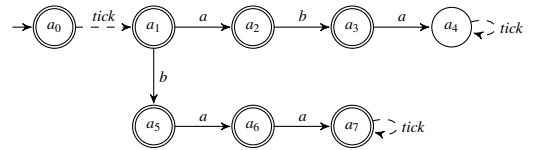
\begin{figure}[htb]
\centering
%\begin{subfigure}[b]{0.4\textwidth}
%\centering
\begin{tikzpicture}[>=stealth',,shorten >=0.8pt,auto,node distance=2.2cm,scale = 0.6, transform shape]

\node[initial,initial text={},state,accepting]           (A)                                    {$a_0$};
\node[state,accepting]         (B) [right of=A]                       {$a_1$};
\node[state,accepting]                   (C) [right of=B]                       {$a_2$};
\node[state,accepting]                   (D) [right of=C]                       {$a_3$};
\node[state]                   (E) [right of=D]                       {$a_4$};

%\node[state]                   (E0) [right of=E]                       {$a_5$};

\node[state,accepting]                   (F) [below of=B]                       {$a_5$};
\node[state,accepting]                   (G) [right of=F]                       {$a_6$};
\node[state,accepting]                   (H) [right of=G]                       {$a_7$};

\path[->] (A) edge [above,dashed]   node [align=center]  {$\tick$} (B)
(B) edge [above]   node [align=center]  {$a$} (C)
(C) edge [above]   node [align=center]  {$b$} (D)
(D) edge [above]      node [align=center]  {$a$} (E)
%(E) edge [above]   node [align=center]  {$a$} (E0)
(E) edge [loop right,dashed]      node [align=center]  {$\tick$} (E)

(B) edge [right]   node [align=center]  {$b$} (F)
(F) edge [above]   node [align=center]  {$a$} (G)
(G) edge [above]      node [align=center]  {$a$} (H)

%(D) edge [loop above,dashed]      node [align=center]  {$\tick$} (D)
(H) edge [loop right,dashed]      node [align=center]  {$\tick$} (H);
\end{tikzpicture}
\caption{Plant for Example \ref{exp:ME2}.}\label{fig:ME2P}
%\end{subfigure}
%\hfill
%\begin{subfigure}[b]{0.4\textwidth}
%\centering
%\begin{tikzpicture}[>=stealth',,shorten >=0.8pt,auto,node distance=2.4cm,scale = 0.5, transform shape]

%\node[initial,initial text={},state,accepting]           (A)                                    {$s_0$};
%\node[state,accepting]         (B) [right of=A]                       {$s_1$};
%\node[state,accepting]                   (C) [right of=B]                       {$s_2$};
%\node[state,accepting]                   (D) [right of=C]                       {$s_3$};

%\node[state,accepting]                   (F) [below of=B]                       {$s_4$};
%\node[state,accepting]                   (G) [right of=F]                       {$s_5$};
%\node[state,accepting]                   (H) [right of=G]                       {$s_6$};
%\node[state,accepting]                   (I) [right of=H]                       {$s_7$};

%\path[->] (A) edge [above]   node [align=center]  {$\tick$} (B)
%(B) edge [above]   node [align=center]  {$a$} (C)
%(C) edge [above]   node [align=center]  {$b$} (D)
%(B) edge [right,dashed]   node [align=center]  {$\tick$} (F)
%(F) edge [above]   node [align=center]  {$b$} (G)
%(G) edge [above]      node [align=center]  {$a$} (H)
%(H) edge [above]      node [align=center]  {$a$} (I)
%(I) edge [loop right,dashed]      node [align=center]  {$\tick$} (I);
%\end{tikzpicture}
%\caption{Conventional Supervisor}\label{fig:ME2S}
%\end{subfigure}
%\caption{Plant and Conventional Supervisor for Example \ref{exp:ME2}.}\label{fig:ME2}
\end{figure}
\end{example}

\begin{example}[Control Delay]
\label{exp:ME3}
Consider the plant depicted in Figure \ref{fig:ME3P}.
To be maximally permissive, $S$ must not disable $a$ at $a_1$, and to be nonblocking $S$ must disable $a$ at $a_3$.
Now, assume that control commands are received by $G$ after %some time delay (integer number of \tick{s}), say 
one \tick. 
Starting from $a_0$, $S$ does not disable $a$ after one \tick (when $G$ is at $a_1$). However, the command is received by $G$ after the passage of one \tick (due to the control delay) when $G$ is at $a_3$. So, by executing $a$ at $a_3$, $G$ goes to $a_4$ which is blocking.
%\MF{I do not agree that it is to satisfy nonblockingness that S must enable a in a1; it is to be maximally permissive. If S disabled a all over, it would be nonblocking. In Ex 2 it was said what S \emph{cannot} do to be nonblocking, that is fine, but that is not the same case as here.}\AR{edited}
%\hfill $\blacksquare$

\begin{figure}[htb]
\centering
%\begin{subfigure}[b]{0.42\linewidth}
%\centering
\begin{tikzpicture}[>=stealth',,shorten >=0.8pt,auto,node distance=2.2cm,scale = 0.6, transform shape]

\node[initial,initial text={},state,accepting]           (A)                                    {$a_0$};
\node[state,accepting]         (B) [right of=A]                       {$a_1$};
\node[state,accepting]                   (C) [right of=B]                       {$a_2$};
\node[state,accepting]                   (D) [below of=B]                       {$a_3$};
\node[state]                   (E) [right of=D]                       {$a_4$};

\path[->] (A) edge [above,dashed]   node [align=center]  {$\tick$} (B)
(B) edge [above]   node [align=center]  {$a$} (C)
(B) edge [right,dashed]   node [align=center]  {$\tick$} (D)
(D) edge [above]      node [align=center]  {$a$} (E)
(D) edge [loop left,dashed]      node [align=center]  {$\tick$} (D)
(E) edge [loop right,dashed]      node [align=center]  {$\tick$} (E)
(C) edge [loop right,dashed]      node [align=center]  {$\tick$} (C);
\end{tikzpicture}
\caption{Plant for Example \ref{exp:ME3}.}\label{fig:ME3P}
%\end{subfigure}
%\hfill
%\begin{subfigure}[b]{0.56\linewidth}
%\centering
%\begin{tikzpicture}[>=stealth',,shorten >=0.8pt,auto,node distance=2.4cm,scale = 0.5, transform shape]
%\node[initial,initial text={},state,accepting]           (A)                                    {$s_0$};
%\node[state,accepting]         (B) [right of=A]                       {$s_1$};
%\node[state,accepting]                   (C) [right of=B]                       {$s_2$};
%\node[state,accepting]                   (D) [below of=B]                       {$s_3$};

%\path[->] (A) edge [above,dashed]   node [align=center]  {$\tick$} (B)
%(B) edge [above]   node [align=center]  {$a$} (C)
%(B) edge [right,dashed]   node [align=center]  {$\tick$} (D)
%(D) edge [loop left]      node [align=center]  {$\tick$} (D)
%(C) edge [loop right,dashed]      node [align=center]  {$\tick$} (C);
%(E) edge [loop right]      node [align=center]  {$\tick$} (E);
%\end{tikzpicture}
%\caption{Conventional Supervisor}\label{fig:ME3S}
%\end{subfigure}
%\caption{Plant and Conventional Supervisor for Example \ref{exp:ME3}.}\label{fig:ME3}
\end{figure}
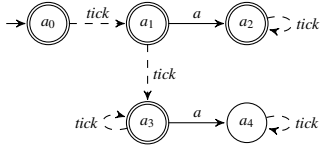
\end{example}

%\MR{Do not see why uncontrollable event $u$ is relevant here.} \AR{Discuss with M.R., what happens to a control command if it reaches the plant, but it executes an uncontrollable event $u$ instead ... will the command remain there or be deleted after the execution of $u$}.
%Although $S$ works perfectly in the conventional setting, it fails in the following cases:
%\begin{itemize}
   % \item suppose that events are not immediately observed by $S$ but after some time (integer number of \emph{tick}s). Then, it could be the case that some word observed by $S$ is not included in $L(S)$ at all, and hence it cannot make a decision for what it observes. For instance, if the word $tuatt \in L(G)$ occurs and events are observed after the passage of one \emph{tick}, then $S$ will actually receive the observation as $ttuat\notin L(S)$.
  %  \item suppose that there is no delay, but consecutive events are observed by $S$ in a different order as they have been executed in $G$. Then, for the same reason, it is possible that a word observed by $S$ is not included in its language. For instance, consider the word $tutua\in L(G)$ being observed as $tutau\notin L(S)$.
  %  \item suppose that control commands sent by $S$ will arrive after some time (integer number of \emph{tick}s) let say one \emph{tick}. Then, it is possible that although $S$ enables/allows $a$ at state $a_2$, the command is received at state $a_4$ leading $G$ to $a_6$.
%\end{itemize}

\begin{remark}
Conventional supervisory control synthesis of a TDES guarantees nonblockingness~\cite{Wonham:19}. %\AR{I think now it is clear what we mean by conventional, right?\MF{In the intro we sort of defined it to mean SCT with the assumption of synchronicity, right?}}
However, as can be seen in Example~\ref{exp:ME2}, it cannot guarantee time-lock freeness; $a_3$ is not TLF, and it is not removed by $S$. This is not an issue in~\cite{Wonham:19} since a TDES is assumed to satisfy the ALF condition. Here, to guarantee time progress, the TLF property must be considered in synthesis.
%\MF{Again, "we" is used as a general term meaning "the authors, the readers, anyone really". I think that should be avoided, "We" should always mean "we, the authors".}
\end{remark}
%\AR{@Martin, is it ok if we do not talk about TLF in the Example itself and just giving information here as you see in ()?}

As \MFedit{it}{}is clear from the examples, a supervisor is required that can deal with the problems caused by communication delays and disordered observations. To achieve such a supervisor, first, the networked supervisory control framework is established.
%\MF{"network-based communication setting"? Previously, and below, this is called "networked supervisory control framework", which I think is a much more to the point. Also, the sentence above is not really proper English.}
%In order to achieve a supervisor satisfying both nonblockingness and time-lock freeness for TDES, we consider the TLF condition in the synthesis algorithm. In this case, time-lock states are also considered as bad states to be removed by the supervisor.
%\AR{Should we provide the synthesis algorithm here as well or describe it in words? "In this case, the set of bad states that needs to be prevented by the supervisory includes the blocking states, time-lock states, and the states leading to bad states only through uncontrollable events. The synthesis is an iterative algorithm, and the set of bad states are computed at the end of each iteration. Note that the event \emph{tick} is uncontrollable at any state unless there exists a forcible event enabled at the state. In case that both the forcible event and \emph{tick} are disabled through synthesis, then the state will be considered a bad state."}

\subsection{NSC Framework}
\begin{figure}[htbp]
\centering
\includegraphics[width=55mm]{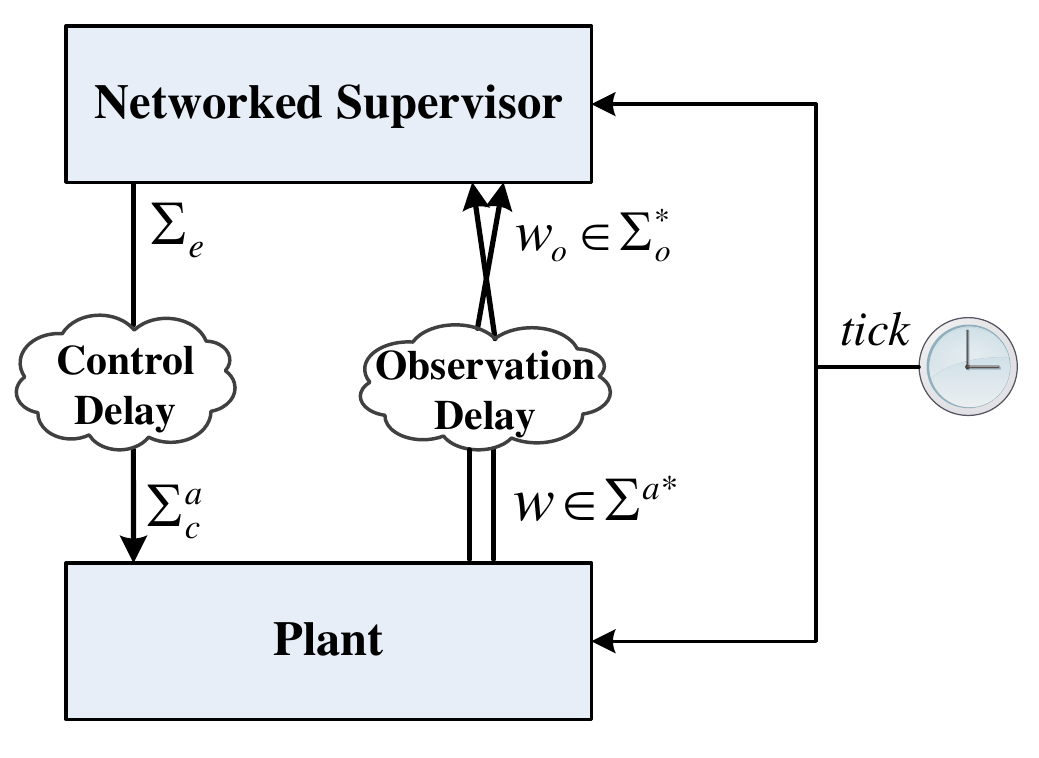}
\caption{NSC framework~\cite{Rashidinejad18}.}
\label{fig:NSCFW}
\end{figure}

In the presence of delays in the control and observation channels, enabling, executing and observing events do not happen at the same time. Figure \ref{fig:NSCFW} depicts the networked supervisory control (NSC) framework that is introduced in this paper.
To recognize the differences between the enablement and observation of events and their execution in the plant, as in \cite{{Rashidinejad18,Rashidinejad:19}}, a set of \emph{enabling events} $\Sigma_e$ and a set of \emph{observed events} $\Sigma_o$ are introduced. 

%Moreover, we assume that there is a global digital clock in the system measuring time. The occurrence of each event is then related to the number of \emph{tick} of the clock. \MR{Last sentence I do not understand.} \MR{It is unclear what you mean by occurrence of an event? I also do not understand how the number of ticks is established? I would simply remove the sentence or state it differently.}

\begin{defn}[Enabling and Observed Events]\label{dfn:Sigmao&Sigmae}
Given a plant $G$, 
to each controllable active event $\sigma\in\Sigma_c$ an enabling event $\sigma_{e}\in\Sigma_{e}$, and to each active event $\sigma\in\Sigma_a$ an observed event $\sigma_o\in\Sigma_o$ are associated such that $\Sigma_e\cap\Sigma_a=\varnothing$ and $\Sigma_o\cap\Sigma_a=\varnothing$ (clearly $\Sigma_e\cap\Sigma_o=\varnothing$).
%\MR{Why are we using subscripts everywhere, but not for active event set? Maybe because we also use $\Sigma^{a}_{c}$? Is it defined?}
%\AR{No, we do not use $\Sigma_a_c$ anymore ...}
%\MF{Surely we must say that the respective alphabets are disjoint, no?}
\hfill $\blacksquare$
\end{defn}
%\MF{Just double checking here... $\Sigma_e\cap\Sigma_o\cap\Sigma_a=\varnothing$ allows them to pair-wise overlap, as long as all thre sets do not share any event. Is that the intent? }

Note that all events executed in the plant are supposed to be observable so that the observed event $\sigma_o$ is associated to any $\sigma\in \Sigma_a$.
However, not all the events are supposed to be controllable. Uncontrollable events such as disturbances or faults occur in the plant spontaneously.
In this regard, enabling events $\sigma_e$ are associated only to events from $\Sigma_c$.
%\MF{Why "channels", plural? There is only one observation channel, or? In addition, the event that occurs in the plant and the observed event are not really the same event, right? So, strictly speaking, "events executed in the plant" are not "sent through communication channels", but their corresponding "observed events" are; not? And if so, probably this last sentence comes in more naturally \emph{after} the definition.}\AR{communication channels was referring to observation and control channels}

Considering Figure \ref{fig:NSCFW}, a networked supervisor for $G$ that fits in the proposed framework is a TDES given as:
\begin{equation*}
\label{eq:NS}
\NS = (Y, \SigmaNS, \deltaNS, y_{0}, Y_{m}),
\end{equation*}
for which the event set $\SigmaNS=\Sigma_e\cup\Sigma_o\cup\{\tick\}$, and the event \tick is produced by the global clock in the system so that $\SigmaNS\cap\Sigma_G=\{\tick\}$.

For the proposed NSC framework, the behavior of the plant under the control of a networked supervisor is achieved through \emph{asynchronous composition}.
To define asynchronous composition, we first need to consider the effects of delays on events sent through the control and observation channels. In this paper, it is assumed that the control (observation) channel has a finite capacity denoted by $\Lmax$ ($\Mmax$), which introduces a constant amount of delay represented by a natural number $N_c$ ($N_o$).
Since the control channel is supposed to be FIFO, a list or sequence is used to consider the journey of events through the control channel. As given in Definition \ref{dfn:list} below, $l\in (\Sigma_c \times [0,N_c])^*$ provides us with the current situation of the control channel.
%\AR{I used $\Sigma_c $ in the definition of $l$ instead of $\Sigma$ as we had before}
The interpretation of $l[i]=(\sigma,n)$ is that the $i^{th}$ enabling event present in the control channel is $\sigma_e$ which still requires $n$ \ticks before being received by the plant.
%Then, the corresponding (active controllable) event $\sigma$ can be executed in the plant.

%time that a command (indicated by an enabling event) still needs to reach the plant.
%\MF{There is an issue here similar to what I mentioned around line 577. The event that is executed is not the event that is sent; what is sent is the \emph{enabling} event (and it should be called exactly that), but what is (possibly but not necessarily?) executed is the corresponding (active controllable) event. Right?}
%\AR{please check if the text is improved regarding this comment}

\begin{defn}[Control Channel Representation]
\label{dfn:list}
The control channel is represented by the set $L=(\Sigma_c \times [0,N_c])^*$.
Moreover, we define the following operations for all $\sigma \in \Sigma_c$, the time counter $n \in [0,N_c]$ and $l \in L$:
\begin{itemize}
\item $\emptyseq$ denotes the empty sequence.
  \item $\mathit{app}(l,(\sigma,n))$ adds the element $(\sigma,n)$ to the end of $l$ if $|l|<\Lmax$ (the channel is not full), otherwise $l$ stays the same.
  %\MR{app is hardly used. Perhaps remove it and write $(\sigma,n)~l$ instead (as we do in the next item)?} \AR{$l~(\sigma,n)$? \MR{Sure, but do we need to introduce it? We use it all the time for words (which are really similar things).}} \AR{If we do not introduce a function, then, how can we say that the new element cannot be added if the channel is full? I mean if we do not want to have it in the operators definitions but here ...}
  %Often this operation is simply written using juxtaposition, i.e., $l~ (\sigma,n)$.
  %Formally, it denotes the list $l'$ with $l'(i) = l(i)$ for $0\le i \leq \text{length}(l)-1$ and $l'(k)= (\sigma,n)$ for $k = \text{length}(l)$.
  \item $\head(l)$ gives the first element of $l$ (for nonempty lists). Formally, $\head((\sigma,n)~l) = (\sigma,n)$ and $\head(\emptyseq)$ is undefined.
  \item $\mathit{tail}(l)$ denotes the list after removal of its leftmost element. Formally, $\mathit{tail}((\sigma,n)~l) = l$ and $\mathit{tail}(\emptyseq)$ is undefined. 
  %, i.e., the list $l'$ with $l'(i) = l(i+1)$ for $0 \leq i \leq \text{length}(l)-1$.
  \item $l-1$ decreases the natural number component of every element in $l$ by one (if possible).
  %Formally, it denotes the function $l'$ for which $(\sigma,n)\in l'\iff (\sigma,n+1)\in l$ for all $n < N_c$, $(\sigma,N_c)\notin l'$
  %\AR{@Michel, we already considered the removal of $(\sigma,0)$ in the above definition, didn't we? I think we discussed this before, right? \MR{Right}}
  It is defined inductively as follows $\emptyseq -1 = \emptyseq$, $((\sigma,0) ~l) -1 = ~l-1$, and $((\sigma,n+1)~l) -1 = (\sigma,n)~(l-1)$.
  \hfill $\blacksquare$
%\item $(\sigma,n) \in l$ denotes $\head(l) = (\sigma,n)$. 
\end{itemize}
\end{defn}

Due to the assumption that the observation channel is non-FIFO, we use a multiset to consider the journey of each event through the observation channel. As given in Definition \ref{dfn:medium} below, the multiset $m:\Sigma_a \times [0,N_o]\rightarrow \mathbb{N}$
%\AR{Similarly, $\Sigma_a$ instead of $\Sigma$ for $m$}
provides us with the current situation of the observation channel.
%provides us with the time that each event executed in the plant still needs in the channel to be observed by the supervisor. 
The interpretation of $m(\sigma,n) = k$ is that currently there are $k$ events $\sigma$ in the observation channel that still require $n$ \ticks before reaching the (networked) supervisor.
%\MF{"the time that each event executed in the plant still needs in the channel to be observed by the supervisor". Should that be "each event  {\bf that has been} executed in the plant"? And then I also wonder, given $m$, for a specific event $\sigma'$ that has been executed in the plant, how do I use $m(\sigma', n) = k$ to calculate the time that $\sigma'$ still needs in the channel before being observed by $S$? $m$ does not really provide with the time, it provides us with number of events with a specific property. If I want to know the time (or number of ticks) for a certain event $\sigma'$, do I have to call $m(\sigma',0)=k, m(\sigma',1)=k, \dots$, until I get $k\neq 0$?}
%\AR{That's right. I suggest to remove the sentence as the sentence comes right after it is describing the same thing in other words \MF{Fine.} }
%\AR{As given in Definition \ref{dfn:medium} below, for each $(\sigma,n)\in\Sigma\times [0,N_o]$, the multiset \MFedit{$m$}{$m:\Sigma \times [0,N_o]\rightarrow \mathbb{N}$} provides us with the number of events $\sigma$ that have been executed in the plant and they are still in the observation channel to be observed after the passage of $n$ \ticks.}

\begin{defn}[Observation Channel Representation]
\label{dfn:medium}
The observation channel is represented by the set 
$M=\{m\, |\, m:\Sigma_a \times [0,N_o]\rightarrow \mathbb{N}\}$.
Moreover, we define the following operations for all $m\in M$, $\sigma,\sigma' \in \Sigma_a$ and the time counters $n,n' \in [0,N_o]$:
\begin{itemize}

%\item $m(\sigma,n)$ denotes the number of elements $(\sigma,n)$ in $m$. \MR{Actually this is not a new definition, but an interpretation of the above. Also, it is already explained above in the text.}
  \item $[]$ denotes the empty multiset, i.e., the function $m$ with $m(\sigma,n)=0$. 
\item $|m|=\sum_{(\sigma,n)\in\Sigma_a\times[0,N_o]}m(\sigma,n)$ denotes the number of events in the observation channel represented by $m$.
%\MR{Replace $\mathbb{N}$ here by $[0,N_o]$?}
%\MF{"number of elements in $m$"? Is $m$ a set here now? And should it not be $(\sigma,n)\in\Sigma\times\mathbb{N_o}$? \MR{$m$ is a function (representing a multi-set, i.e., a set in which element may occur multiple times. The $|...|$ is introduced to denote the number of events in the channel (and corresponds intuitively with the number of elements in the multiset). Would it be clearer to state that $|m|$ is the number of events in the channel represented by $m$ (avoiding the use of the word set)?} Yes, I think it is better now.}
  \item $m\uplus [(\sigma,n)]$ inserts $(\sigma,n)$ to $m$ if $|m|<\Mmax$ (the observation channel is not full). Formally, it denotes the function $m'$ for which $m'(\sigma,n)=m(\sigma,n)+1$ and $m'(\sigma',n') = m(\sigma',n')$ otherwise. If $|m|=\Mmax$ (the observation channel is full), then the channel stays the same, i.e., $m'=m$.
%  \MF{There are two typos above here: ``$|m|<\Sigmauc$'' and ``$|m|\geq $''. I am not sure what $|m|$ really should be compared to, though ($N_o$?), so I do not dare to edit it :-) And I think this could be reformulated to only mention either ``channel is full'' or ``channel is not empty'', but not both. }
  \item $m\setminus [(\sigma,n)]$ removes $(\sigma,n)$ from $m$ once. Formally, it denotes the function $m'$ for which $m'(\sigma,n)=\max(m(\sigma,n)-1,0)$ and $m'(\sigma',n') = m(\sigma',n')$ otherwise. 
  \item $m-1$ decreases the natural number component of every element by one (as long as it is positive). Formally, it denotes the function $m'$ for which $m'(\sigma,n) = m(\sigma,n+1)$ for all $n < N_o$ and $m'(\sigma,N_o) = 0$.
  %%\MF{Should this really be $n \neq N_o$? Should it not be $n\neq 0$? Hm... maybe not. But still, what is "the natural number component"? Is it $n$ or $k$? }
\item $(\sigma,n) \in m$ denotes that the pair $(\sigma,n)$ is present in $m$, it holds if $m(\sigma,n) > 0$. \hfill $\blacksquare$
\end{itemize}
\end{defn}
In the rest of the paper, a networked supervisor for the plant $G$ is given as the TDES $\NS$ represented by the automaton $(Y,\SigmaNS,\deltaNS,y_0,Y_m)$.
%\end{remark}

Considering the representation of control and observation channels, an asynchronous composition operator is defined to achieve a networked supervised plant.

%\AR{What about this one? Given deterministic $\NS$ and $G$, for any $w,w'\in L(\NSP)$, if $P_{\Sigma_G}(w)=P_{\Sigma_G}(w')$ then $P_{\SigmaNS}(w)=P_{\SigmaNS}(w')$. We may need this property in the proof of timed networked controllability ...}

%\AR{A proper supervisor never enlarges the behavior of a plant, and so the behavior of the supervised plant is always a subset or equal to the behavior of the plant. This property should also hold for a proper networked supervisor. In this case, we need to guarantee that a networked supervised plant achieved from the asynchronous composition operator has this property as well.}\MR{vague}\AR{I should discuss this with M.R.}  We will formalize and prove this as property for the proposed operator.

\begin{defn}[Timed Asynchronous Composition Operator]
\label{dfn:operator}
Given a plant $G$ and a networked supervisor $\NS$ (for $G$),
%through non-FIFO observation channel with maximum capacity $\Mmax$ and FIFO control channel with maximum capacity $\Lmax$ introducing constant delays $N_o$ and $N_c$, respectively.
the asynchronous product of $G$ and $\NS$, denoted by $\NS_{N_c}\|_{N_o}\,G$, is given by the automaton
\begin{equation*}
\NS_{N_c}\|_{N_o}\,G = (Z, \SigmaNSP, \deltaNSP, z_{0}, Z_{m}),
\end{equation*}
where
\begin{equation*}
\begin{aligned}
&Z= A\times Y\times M\times L,\qquad
\SigmaNSP= \SigmaNS\cup\Sigma,\\
&z_{0}=(a_{0},y_0,[],\emptyseq),\hspace{1.30cm}
Z_{m}=A_{m}\times Y_{m}\times M\times L.
\end{aligned}
\end{equation*}

Moreover, for $a\in A$, $y\in Y$, $m\in M$, and $l\in L$, $\deltaNSP:Z\times\SigmaNSP\rightarrow Z$ is defined as follows:
\begin{enumerate}
\item When an event $\sigma_e \in \Sigma_e$ occurs in $\NS$, it is sent through the control channel. This is represented by adding $(\sigma,N_c)$ to $l$ where $N_c$ is the remaining time for $\sigma_e$ until being received by $G$. If $\deltaNS(y,\sigma_e)!$:
\begin{equation*}
\deltaNSP((a,y,m,l),\sigma_{e})=(a,\deltaNS(y,\sigma_e),m,\mathit{app}(l,(\sigma,N_c))).
%\textit{app}
\end{equation*}
%\begin{multline*}
%\deltaNSP((a,v,m,l),\sigma_{e})=\\(a,\deltaNS(v,\sigma_e),m,\textit{app}(l,(\sigma,N_c))).
%\end{multline*}
%where $\textit{app}(l,(\sigma,N_c))$ adds the element $(\sigma,N_c)$ to the end of $l$ to keep the ordering.
%\MF{Should the last $\sigma$ above be $\sigma_e$? The text seems to say that.}
%\AR{$(\sigma,N_c)$ will be stored in $l$ to show that the corresponding command $\sigma_e$ has been sent for the event $\sigma$, and so it can be executed in the plant wherever it is enabled after $N_c$ ticks}
%\MF{But the text says "When an event $\sigma_e \in \Sigma_e$ is enabled by $\NS$, it will be stored in $l$". In the comment you are saying that "$(\sigma,N_c)$ will be stored in $l$". That's different. The text should probably say something like "When an event $\sigma_e \in \Sigma_e$ is enabled by $\NS$, $\sigma$ will be stored in $l$".}
%\MF{And is $(\sigma,N_c)$ really stored in $l$ when it is \emph{enabled}? Is it not when it occurs?}

\item An active controllable event $\sigma\in \Sigma_c$ can occur if the plant enables it, and the corresponding control command (enabling event) is received by the plant as $(\sigma,0)$ (as the enabling event finished its journey through the control channel). When $\sigma$ occurs, it will be stored in $m$ with the remaining time $N_o$ until being observed by $\NS$. If $\delta_G(a,\sigma)!$ and $\head(l)=(\sigma,0)$:
\begin{equation*}
\deltaNSP((a,y,m,l),\sigma)=(\delta_G(a,\sigma),y,m\uplus [(\sigma,N_{o})],\mathit{tail}(l)).
\end{equation*}
%\MF{"In addition"? Don't you mean something like "when it occurs"?}

\item An uncontrollable event $\sigma \in \Sigmauc$ can occur if it is enabled in $G$. When $\sigma$ occurs, it will be stored in $m$ with the remaining time $N_o$ until being observed by $\NS$. If $\delta_G(a,\sigma)!$:
\begin{equation*}
\deltaNSP((a,y,m,l),\sigma)=(\delta_G(a,\sigma),y,m\uplus [(\sigma,N_{o})],l).
\end{equation*}
%\MF{"In addition"? Same as above...? And should it not be "after at most $N_o$ \ticks"?}
%\AR{\item Event \emph{tick}, can be executed in $\mathit{NSP}$ if it is enabled at a state of $G$ and it cannot be preempted because there is no event ready to be observed by $\NS$ and also there is no forcible event enabled at that state.}

\item Event \tick can occur if both $\NS$ and $G$ enable it, and there is no event ready to be observed by $\NS$.
Upon the execution of \tick, all the time counters in $m$ and $l$ are decreased by one. 
%\MF{Here is the first time that the n-elements are called "time counters". I think this is crucial info that should be given much earlier, up where the channels are defined.}
If $\delta_G(a,\tick)!$, $\deltaNS(y,\tick)!$, $(\sigma,0)\notin m$ for all $\sigma\in \Sigma_a$
%\MF{For some reason I cannot help to wonder what happens if this does not hold for all $\sigma \in \Sigma_a$...}\AR{(if there exists some $(\sigma,0)\in m$, then \tick can occur only after the execution of $\sigma_o$ in item 5))}
\begin{multline*}
\deltaNSP((a,y,m,l),\tick)=\\
(\delta_G(a,\tick),\deltaNS(y,\tick),m-1,l-1).
\end{multline*}
%\AR{Is it the right condition for \emph{tick} execution? Also, now if $\sigma_e$ is expired, there will be $(\sigma,\perp)\in l$}
%\AR{$(\sigma,0)\in l$ as a condition for enabling \tick cannot be right ... I think the condition should be changed to $\nexists \sigma\in\Sigma_e\cup\Sigma_o, \deltaNS(v,\sigma)!$ ... }
%Moreover, if there is an event ready to be received by $G$, it will be removed from $l$ since the time of its execution has been expired. \MR{Why do you want this change with removing events from the control channel? What if the plant alows the event in the control channel and a tick? What does it then mean to enable a forcible event (if it may be skipped by the plant)?}
%\AR{what if time goes out for a controllable event and it is not executed, shall we remove it from $l$? we can say whenever tick is executed and there is $(\sigma,0)\in l$. It will be removed? for observation if there is $(\sigma,0)\in m$, then tick is disabled (sounds like $\Sigma_o$ events are uncontrollable forcible events which always preempt \emph{tick}?)}

%\AR{What about uncontrollable \emph{tick}, shall we separate it?}

\item The observation of an active event $\sigma \in \Sigma_a$ can occur when it finishes its journey through the observation channel (and so it is received by $\NS$), and $\sigma_o$ is enabled by $\NS$. When $\sigma_o$ occurs, $(\sigma,0)$ is removed from $m$. If $\deltaNS(y,\sigma_o)!$ and $(\sigma,0)\in m$:
\begin{equation*}
\deltaNSP((a,y,m,l),\sigma_{o})=(a,\deltaNS(y,\sigma_o),m\setminus[(\sigma,0)],l). \hspace*{0.55cm} \blacksquare
\end{equation*}
%\MR{Should we consider the observations as uncontrollable? In some sense, the supervisor should simply deal with these and not restrict them?}
%\AR{We assume that they are uncontrollable when we do synthesis ... so we start talking about them in synthesis section! Do you mean removing the condition: $(\sigma,0)\in m$?}
\end{enumerate}
\end{defn}

In the rest of the paper, the asynchronous composition $\NS_{N_c}\|_{N_o}\,G$ of the plant $G$ and the networked supervisor $\NS$ (for that plant) is assumed to be the TDES $\NSP$ represented by the automaton $(Z, \SigmaNSP, \deltaNSP, z_{0}, Z_{m})$.
%\end{remark}

%\AR{Here, I really prefer to call Property \ref{property:NSP&P} a property, and Property \ref{property:NSP} a Lemma since the first one is something important that should be held for any operator that anyone presents. But, the second is just a minor property that (only) will help us in the proof ... can I name anything that we introduce to only help us in the proofs as Lemma?}
%\MF{Yes. There is really no universal terminology for this, so using it your own way is fine.}
%\MR{I do not like the name of the lemma}
%\AR{I will either remove/change them at the end}

%\begin{lemma}[\NSP Transitions]
%\label{lemma:NSP}

Note that the networked supervised plant models the behavior of a plant controlled by a networked supervisor, and so for the proposed operator, we need to prove that the result does not enlarge the behavior of 
the plant.
%This is formulated as Property \ref{property:NSP&P}.

\begin{property}[\NSP and Plant]
\label{property:NSP&P}
Given a plant $G$ and networked supervisor $\NS$ (for that plant):
 $P_{\Sigma_G}(L(\mathit{\NSP})) \subseteq L(G)$.
%\hfill$\blacksquare$
\end{property}

\begin{proof}
See Appendix \ref{proof:NSP&P}.
\hfill $\blacksquare$
\end{proof}

%As previously stated, we assume that some events of the plant are uncontrollable while all events are observable.
%Hence, a networked supervisor is acceptable for which we can guarantee the controllability of the networked supervised plant.

%\MF{What does the above sentence mean, "acceptable", in what way? Also, the word order is strange. Is not the next sentence enough? And why is observability mentioned here? Is it not better to only deal with controllability here (this would remove also the confusing mentioned in your comments)}
A networked supervisor is controllable with respect to a plant if it never disables any uncontrollable event that can be executed by the plant.
%\AR{Furthermore, at every state, $\NS$ sends commands (make decisions) for words in its time-window (as defined in Definition \ref{dfn:TW}). Therefore, we can say that a networked supervisor is observable if whenever it cannot differentiate between any two words in its time-window, then it applies the same control command to them.}
To have a formal representation of controllability in the NSC framework, Definition \ref{dfn:cont.TDES} is adapted to \emph{timed networked controllability}.
%We also provide a formal definition of safety which guarantees that the behavior of the networked supervised plant is restricted to the desired behavior given as safety(control) requirements.
%\MR{I think we should mention the small difference with normal controllability, as provided previously in the paper (only the composition).}
%\AR{Definition \ref{dfn:NScont} is changed}
%\MR{Still the differences are not properly explained.}
\begin{comment} 
\begin{defn}
[Timed Networked Controllability]
\label{dfn:NScont}
\AR{I am still wondering if this is the right definition ... $wu\in P_{\Sigma_G}(L(\NSP))$ only guarantees that there exists a word in $\NSP$ with the same projection on $\Sigma$ followed by $u$?}
Consider a plant $G$ and a networked supervisor $\NS$ for that plant with control and observation delays $N_c$ and $N_o$, respectively.
Then, \NSP is timed networked controllable for $G$ if for any $w\in P_{\Sigma_G}(L(\NSP))$ and 
\begin{equation*}
u\in \Sigmauc\cup \begin{cases}
       \varnothing &\quad\text{if}\quad \exists_{\sigma\in\SigmaFor}~w\sigma\in P_{\Sigma_G}(L(\NSP))  \\
       \{tick\} &\quad\text{otherwise,}
\end{cases}
\end{equation*} 
whenever $wu\in L(G)$, then $wu\in P_{\Sigma_G}(L(\NSP))$.
\hfill$\blacksquare$
\end{defn}
\end{comment} 

\begin{defn}[Timed Networked Controllability]
\label{dfn:NScont}
Given a plant $G$ with uncontrollable events $\Sigmauc$ and forcible events $\SigmaFor$, a networked supervisor $\NS$, is controllable w.r.t.\ $G$ if for all
$w\in L(\NSP)$ and $\sigma\in \Sigmauc\cup\{tick\}$, whenever $P_{\Sigma_G}(w)\sigma\in L(G)$: 
\begin{enumerate}
    \item \label{defCtrlStandard}$w\sigma\in L(\NSP)$ , or
    \item \label{defCtrlForcible}$\sigma=\tick$ and $w\sigma_f\in L(\NSP)$ for some $\sigma_f\in \hatSigmaFor\cup\Sigma_o$, where $\hatSigmaFor=\SigmaFor\cup\Sigma_e$.\hfill
$\blacksquare$
\end{enumerate}
\end{defn}
When there is no network, i.e., $\SigmaNS=\Sigma_G$, timed networked controllability coincides with  conventional controllability for TDES (Definition \ref{dfn:cont.TDES}).

\begin{remark}
Considering Definition \ref{dfn:operator}, \tick does not occur if there is an event ready to be observed ($(\sigma,0)\in m$).
In other words, observed events always preempt \tick since they occur once they finish their journey in the observation channel.
The enabling events are assumed to be forcible as well.
This gives the opportunity to the networked supervisor to preempt \tick by enabling an event whenever it is necessary.
In Section \ref{section:PV}, we discuss other possible cases.
%The commands sent by the networked supervisor reach the plant after some control delay. If time passes while the command is not sent, then it can never reach the plant on time.
%To empower the networked supervisor to send the control commands soon enough to reach the plant on time, the enabling events are assumed to be forcible.
%\MR{Why are the $\Sigma_o$ events observable?}
%\AR{Because those are what actually $\NS$ observes? Do you mean forcible? because they always preempt \tick as we consider this in our definition of asynchronous composition.}
%\MF{This text needs to be rewritten}
%\AR{note that this is the definition of controllability w.r.t. the plant. the observed events as well as the plant events are uncontrollable events that just occur in NSP, and NS cannot disable them ...}
\end{remark}

A networked supervisor $NS$ is called proper in NSC framework if is timed networked controllable, nonblocking, and TLF. Similar to controllability, the definition of maximal permissiveness (in the conventional setting) is adapted to \emph{timed networked maximal permissiveness} (for NSC Framework).

%\AR{please check the following definition \MR{The label says timed networked max. perm, but you actually define networked max.perm. Confusing!}}

\begin{defn}[Timed Networked Maximal Permissiveness]
A proper networked supervisor $NS$ is timed networked maximally permissive for a plant $G$, if for any other proper networked supervisor $NS'$ in the same NSC framework (with event set $\SigmaNS$): $P_{\Sigma_G}(L(\NS'_{N_c}\|_{N_o}\,G))\subseteq P_{\Sigma_G}(L(\NSP)$. In other words, $\NS$ preserves the largest admissible behavior of $G$.
\hfill$\blacksquare$
\end{defn}
Again, when there is no network, this notion coincides with conventional maximal permissiveness (Definition \ref{dfn:MaxPer}).

%\AR{I am thinking if we can say for any $NS'$ with event set $\SigmaNS$ and $L(\NS')\nsubseteq P_{\SigmaNS}(L(\NP))$, it cannot be proper (causes blocking) for sure ...}
%\AR{we only compare $\NS$ with any other proper $\NS'$ such that $L(\NS')\subseteq P_{\SigmaNS}(L(\NP))$ because otherwise $\NS'$ may only add extra (same) enabling events which are not executed in the plant and so do not change $P_{\Sigma}(L(NS'P))$ ... (if $L(\NS')\subseteq P_{\SigmaNS}(L(\NP))$ with extra \tick or $\Sigma_o$ transitions, then $\NS'$ is not proper)}
%\MR{If I compare this to the conventional setting, there is no such restriction. If the S' would allow more behaviour, the composition with the plant will simply guarantee that it does not occur. Why is the situation different here? I think t is because the moment that behaviour is filtered out by the plant is later in time, when the supervisor already enabled the events. This will most likely lead to blocking?}

%\AR{It seems that we need the condition because 1. Lemma 7 cannot be relaxed to all cases of $NS$ without the condition $L(\NS)\subseteq P_{\SigmaNS}(L(\NP))$; counter example: $L(G)=\{tat*\}$ and $L(NS')=\{a_e a_e t t a_o\}$. Then, $L(NS'G)\neq L(NS'||NP)$. 2. maximally permissiveness cannot be proved without Lemma 7 ...}

\subsection{Problem Formulation}

The \emph{Basic NSC Problem} is defined as follows. Given a plant model $G$ as a TDES, observation (control) channel with delay $N_o$ ($N_c$) and maximum capacity $\Mmax$ ($\Lmax$), provide a networked supervisor $\NS$ such that 
%\MF{Suggestion: Write the networked synch like \NSP. I really think that this aids readability. And I would create a macro for this, to aid in writeability; see the "newcommand" on line 914. And maybe $N_o$ should be on the left, depends your ide of what is (most) logical.}

\begin{itemize}
\item \NSP is nonblocking,
\item \NSP is time-lock free
\item \NS is timed networked controllable for $G$, and
\item \NS is timed networked maximally permissive.
\end{itemize}

%\AR{People have worked on delay observability and delay controllability to find a deterministic NS! should we mention anything about determinism? our approach gives a deterministic $OP,S,NS, NSP$}
%\AR{and $\NSP$ is deterministic? (It can be proved for our method based on the fact that $S$ makes observable decisions for $OP$)}
%Definitions of controllability, observability, and safety are given in Definition \ref{dfn:cont,obs,safety}.
%\AR{NSC Controllability? People have already introduced the term networked controllability/observability ...}
%The definition of timed networked controllability is given in Definition \ref{dfn:NScont}. The definition of nonblockingness is general and remains the same for any automaton in any setting. \MR{Remove the last (and possibly last two) sentences?}

\section{Networked Supervisory Control Synthesis}
\label{section:synthesis}
\begin{figure}[h]
\centering
\includegraphics[width=60mm]{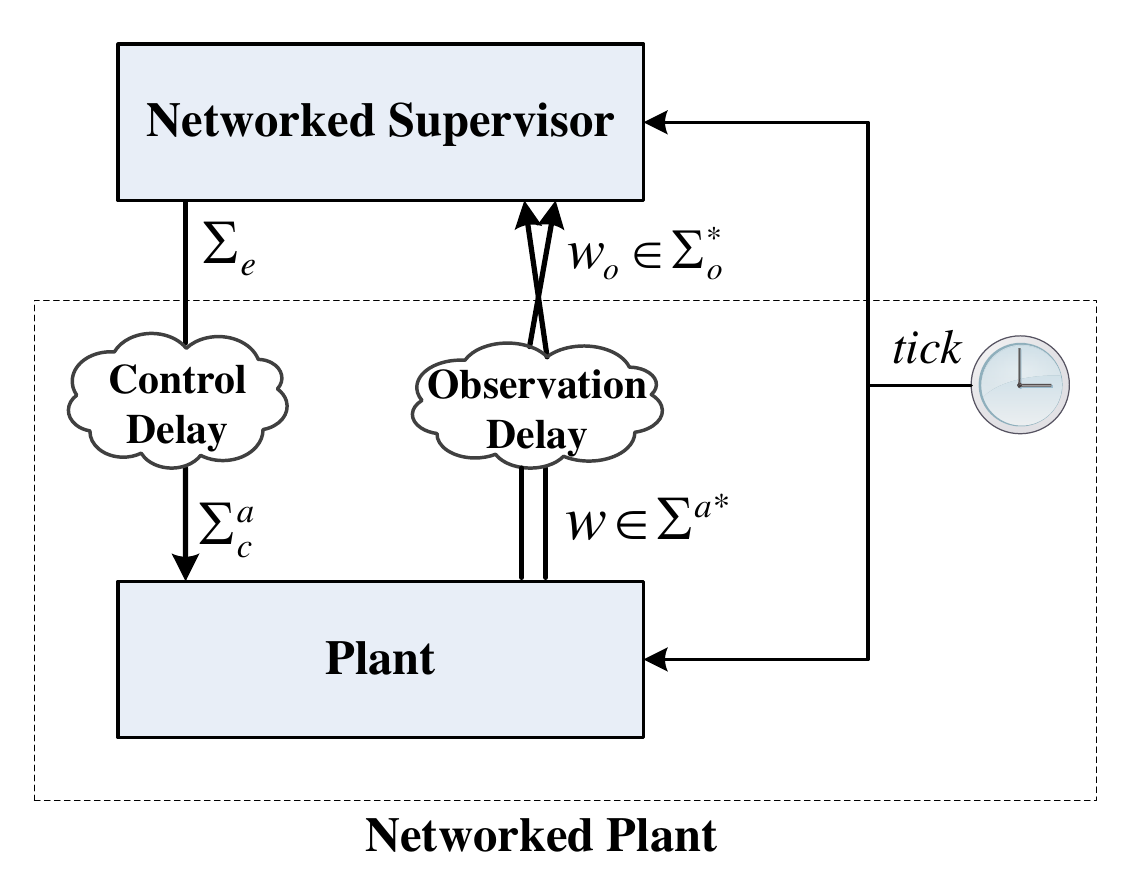}
\caption{Networked plant.}
    \label{fig:solution1}
\end{figure}

To achieve a proper and maximally permissive networked supervisor (in the NSC framework), the synthesis is applied on the ``networked plant", as indicated in Figure \ref{fig:solution1}.
The networked plant is a model for how events are executed in the plant according to the enabling events, and how the observations of the executed events may occur in a networked supervisory control setting.
Based on the networked plant, a synthesis algorithm is proposed to obtain a networked supervisor, which is a solution to the basic NSC problem.
Example \ref{exp:BusPed} is used to illustrate each step of the approach.

\begin{example} (Endangered Pedestrian)
\label{exp:BusPed}
Let us consider the endangered pedestrian example from~\cite{Wonham:19}. 
The plant $G$ is depicted in Figure \ref{fig:BusPed}. Both the bus and pedestrian are supposed to do single transitions indicated by \emph{p} for passing and \emph{j} for jumping. The requirement considered in~\cite{Wonham:19} is that the pedestrian should jump before the bus passes. However, since we do not consider requirements here (yet), we adapt the plant from~\cite{Wonham:19}  such that if the bus passes before the pedestrian jumps, then $G$ goes to a blocking state.
The control channel is FIFO, the observation channel is non-FIFO, $N_c=N_o=1$, $\Lmax=1$, and $\Mmax=2$. We aim to synthesize a proper and maximally permissive networked supervisor for $G$.
%\hfill $\blacksquare$
%\MR{What would happen if, in violation of Assumption 1, there is also a $j$ from $a_0$? Perhaps good example to discuss removing Assumption 1?}
%\AR{Yes, since $j$ is controllable, it does not occur in $\NP$ at all! I will check if any problem occurs if we do not have this assumption!}

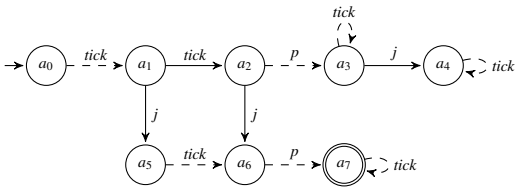
\begin{figure}[htbp]
\centering
\begin{tikzpicture}[>=stealth',,shorten >=1pt,auto,node distance=2.2cm,scale = 0.6, transform shape]

\node[initial,initial text={},state]           (A)                                    {$a_0$};
\node[state]         (B) [right of=A]                       {$a_1$};
\node[state]                   (C) [right of=B]                       {$a_2$};
\node[state]                   (D) [right of=C]                       {$a_3$};
\node[state]                   (E) [right of=D]                       {$a_4$};
\node[state]                   (F) [below of=B]                       {$a_5$};
\node[state]                   (G) [right of=F]                       {$a_6$};
\node[state,accepting]                   (H) [right of=G]                       {$a_7$};
%\node[state,accepting]         (D) [below of=B]                       {$s_3$};

\path[->] (A) edge [above,dashed]   node [align=center]  {$\tick$} (B)
      (B) edge [above]      node [align=center]  {$ \tick $}    (C)
      (C) edge [dashed,above]      node [align=center]  {$ p $}    (D)
    (D) edge [above]      node [align=center]  {$ j $}    (E)
      (B) edge [right]      node [align=center]  {$j$} (F)
      (C) edge [right]      node [align=center]  {$j$} (G)
      %(D) edge [right]      node [align=center]  {$j$} (G)
     % (E) edge [above]      node [align=center]  {$\tick$} (F)
      (F) edge [above,dashed]      node [align=center]  {$ \tick $}    (G)
      (G) edge [dashed,above]      node [align=center]  {$ p $}    (H)
      (E) edge [loop right,dashed]      node [align=center]  {$\tick$} (E)
      (H) edge [loop right,dashed]      node [align=center]  {$\tick$} (H)
      (D) edge [loop above,dashed]      node [align=center]  {$\tick$}    (D);
\end{tikzpicture}
\caption{Endangered pedestrian from Example \ref{exp:BusPed}.}
\label{fig:BusPed}
\end{figure}
\end{example}

%\subsection{Observed Plant Automaton}
%\label{subsection:OP}
\subsection{Networked Plant}
The behavior of the plant communicating through the control and observation channels is captured by the \emph{networked plant}.
%As it is clear from Figure \ref{fig:solution1}, if we do not consider enabling and observation of events, the behavior of the plant itself should be preserved in the networked plant. This property will be formalized and proven.
%\AR{To have the full control over the plant, we need to make sure that $\NP$ does not exclude some plant behavior due to delays (that is why we look ahead and determine $\Sigma_e$) or channel capacities (we need to have the proper control platform in terms of maximum required capacities of channels) to satisfy the property: $P_{\Sigma_G}(NP)=P$.}
As is clear from Figure~\ref{fig:solution1}, if we do not consider enabling and observation of events, what is executed in the networked plant is always a part of the plant behavior.
%the behavior of the plant itself should be preserved by the networked plant.
%\AR{As is clear from Figure~\ref{fig:solution1}, the networked plant is actually the plant augmented with the enabling events received from the control channel to execute the relevant events in the plant and the observed events sent through the observation channel related to the events executed in the plant.}
Let us denote by $\NP$ the networked plant automaton, then $P_{\Sigma_G}(L(\NP))\subseteq L(G)$. 

%This will be given and proven as a property of the networked plant (Property \ref{property:NPLE}).
Moreover, note that a networked supervisor is synthesized for a plant on the basis of the networked plant. The networked plant should represent all the possible behavior of the plant in the networked supervisory control setting, and it is only the networked supervisor that may prevent the occurrence of some plant events by disabling the relevant enabling event. This means that $\NP$ should be such that $L(G)\subseteq P_{\Sigma_G}(L(\NP))$. 
%This is given and proven as a property for the networked plant (Property \ref{property:NPLE2}).
The latter property relies on the following assumptions.

\textbf{Assumption 1:} The plant enables enough \ticks in the beginning; there are at least $N_c$ \ticks (there can be uncontrollable events occurring between \ticks) enabled before the first controllable event. 

\textbf{Assumption 2:} The control channel provides enough capacity for all enabling commands being sent to the plant. Imagine that $\tick\,\sigma\,\tick^*\in L(G)$, and $\Lmax=0$. Then, $\sigma_e$ may occur in $\NP$, but the plant will never execute $\sigma$ as it does not receive the relevant enabling command.
To avoid this situation, the size of the control channel should be such that it always has the capacity for all enabling events. An enabling event will be removed from the control channel after $N_c$ \ticks.
So, considering all substrings $w$ that can appear in the plant (after an initial part $w_0$) which are no longer (in the time sense) than $N_c$ \ticks, then the control channel capacity should be at least equal to the number of controllable events occurring in $w$; $\Lmax\geq \max_{w\in W}|P_{\Sigma_c}(w)|$ where $W=\{w\in\Sigma^*_G\mid \exists w_0w\in L(G), |P_{\{tick\}}(w)| \leq N_c-1\}$.

To obtain the networked plant, we present the function $\Pi$ in Definition \ref{dfn:NP}. In order to determine enabling commands
%we look to 1. what the supervisor observes through the observation channel as we already have it in $OP'=P_{\Sigma_{OP}\setminus\Sigma}(\OP)$ and
we look $N_c$ \ticks ahead for only the controllable active events enabled in $G'=P_{\Sigma_G\setminus\Sigma_u}(G)$. 
We use a list $L$ to store the controllable events that have been commanded and a medium $M$ to store the events that were executed.

%Note that to guarantee a finite representation of the networked plant in case of the present of an event-loop, the capacity of the control and observation channels are assumed to be limited to $\Lmax<+\infty$ and $\Mmax<+\infty$, respectively.
%\MR{I thought we already indicated they are natural numbers didn't we? Then this is always the case (since $\infty \notin N$.}

\begin{defn}[Networked Plant Operator]
\label{dfn:NP}
For a given plant, $G$,
%control delay $N_c$, observation delay $N_o$, maximum observation channel capacity $\Mmax$, and maximum control channel capacity $\Lmax$,
%control and observation channels,
$\Pi$  gives the networked plant as:
%\MF{1. It is strange to me that $\Pi$ does not need $\Sigma_e$ or $\Sigma_o$.
%\AR{$\Pi$ is a function that for the inputs: $G,N_c,N_o,\Lmax,\Mmax$, gives $\NP$ by determining when $\sigma_e$ and $\sigma_o$ occur. So, $\sigma_e$ and $\sigma_o$ both belong to $\Sigma_{NSP}$ are link of output of $\Pi$ and not the input\MF{But if $\Sigma_e$ and $\Sigma_o$ are not given, how does $\Pi$ know what they are?}}
%2. Is $\Pi$ ever used? If not, then there is no reason to define it. It simply generates what has already been called NP, right? \AR{We introduce $\Pi$ as the general function, and then we use the acronym $\NP$ to refer to $\Pi$ with $G,...$ as input}}
\begin{equation*}
\Pi(G,N_c,N_o,\Lmax,\Mmax)=(X,\SigmaNSP,\deltaNP,x_{0},X_{m}),
\end{equation*}
Let $G'=P_{\Sigma_G\setminus\Sigma_u}(G)=(A', \Sigma_G, \delta'_G, a'_{0}, A'_{m})$, and 
\begin{equation*}
\begin{aligned}
%\Sigma_{AP}&= \Sigma_{OP}\cup \Sigma_{e},\\
X &= A\times  A'\times M \times L,\qquad
%\Sigma_{PS}&=\Sigma_{e}\cup\Sigma_{o}\cup\{\tick\}\\
x_{0} =(a_{0},\delta'_G(a'_0,\tick^{N_c}),[],\varepsilon),\\
X_{m}&=A_{m}\times A'\times M \times L.
%L_{m}&=\{l=\delta_{PS}(q,q')|\exists \sigma_{OP} \in \Sigma_{OP} s.t \delta_{OP}(q,\sigma_{OP})\in Q_{m}\},
\end{aligned}
\end{equation*}
For $a \in A$, $a'\in A'$, $m\in M$ and $l\in L$, the transition function $\deltaNP:X\times\SigmaNSP\rightarrow X$ is defined as follows:

\begin{enumerate}

\item 
If $\delta'_G(a',\sigma)!$, $\sigma\in\Sigma_c$
% and $|l|<\Lmax$
\begin{equation*}
\deltaNP((a,a',m,l),\sigma_{e})=(a,\delta'_G(a',\sigma),m,\mathit{app}(l,(\sigma,N_c))).
%app
\end{equation*}

\item
If $\delta_G(a,\sigma)!$,  $\head(l)=(\sigma,0),\sigma \in \Sigma_c$
% and $|m|<\Mmax$
\begin{equation*}
\deltaNP((a,a',m,l),\sigma)=(\delta_G(a,\sigma),a',m\uplus[(\sigma,N_o)],\mathit{tail}(l)).
\end{equation*}

\item 
If $\delta_G(a,\sigma)!,\sigma\in\Sigmauc$
% and $|m|<\Mmax$
\begin{equation*}
\deltaNP((a,a',m,l),\sigma)=(\delta_G(a,\sigma),a',m\uplus[(\sigma,N_o)],l).
\end{equation*}

\item
If $\delta_G(a,\tick)!$, $\neg\delta'_G(a',\sigma)!$ for all $\sigma\in \Sigma_c$, and $(\sigma',0)\notin m$ for all $\sigma, \sigma'\in \Sigma_a$
\begin{multline*}
\deltaNP((a,a',m,l),\tick)=\\
\begin{cases}
    (\delta_G(a,tick),\delta'_G(a',\tick),m-1,l-1) &\text{if $\delta'_G(a',\tick)!$,}\\
    (\delta_G(a,tick),a',m-1,l-1) &\text{otherwise.}
    \end{cases}
\end{multline*}
%\AR{Imagine that $G$ is not TLF, lets say $\tick\tick\sigma\in L(G)$, then we get $\tick\sigma_e\in L(\NP)$ and then \tick is blocked by $G'$! so even then we do not have the second property for $\NP$ :( ... I am thinking if we can fix this ... what if we say \tick occurs in NP only if $G$ enables it then if $G'$ also enables it, its state changes. Otherwise, $G'$ stays in the same state? If this it makes sense to apply this change, then this proof and a few more become simpler as well} and due to Property \ref{lemma:projectionTLF}
%\begin{multline*}
%\deltaNP((a,a',m,l),\tick)=(\delta(a,tick),\delta'_G(a',\tick),\\m-1,l-1).
%\end{multline*}
%\AR{I think we do not need any condition on medium or list here, right?}
%\AR{I believe that the condition on having $(\sigma,0)\in m$ is already considered in the definition of $OP$.(should observed events be kind of uncontrollable forcible event for $\NP$?)}
%\AR{What about if there is $(\sigma,0)\in l$, and \emph{tick} is enabled? then is $(\sigma,0)$ removed after tick?}
%\item
%If $\delta(a,\tick)!$, $\delta_{OP'}(q',\tick)!$, $\delta'_G(a',\tick)!$, and $(\sigma,0)\in l$
%\begin{multline*}
%\delta_{AP}((a,q',a',m,l),\tick)=(\delta(a,tick),\delta_{OP'}(q',\tick),\\
%\delta'_G(a',\tick),m-1,l\setminus(\sigma,0)-1).
%\tag*{$\blacksquare$}   
%\end{multline*}

\item 
If $(\sigma,0)\in m$
\begin{equation*}
\deltaNP((a,a',m,l),\sigma_{o})=(a,a',m\setminus[(\sigma,0)],l). \hspace*{1.65cm} \blacksquare
\end{equation*}
%\hfill$\blacksquare$ 
\end{enumerate}
%\AR{Note: there is no need for the second item here ( if $(\sigma,0)\in m$, $\neg\delta_{OP'}(q',\sigma_{o})!$ like CASE) since everything is synchronous now! Then, do we need to use $m$ here at all?}
%\item 
%If $(\sigma,0)\in m$, $\neg\delta_{OP'}(q',\sigma_{o})!$ 
%\begin{equation*}
%\deltaNP((a,q',a',m,l),\sigma_{o})=(a,q',a',m\setminus[(\sigma,0)],l).
%\end{equation*}
\end{defn}

Note that due to Assumption 1, $\delta'_G(a'_0,\tick^{N_c})$ is always defined. 
In the rest of the paper, the networked plant of the plant $G$ is assumed to be the TDES $\NP$ represented by the automaton $(X,\SigmaNSP,\deltaNP,x_{0},X_{m})$.

%%MF I moved a short piece from here to after the def of Property 2, where I think it fits better.

%\begin{lemma}[NP Transitions]
%\label{lemma:NP}

%\begin{lemma}[NP Enabling Commands]
%\label{lemma:ontime}

%Assume that $\tick\sigma\tick\in L(G)$, $N_c=1$, and Lemma \ref{lemma:ontime} does not hold. For instance $\sigma_e$ is enabled in $\NP$ after the first \tick, then $\tick\sigma_e\sigma\notin L(\NP)$ ($\sigma$ cannot occur) which makes $\NP$ not acceptable as it does not satisfy the mentioned properties.

%Lemma \ref{lemma:NP} and Lemma \ref{lemma:ontime} are used in the proofs to come.

%\begin{lemma}[NSP and NP]
%\label{lemma:m,l}

\begin{property}[\NP and Plant]
\label{property:NPLE}
For any plant $G$:
%observation and control channels \MR{What is this about channels??? Do you want to introduce the constants (which you should)?}\AR{I gave it once as a remark (Remark 5) for the rest of the paper to avoid repetition and saving space \MR{Yes, but what are these words doing here then? What do you want to say about observation and control channels}}
\begin{enumerate}
    \item\label{NP&P} $P_{\Sigma_G}(L(\NP))\subseteq L(G)$, and
    \item\label{P&NP} $L(G)\subseteq P_{\Sigma_G}(L(\NP))$ whenever assumptions 1 and 2 hold.
\end{enumerate}
%\hfill$\blacksquare$
\end{property}

\begin{proof}
See Appendix \ref{proof:NPLE}.
\hfill $\blacksquare$
\end{proof}

\begin{example}
\label{exp:BusPedOP}
For the endangered pedestrian from Example \ref{exp:BusPed}, $G'$ and $\NP$ are given in Figure \ref{fig:P'4BusPed} and Figure \ref{fig:NP4BusPed}, respectively.
%\hfill $\blacksquare$

\begin{figure}[htbp]
\centering
\begin{tikzpicture}[>=stealth',,shorten >=1pt,auto,node distance=2.2cm,scale = 0.6, transform shape]

\node[initial,initial text={},state]           (A)                                    {$a'_0$};
\node[state]         (B) [right of=A]                       {$a'_1$};
\node[state]                   (C) [right of=B]                       {$a'_2$};
%\node[state]                   (D) [right of=C]                       {$a_3$};
%\node[state]                   (E) [right of=D]                       {$a_4$};
\node[state]                   (F) [below of=B]                       {$a'_3$};
\node[state,accepting]                   (G) [right of=F]                       {$a'_4$};
%\node[state,accepting]                   (H) [right of=G]                       {$a_7$};
%\node[state,accepting]         (D) [below of=B]                       {$s_3$};

\path[->] (A) edge [above,dashed]   node [align=center]  {$\tick$} (B)
      (B) edge [above]      node [align=center]  {$ \tick $}    (C)
%      (C) edge [dashed,above]      node [align=center]  {$ p $}    (D)
%    (D) edge [above]      node [align=center]  {$ j $}    (E)
      (B) edge [right]      node [align=center]  {$j$} (F)
      (C) edge [right]      node [align=center]  {$j$} (G)
      %(D) edge [right]      node [align=center]  {$j$} (G)
     % (E) edge [above]      node [align=center]  {$\tick$} (F)
      (F) edge [above,dashed]      node [align=center]  {$ \tick $}    (G)
%      (G) edge [dashed,above]      node [align=center]  {$ p $}    (H)
%      (E) edge [loop right]      node [align=center]  {$\tick$} (E)
      (G) edge [loop right,dashed]      node [align=center]  {$\tick$} (G)
      (C) edge [loop right,dashed]      node [align=center]  {$\tick$}    (C);
\end{tikzpicture}
\caption{$G'$ for the endangered pedestrian from Example \ref{exp:BusPed}.}
\label{fig:P'4BusPed}
\end{figure}
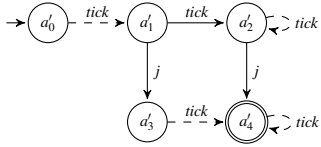

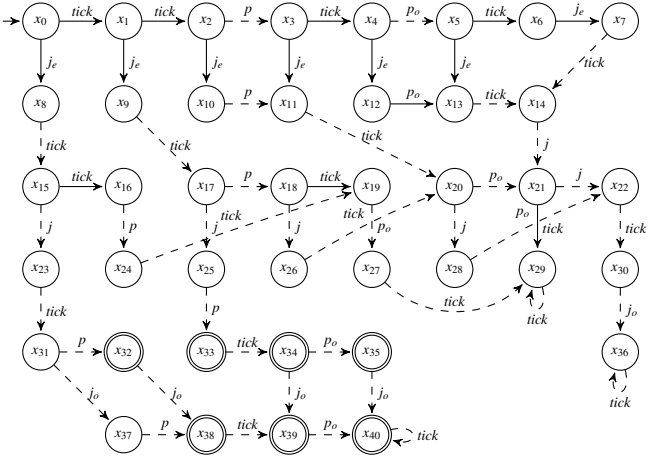
\begin{figure}[htbp]
\centering
\begin{tikzpicture}[>=stealth',,shorten >=1pt,auto,node distance=2.0cm,scale = 0.55, transform shape]

\node[initial,initial text={},state]           (x0)                                    {$x_0$};
\node[state]                   (x1) [right of=x0]                       {$x_1$};
\node[state]                   (x2) [right of=x1]                       {$x_2$};
\node[state]                   (x3) [right of=x2]                       {$x_3$};
\node[state]                  (x4) [right of=x3]                        {$x_4$};
\node[state]                   (x5) [right of=x4]                       {$x_5$};
\node[state]                   (x6) [right of=x5]                       {$x_6$};
\node[state]                   (x7) [right of=x6]                       {$x_7$};
\node[state]                   (x8) [below of=x0]                       {$x_8$};
\node[state]                   (x9) [right of=x8]                       {$x_9$};
\node[state]                   (x10) [right of=x9]                       {$x_{10}$};
\node[state]                   (x11) [right of=x10]                       {$x_{11}$};
\node[state]                   (x12) [right of=x11]                       {$x_{12}$};
\node[state]                   (x13) [right of=x12]                       {$x_{13}$};
\node[state]                   (x14) [right of=x13]                       {$x_{14}$};
\node[state]                   (x15) [below of=x8]                       {$x_{15}$};
\node[state]                   (x16) [right of=x15]                       {$x_{16}$};
\node[state]                   (x17) [right of=x16]                       {$x_{17}$};
\node[state]                   (x18) [right of=x17]                       {$x_{18}$};
\node[state]                   (x19) [right of=x18]                       {$x_{19}$};
\node[state]                   (x20) [right of=x19]                       {$x_{20}$};
\node[state]                   (x21) [right of=x20]                       {$x_{21}$};
\node[state]                   (x22) [right of=x21]                       {$x_{22}$};
\node[state]                   (x23) [below of=x15]                       {$x_{23}$};
\node[state]                   (x24) [right of=x23]                       {$x_{24}$};
\node[state]                   (x25) [right of=x24]                       {$x_{25}$};
\node[state]                   (x26) [right of=x25]                       {$x_{26}$};
\node[state]                   (x27) [right of=x26]                       {$x_{27}$};
\node[state]                   (x28) [right of=x27]                       {$x_{28}$};
\node[state]                   (x29) [right of=x28]                       {$x_{29}$};
\node[state]                   (x30) [right of=x29]                       {$x_{30}$};
\node[state]                   (x31) [below of=x23]                       {$x_{31}$};
\node[state,accepting]                   (x32) [right of=x31]                       {$x_{32}$};
\node[state,accepting]                   (x33) [right of=x32]                       {$x_{33}$};
\node[state,accepting]                   (x34) [right of=x33]                       {$x_{34}$};
\node[state,accepting]                   (x35) [right of=x34]                       {$x_{35}$};
\node[state]                   (x36) [below of=x30]                       {$x_{36}$};
\node[state]                   (x37) [below of=x32]                       {$x_{37}$};
\node[state,accepting]                   (x38) [right of=x37]                       {$x_{38}$};
\node[state,accepting]                   (x39) [right of=x38]                       {$x_{39}$};
\node[state,accepting]                   (x40) [right of=x39]                       {$x_{40}$};

\path[->] (x0) edge [above]   node [align=center]  {$\tick$} (x1)
      (x0) edge [right]   node [align=center]  {$j_e$} (x8)
      (x1) edge [above]      node [align=center]  {$\tick $}    (x2)
      (x1) edge [right]      node [align=center]  {$j_e $}    (x9)
      (x2) edge [above,dashed]      node [align=center]  {$ p $}    (x3)
      (x2) edge [right]      node [align=center]  {$j_e $}    (x10)
      (x3) edge [above]      node [align=center]  {$\tick $}    (x4)
      (x3) edge [right]      node [align=center]  {$j_e $}    (x11)
      (x4) edge [above,dashed]      node [align=center]  {$p_o$}    (x5)
      (x4) edge [right]      node [align=center]  {$j_e $}    (x12)
      (x5) edge [above]      node [align=center]  {$\tick $}    (x6)
      (x5) edge [right]      node [align=center]  {$j_e $}    (x13)
      (x6) edge [above]      node [align=center]  {$j_e$}    (x7)
      (x7) edge [right,dashed]      node [align=center]  {$\tick $}    (x14)
      (x8) edge [right,dashed]      node [align=center]  {$\tick $}    (x15)
      (x9) edge [right,dashed]      node [align=center]  {$\tick $}    (x17)
      (x10) edge [above,dashed]      node [align=center]  {$p$}    (x11)
      (x11) edge [above,dashed]      node [align=center]  {$\tick$}    (x20)
      (x12) edge [above]      node [align=center]  {$p_o$}    (x13)
      (x13) edge [above,dashed]      node [align=center]  {$\tick$}    (x14)
      (x14) edge [right,dashed]      node [align=center]  {$j$}    (x21)
      (x15) edge [above]      node [align=center]  {$\tick$}    (x16)
      (x15) edge [right,dashed]      node [align=center]  {$j$}    (x23)
      (x16) edge [right,dashed]      node [align=center]  {$p$}    (x24)
      (x17) edge [above,dashed]      node [align=center]  {$p$}    (x18)
      (x17) edge [right,dashed]      node [align=center]  {$j$}    (x25)
      (x18) edge [above]      node [align=center]  {$\tick$}    (x19)
      (x18) edge [right,dashed]      node [align=center]  {$j$}    (x26)
      (x19) edge [right,dashed]      node [align=center]  {$p_o$}    (x27)
      (x20) edge [above,dashed]      node [align=center]  {$p_o$}    (x21)
      (x20) edge [right,dashed]      node [align=center]  {$j$}    (x28)
      (x21) edge [right]      node [align=center]  {$\tick$}    (x29)
      (x21) edge [above,dashed]      node [align=center]  {$j$}    (x22)
      (x22) edge [right,dashed]      node [align=center]  {$\tick$}    (x30)
      (x23) edge [right,dashed]      node [align=center]  {$\tick$}    (x31)
      (x24) edge [bend left=2,dashed]      node [align=center]  {$\tick$}    (x19)
      (x25) edge [right,dashed]      node [align=center]  {$p$}    (x33)    
      (x26) edge [bend left=5,dashed]      node [align=center]  {$\tick$}    (x20)
      (x27) edge [bend right=45,dashed]      node [align=center]  {$\tick$}    (x29)
      (x28) edge [bend left=5,dashed]      node [align=center]  {$p_o$}    (x22)
      (x29) edge [loop below,dashed]      node [align=center]  {$\tick$}    (x29)
      (x30) edge [right,dashed]      node [align=center]  {$j_o$}    (x36)
      (x31) edge [above,dashed]      node [align=center]  {$p$}    (x32)
      (x31) edge [right,dashed]      node [align=center]  {$j_o$}    (x37)
      (x32) edge [right,dashed]      node [align=center]  {$j_o$}    (x38)
      (x33) edge [above,dashed]      node [align=center]  {$\tick$}    (x34)
      (x34) edge [above,dashed]      node [align=center]  {$p_o$}    (x35)
      (x34) edge [right,dashed]      node [align=center]  {$j_o$}    (x39)  
      (x35) edge [right,dashed]      node [align=center]  {$j_o$}    (x40)      
      (x36) edge [loop below,dashed]      node [align=center]  {$\tick$}    (x36)
      (x37) edge [above,dashed]      node [align=center]  {$p$}    (x38)      
      (x38) edge [above,dashed]      node [align=center]  {$\tick$}    (x39)      
      (x39) edge [above,dashed]      node [align=center]  {$p_o$}    (x40)
      (x40) edge [loop right,dashed]      node [align=center]  {$\tick$}    (x40);
\end{tikzpicture}
\caption{Networked plant for the endangered pedestrian from Example \ref{exp:BusPed} ($N_c=1,N_o=1$).}
\label{fig:NP4BusPed}
\end{figure}
\end{example}

%\begin{remark}
%\AR{As clear form Figure \ref{fig:solution1}, enabling events are the only controllable events for $\NS$ that can be disabled. All other events in $\NP$ (plant active events and observed events) are considered to be uncontrollable in synthesis. Moreover, controllability of \tick now depends on the plant forcible events and also enabling events as we assumed that they are forcible as well. To clarify, uncontrollable events are indicated by dashed lines in Figure \ref{fig:NP4BusPed}. Also, note that only the observed events are observable to the networked supervisor.}
%\end{remark}

\subsection{Synthesis}
\label{subsection:S}
As is clear from Figure~\ref{fig:solution1}, enabling events are the only controllable events that can be disabled by the networked supervisor. All other events in the networked plant (active events and observed events) are uncontrollable. Moreover, controllability of \tick depends on the forcible events of the plant as well as the enabling events (as we assume that they are forcible). To clarify, uncontrollable events are indicated by dashed lines in Figure \ref{fig:NP4BusPed}. Note also that the observed events are observable to the networked supervisor. Also, events from $\Sigma_e$ are \MFedit{supposed to be}{}observable, as the networked supervisor knows about the commands that it sends to the plant. However, the events from $\Sigma_a$ are now unobservable to the networked supervisor.
%Although all events are supposed to be observable in $G$, the set of active events $\Sigma$ becomes unobservable in $\NP$ due to the observation delay (this can be seen in Figure \ref{fig:solution1}).
%Events from $\Sigma_e$ are supposed to be observable as the networked supervisor knows about the commands that it sends to the plant.
%Moreover, enabling events are the only controllable events in $\NP$ since events from $\Sigma_a$ cannot be preempted if they occur in the plant. 
%Moreover, events from $\Sigma_o$ merely represent an observation of an event that has already been executed in the plant.
To consider these issues in the current step of the approach, the sets of unobservable events $\hatSigmauo$, observable events $\hat{\Sigma}_o$, uncontrollable active events $\hatSigmauc$, and controllable active events $\hat{\Sigma}_c$ of the networked plant are given by 
$\hatSigmauo=\Sigma_a$,
$\hat{\Sigma}_o=\Sigma_e\cup\Sigma_o\cup\{\tick\}$,
$\hatSigmauc=\Sigma_a\cup \Sigma_o$, $\hat{\Sigma}_c=\Sigma_e$. 
Also, as mentioned before $\hatSigmaFor=\SigmaFor\cup\Sigma_e$.
The event \tick is always observable to the networked supervisor. Moreover, it is uncontrollable unless there exists an event from $\hatSigmaFor$ enabled in parallel to \tick.
Regarding the new sets of events, the synthesis algorithm takes into account the TDES conventional controllability (in Definition \ref{dfn:cont.TDES})  and is inspired from the weak observability condition introduced in~\cite{Takai:06,Cai:16}.

Algorithm \ref{algo} presents the synthesis procedure in which we use the following additional concepts and abbreviations:
\begin{itemize}
\item $\BS(\NS)=\BLock(\NS)\cup\TLock(\NS)$ where $\BLock(\NS)$ gives the set of blocking states of $\NS$, and $\TLock(\NS)$ gives the set of time-lock states of $\NS$.
\item  Due to the fact that events from $\Sigma_a$
%\AR{this should be active events from $\Sigma_G$. Also in $OBS$ definition, the projection should be on $\hat{\Sigma}_o=\Sigma_e\cup\Sigma_o\cup\{\tick\}$, right? so I defined $\hat{\Sigma}_o$ to refer to that}
are unobservable in the networked plant, one should be careful that the same control command is applied on the states reachable through the same observations. To take this issue into account, the following function is used in the synthesis algorithm; $\OBS(x)=\{ x'\in X \mid \exists w,w'\in \Sigma^*_{NP},
\deltaNP(x_0,w) = x \land 
\deltaNP(x_0,w') = x' \land P_{\hat{\Sigma}_o}(w)=P_{\hat{\Sigma}_o}(w') \}$ gives the set of states observationally equivalently reachable as $x$.
The function $\OBS$ can be applied on a set of states $X'\subseteq X$ as well such that $\OBS(X')=\bigcup_{x\in X'} \OBS(x)$.
%\item $E(x') = \{ x \in X \mid \exists w,w'\in \Sigma^*_{NP},\deltaNP(x_0,w) = x \land \deltaNP(x_0,w') = x' \land P_{\Sigma_G}(w)=P_{\Sigma_G}(w') \}$ is the set of states that indicate the same state of $G$.
\item $F(y)=\{\sigma\in\hatSigmaFor \,|\, \deltaNS(y,\sigma)!\}$ is the set of forcible events enabled at state $y$.
\item Besides blocking and time-lock states, we should take care of states from which a state from $\BS(\NS)$ can be reached in an uncontrollable way, taking preemption of \tick events into account. $\Uncon(\BS(\NS))$ gives a set of states, called \emph{bad states}, such that 
\begin{enumerate}
\item $\BS \subseteq \Uncon(\BS(\NS))$;
\item if $\deltaNS(y,\sigma) \in \Uncon(\BS(\NS))$ for some $y \in Y$ and $\sigma \in \hatSigmauc$, then $y \in \Uncon(\BS(\NS))$;
\item if $\deltaNS(y,\tick) \in \Uncon(\BS(\NS))$ for some $y \in Y$ such that for all $y'\in \OBS(y)$, $F(y)\cap F(y')=\varnothing$, then $y \in \Uncon(\BS(\NS))$.  This is to make sure that the supervisor behaves the same towards all observationally equivalent transitions.
%\AR{Should we have the condition at line 6 that y is not bad itself but just leading to a bad state through a controllable event/tick? and then we do not need the if condition at line 8? \MR{If you are convinced that that is sufficient (or better the same in the end) then that would indeed be nicer.}}
%if $\deltaNS(y,\tick) \in \mathit{Uncon}_{NS}(\BS)$ for some $y \in Y$ and $\nexists_{\sigma \in \hat{\Sigma}_f}~  \deltaNS(y,\sigma)!$, then $y \in \mathit{Uncon}_{NS}(\BS)$. 
\end{enumerate}
\item $\mathit{\BPre}(\NS) = \{ y \in Y \mid F(y)=0$  $\land$  $\neg\deltaNS(y,\tick)!$ $\land$ $\deltaNP(y,\tick)! \}$
%\AR{Shall we define $\delta_{NS(0)}$ as well? \MR{I do not understand the question? In line 2 $\ns(0)$ is defined to be $NP$, therefore, I hope and expect, $\delta_{NS(0)}=\delta_{NP}$}  $\deltaNP(y,\tick)$ should actually be $\delta_{NS(0)}(y,\tick)$?}
contains states (still in \NS) from which no forcible events and no $\tick$ are enabled while there was a $\tick$ event enabled in the networked plant. 
%\item \AR{$\mathit{\BPre2}(S) =\{y'\in Y \mid \exists y\in Y, u\in\Sigmauc, w_c\in{\ut{\Sigma}_c}^*,\delta_S(y,w_c(1))=y', \delta_S(y,w_c)\in y_m, \delta_S(y,u)!\land\neg\delta_S(y,uw_c)! \lor \delta_S(y,uw_c)\notin Y_m\}$}
\item $\mathit{Reach}(\NS)$ restricts an automaton to those states that are reachable from the initial state.
\end{itemize}

\begin{comment}
\renewcommand{\algorithmicrequire}{\textbf{Input: }}
\renewcommand{\algorithmicensure}{\textbf{Output: }}
\begin{algorithm}
\caption{Networked supervisory control synthesis\\
\algorithmicrequire $\NP=(X, \SigmaNSP, \deltaNP, x_{0}, X_{m})$, $\hat{\Sigma}_{uo}$, $\hat{\Sigma}_{uc}$, $\hat{\Sigma}_c$, $\hatSigmaFor$\\
\algorithmicensure{$\NS=(Y, \SigmaNS, \deltaNS, y_{0}, Y_{m})$}}\label{algo}
\begin{algorithmic}[1]
\State \textcolor{red}{$i\gets 0$}
\State $\ns(0)\gets NP$
\State $\BS\gets \BLock(\ns(0))\cup \TLock(\ns(0))$
%\State \textcolor{red}{$\ns(0)\gets NS$}
\While{$x_0\notin BS \wedge BS \neq \varnothing$}
\State \textcolor{red}{$\NS\gets NS(0)$}
\For{$y \in Y \And \sigma \in \hat{\Sigma}_{c}$}
\If{$\deltaNS(y,\sigma) \in \Uncon(\BS)$}
%\For{$y'\in OBS(y)$}
\If{$\sigma\neq\{tick\}$}
\For{$y'\in \OBS(y)$}
\State $\deltaNS(y',\sigma) \gets \textbf{undefined}$ \label{line:disabling1}
\EndFor
\EndIf
\If{$\sigma=\{tick\}\And F(y)\cap F(y')\neq\varnothing$}
\For{$y'\in E(\OBS(y))$}
\State $\deltaNS(y',\sigma) \gets \textbf{undefined}$ \label{line:disabling2}
\EndFor
\EndIf
%\EndFor
\EndIf
\EndFor
\State \textcolor{red}{$i\gets i+1$}
\State \textcolor{red}{ $\ns(i) \gets \mathit{Reach}(\NS)$} \label{line:reachable}
%\State \textcolor{red}{$\ns(i)\gets NS$}
\State $\BS\gets \mathit{\BPre}(\ns(i)) \cup\BLock(\ns(i))\cup\TLock(\ns(i))$ \label{line:BPre}
\EndWhile
\If {$x_0\in BS$}
\State{no result}
\EndIf
\State \textcolor{red}{$\NS \gets P_{\SigmaNSP\setminus\Sigma}(\ns(i))$}\label{line:project}
\end{algorithmic}
\end{algorithm}
\end{comment}

\renewcommand{\algorithmicrequire}{\textbf{Input: }}
\renewcommand{\algorithmicensure}{\textbf{Output: }}
\begin{algorithm}
\caption{Networked supervisory control synthesis\\
\algorithmicrequire $\NP=(X, \SigmaNSP, \deltaNP, x_{0}, X_{m})$, $\hat{\Sigma}_{uo}$, $\hat{\Sigma}_{uc}$, $\hat{\Sigma}_c$, $\hatSigmaFor$\\
\algorithmicensure{$\NS=(Y, \SigmaNS, \deltaNS, y_{0}, Y_{m})$}}\label{algo}
\begin{algorithmic}[1]
\State \textcolor{red}{$i\gets 0$}
\State $\ns(0)\gets NP$
\State $\bs(0)\gets BS(\ns(0))$
%\State \textcolor{red}{$\ns(0)\gets NS$}
\While{$y_0\notin \Uncon(\bs(i)) \wedge \bs(i) \neq \varnothing$}\label{line:while}
\For{$y \in Y\setminus\Uncon(\bs(i))$ and $\sigma \in \hat{\Sigma}_{c}\cup\{tick\}$}
%\AR{$Y\setminus\OBS(\Uncon(\bs(i)))$ or $Y\setminus\Uncon(\bs(i))$?}
\label{line:ctrl}
\If{$\deltaNS(y,\sigma) \in \OBS(\Uncon(\bs(i)))$}
\label{line:obs.eq}
%\AR{Can we remove OBS here?}
\For{$y'\in \OBS(y)$}
%\If{$\sigma\neq\{tick\}\vee  F(y)\cap F(y')\neq\varnothing$}
\State $\deltaNS(y',\sigma) \gets \textbf{undefined}$ \label{line:disabling1}
%\EndIf
\EndFor
\EndIf
\EndFor
%\AR{I think we need to remove $\Uncon(\OBS(\bs(i)))$ as this appears in nonblockingness proof ... I will try to find an example showing the reason}
\State $Y \gets Y\setminus\Uncon(\bs(i))$
\label{line:removeUncon}
\State \textcolor{red}{$i\gets i+1$}
\State \textcolor{red}{ $\ns(i) \gets \mathit{Reach}(\ns(i-1))$} \label{line:reachable}
%\State \textcolor{red}{$\ns(i)\gets NS$}
\State $\bs(i)\gets \BPre(\ns(i)) \cup BS(\ns(i))$\label{line:BPre}
\EndWhile
\If {$y_0\in \Uncon(\bs(i))$}
\State{no result}
\EndIf
\State \textcolor{red}{$\NS \gets P_{\SigmaNSP\setminus\Sigma}(\ns(i))$}\label{line:project}
\end{algorithmic}
\end{algorithm}

%\AR{The algorithm (line 6-8) and the termination proof (Property 3) are edited regarding Martin's comments ... Please check if they are correct now ... Thanks}

Starting from $\NS=NP$, Algorithm \ref{algo} changes $\NS$ by disabling transitions at line \ref{line:disabling1} and delivering the reachable part at line \ref{line:reachable}.
%In the properties and theorems to come, we use $\ns(i)$ indicating the result of the algorithm at  iteration $i$.
For the proposed algorithm, the following property and theorems hold.

\begin{property}[Algorithm Termination]
\label{property:termination}
The synthesis algorithm presented in Algorithm \ref{algo} terminates.
%\hfill $\blacksquare$
\end{property}

\begin{proof}
See Appendix \ref{proof:termination}.
\hfill $\blacksquare$
\end{proof}

\begin{theorem}[Nonblocking \NSP]
\label{theorem:nonblockingness}
Given a plant $G$ and the networked supervisor $\NS$ computed by Algorithm \ref{algo}: \NSP is  nonblocking.
%\hfill$\blacksquare$
\end{theorem}

\begin{proof}
See Appendix \ref{proof:NBness}.
\hfill $\blacksquare$
\end{proof}

\begin{theorem}[TLF \NSP]
\label{theorem:TLF}
Given a plant $G$ and the networked supervisor $\NS$ computed by Algorithm \ref{algo}: \NSP is  TLF.
%\hfill $\blacksquare$
\end{theorem}

\begin{proof}
See Appendix \ref{proof:TLF}.
\hfill $\blacksquare$
\end{proof}

\begin{theorem}[Controllable \NS]
\label{theorem:controllability}
Given a plant $G$ and the networked supervisor $\NS$ computed by Algorithm \ref{algo}: $\NS$ is  timed networked controllable w.r.t.\ $G$.
%\hfill$\blacksquare$
\end{theorem}

\begin{proof}
See Appendix \ref{proof:controllability}.
\hfill $\blacksquare$
\end{proof}

\begin{theorem}[Timed Networked Maximally Permissive \NS]
\label{theorem:MPness}
For a plant $G$, the networked supervisor $\NS$ computed by Algorithm \ref{algo} is timed networked maximally permissive.
%with event set $\SigmaNS$ and $L(\NS)\subseteq P_{\SigmaNS}(L(\NP))$.
%\AR{I need to give definition of max per for NSC setting}
%\MF{Can we have other supervisor with different event sets that achieves a larger language? Or is this not meant to be a qualifier of the claim, but a consequence?}
%\hfill$\blacksquare$
\end{theorem}

\begin{proof}
See Appendix \ref{proof:MPness}.
\hfill $\blacksquare$
\end{proof}

\subsection{Possible Variants}
\label{section:PV}
The proposed synthesis approach can be adjusted for the following situations.

%\begin{remark}
\subsubsection{Nonblockingness or time-lock freeness}
Algorithm \ref{algo} can easily be adapted to either only provide nonblockingness or time-lock freeness by removing $\TLock(\NS)$ and $\BLock(\NS)$ from $\BS(\NS)$, respectively. 
%\end{remark}

\subsubsection{Unobservable enabling events}
%\begin{remark}
%\MR{Could we also have assumed that a subset of these $\Sigma_e$ events is unobservable? Is more general ...}
We could have assumed that some events from $\Sigma_e$ are unobservable. In this case, $\Sigma_a\subseteq\hat{\Sigma}_{uo}\subseteq\Sigma_a\cup\Sigma_e$, and so there would be more states that become observationally equivalent. Hence, the resulting supervisor could be more restrictive since a control command should be disabled at all observationally equivalent states if it needs to be disabled at one of them.
%\AR{@Martin: Do you think the following can be given as a seperate subsubsection or is better to be here or event to be removed?}
Also if the observation channel does not provide enough capacity, more states become observationally equivalent, resulting in a more conservative solution. To not introduce any observation losses, the observation channel needs to be such that it has the capacity for all observations of events executed in the plant; $\Mmax\geq \max_{w\in W}\{|P_{\Sigma_a}(w)|\}$ where  $W=\{w\in\Sigma^*_G \mid \exists w_0w\in L(G), |P_{\{\tick\}}(w)|\leq N_o\}$ as all events are observed after $N_o$ \ticks.
%\MF{I think that these requirements, both on $\max_o$ and $\max_c$, should be given very explicitly at some point, probably earlier than here. These are claims that are made with little supporting argument, which I think would be better to have up front. Preferably mathematical proofs, but I am not sure what that would look like. Also, as I mention above, it is not clear (to me) exactly which values the $\max$ is taken over.}
%\end{remark}

%\MR{we could also (in the future) provide a synthesis algorithm that given some size of channel guarantees nonblocking. Will indeed be more restrictive. This is interesting because channel size is not always a design choice.}
%\AR{The observation channel size does not hurt maximal permissiveness (as it does not change $P_{\Sigma_G}(L(\NSP))$). But, if we relax Assumption 2 (on control channels capacity), we cannot say $P_{\Sigma_G}(L(\NP))=L(G)$, and so we lose maximal permissiveness.}

%\MR{we could also (in the future) provide a synthesis algorithm that given some size of channel guarantees nonblocking. Will indeed be more restrictive. This is interesting because channel size is not always a design choice.}
%\AR{do you mean in future work saying something about relaxing Assumption 2? we do not have any assumption on observation channel capacity, and we just say if it not big enough, we may end up in a conservative result as more states become observationally equivalent. The control channel capacity should be big enough (Assumption 2) to not lose plant's behavior due to design issue.}

%\begin{remark}
\subsubsection{Non-forcible enabling events}
We could have assumed that some events from $\Sigma_e$ are not forcible. In this case, $\SigmaFor$ $\subseteq$ $\hatSigmaFor\subseteq\SigmaFor\cup\Sigma_e$.
%\AR{The following is not needed if we stay with the new controllability definition}
%In this case, we have $\Sigma_a_f\subseteq\hat{\Sigma}_f\subseteq\Sigma_a_f\cup\Sigma_e$, and the Definition \ref{dfn:NScont} is modified accordingly:
%Consider a plant $G$ and a networked supervisor $\NS$ with observation and control delays $N_o$ and $N_c$, respectively. Then, \NSP is timed networked controllable for $G$ if for any $w\in L(\NSP)$ and 
%\begin{equation*}
%u\in \Sigma_a_{uc}\cup \begin{cases}
       %\varnothing &\quad\text{if}\quad \exists_{\hat{\Sigma}_f\cup\Sigma_o}~w\sigma\in L(\NSP)  \\
%       \{tick\} &\quad\text{otherwise,}
%\end{cases}
%\end{equation*} 
%whenever $P_{\Sigma_G}(w)u\in L(G)$, then $wu\in L(\NSP)$.
Providing less forcible events makes the synthesis result more conservative since if the non-preemptable \tick leads to a bad state, the current state where \tick is enabled must be avoided as well (illustrated by Example \ref{exp:NS4BusPed}).
%We could also assume that events from $\Sigma_e$ are forcible events. In this case, $\hat{\Sigma}_f=\Sigma_a_f\cup\Sigma_e$. Having more forcible events makes the result becomes less conservative since if the event \emph{tick} and an event from $\Sigma_e$ are enabled at a state, the event \emph{tick} becomes controllable and can be disabled if it leads to a blocking state (see Example \ref{exp:NS4BusPed} for more details).
%\end{remark}

\begin{example}
\label{exp:NS4BusPed}
Consider the endangered pedestrian from Example \ref{exp:BusPedOP}. With the assumption that events from $\Sigma_e$ are forcible, the networked supervisor is given in Figure \ref{fig:NS4BusPed}. Without this assumption, there exists no networked supervisor.

\begin{figure}[htbp]
\centering
\begin{tikzpicture}[>=stealth',,shorten >=1pt,auto,node distance=2.2cm,scale = 0.6, transform shape]

\node[initial,initial text={},state]           (y0)                                    {$y_0$};
\node[state]                   (y1) [right of=y0]                       {$y_1$};

\node[state]                   (y2) [below of=y0]                       {$y_2$};
\node[state]                   (y3) [right of=y2]                       {$y_3$};
\node[state]                   (y4) [right of=y3]                       {$y_4$};
\node[state,accepting]                   (y5) [right of=y4]                       {$y_5$};
\node[state,accepting]                   (y6) [right of=y5]                       {$y_6$};

\node[state]                   (y7) [below of=y4]                       {$y_7$};
\node[state,accepting]                   (y8) [right of=y7]                       {$y_8$};
\node[state,accepting]                   (y9) [right of=y8]                       {$y_9$};

\path[->] (y0) edge [above,dashed]   node [align=center]  {$\tick$} (y1)
      (y0) edge [right]   node [align=center]  {$j_e$} (y2)
      (y1) edge [right]      node [align=center]  {$j_e $}    (y3)

      (y2) edge [above,dashed]      node [align=center]  {$\tick $}    (y3)
      (y3) edge [above,dashed]      node [align=center]  {$\tick $}    (y4)
      (y4) edge [above,dashed]      node [align=center]  {$\tick $}    (y5)
      (y5) edge [above]      node [align=center]  {$p_o$}    (y6)
      (y4) edge [right]      node [align=center]  {$j_o$}    (y7)
      (y5) edge [right]      node [align=center]  {$j_o$}    (y8)

      (y7) edge [above,dashed]      node [align=center]  {$\tick$}    (y8)

      (y8) edge [above]      node [align=center]  {$p_o$}    (y9)
      (y6) edge [right]      node [align=center]  {$j_o$}    (y9)
     
      (y9) edge [loop right,dashed] node [align=center]  {$\tick$} (y9)
;
\end{tikzpicture}
\caption{Networked supervisor for the endangered pedestrian from Example \ref{exp:BusPed} ($N_c=1,N_o=1$).}
\label{fig:NS4BusPed}
\end{figure}
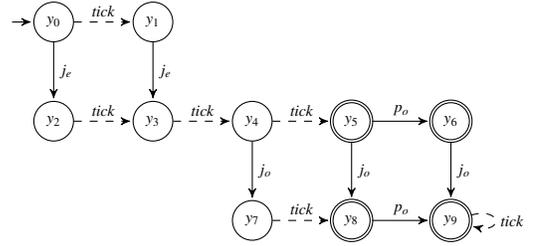
\end{example}

%\AR{WARNING remark with example:}
%\begin{remark}
\subsubsection{Non-FIFO control channel}
\label{remark:nonfifocontrolchannel}
Our proposed framework can easily be extended to the case that the control channel is non-FIFO by applying the following changes. Similar to the observation channel, the control channel is represented by $L=\{l\mid l:\Sigma\times[0,N_c]\rightarrow\mathbb{N}\}$ where $l$ is a multiset. So, for each $l\in L$ and the time counter $n$, we define the operators $l\uplus [(\sigma,n)]$ and $l\setminus[(\sigma,0)]$ instead of
%\mathit{app}
$\mathit{app}(l,(\sigma,n))$ and $\mathit{tail}(l)$, respectively. This affects item 1) of both Definition \ref{dfn:operator} and Definition \ref{dfn:NP} such that $(\sigma,N_c)$ is simply added to $l$ without taking into account the order of elements. Also, in item 2) of both definitions, $head(l)$ is replaced by $\exists (\sigma,0)\in l$. This may change the result pretty much as the enabling events can now be received by $G$ in any possible order. As Example \ref{ex:non-FIFO-CCH} illustrates, this may increase the chance of reaching blocking or time-lock states and result in very conservative solutions for many applications.
%\end{remark}

\begin{example}
\label{ex:non-FIFO-CCH}
Given a plant $G$ indicated in Figure \ref{fig:non-FIFO-CCH-G}, $N_c=N_o=1$, and $\Lmax=\Mmax=1$, $\NP$ is obtained as in Figure \ref{fig:non-FIFO-CCH-NP}. The networked supervisor computed by Algorithm \ref{algo} only disables the event $b_e$ at $x_0$.
Now, assume that the control channel is non-FIFO as well. Then, 
at $x_3=(\delta_G(a_0,\tick),\delta'(a'_0,\tick\,a\,b\,\tick),[],(a,0)(b,0))$, $b$ can be executed as well as $a$.
By executing $b$ at $x_3$, $\NP$ goes to a blocking state. In this case, Algorithm \ref{algo} returns no result since $x_0$ becomes a blocking state and needs to be removed.

\begin{figure}[htb]
\centering
\begin{tikzpicture}[>=stealth',,shorten >=0.8pt,auto,node distance=2.2cm,scale = 0.6, transform shape]

\node[initial,initial text={},state]           (A)                                    {$a_0$};
\node[state]         (B) [right of=A]                       {$a_1$};
\node[state]                   (C) [right of=B]                       {$a_2$};
\node[state,accepting]                   (D) [right of=C]                       {$a_3$};

%\node[state]                   (E0) [right of=E]                       {$a_5$};

\node[state]                   (F) [below of=B]                       {$a_4$};

\path[->] (A) edge [above,dashed]   node [align=center]  {$\tick$} (B)
(B) edge [above]   node [align=center]  {$a$} (C)
(C) edge [above]   node [align=center]  {$b$} (D)
(D) edge [loop right,dashed]      node [align=center]  {$\tick$} (D)

(B) edge [right]   node [align=center]  {$b$} (F)
(F) edge [loop right,dashed]      node [align=center]  {$\tick$} (F);
\end{tikzpicture}
\caption{Plant from Example \ref{ex:non-FIFO-CCH}.}
\label{fig:non-FIFO-CCH-G}
\end{figure}
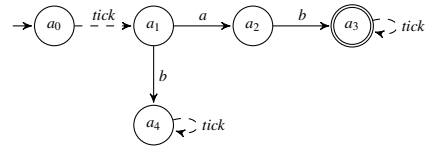

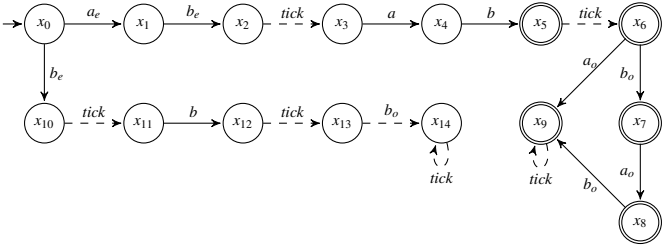
\begin{figure}[htb]
\centering
\begin{tikzpicture}[>=stealth',,shorten >=0.8pt,auto,node distance=2.2cm,scale = 0.6, transform shape]

\node[initial,initial text={},state]           (x0)                                    {$x_0$};
\node[state]         (x1) [right of=x0]                       {$x_1$};
\node[state]                   (x2) [right of=x1]                       {$x_2$};
\node[state]                   (x3) [right of=x2]                       {$x_3$};
\node[state]                   (x4) [right of=x3]                       {$x_4$};
\node[state,accepting]                   (x5) [right of=x4]                       {$x_5$};
\node[state,accepting]                   (x6) [right of=x5]                       {$x_6$};

\node[state,accepting]                   (x7) [below of=x6]                       {$x_7$};
\node[state,accepting]                   (x8) [below of=x7]                       {$x_8$};
\node[state,accepting]                   (x9) [below of=x5]                       {$x_9$};

\node[state]                   (x10) [below of=A]                       {$x_{10}$};
\node[state]                   (x11) [right of=x10]                       {$x_{11}$};
\node[state]                   (x12) [right of=x11]                       {$x_{12}$};
\node[state]                   (x13) [right of=x12]                       {$x_{13}$};
\node[state]                   (x14) [right of=x13]                       {$x_{14}$};

\path[->] (x0) edge [above]   node [align=center]  {$a_e$} (x1)
(x1) edge [above]   node [align=center]  {$b_e$} (x2)
(x2) edge [above,dashed]   node [align=center]  {$\tick$} (x3)
(x3) edge [above]      node [align=center]  {$a$} (x4)
(x4) edge [above]   node [align=center]  {$b$} (x5)
(x5) edge [above,dashed]   node [align=center]  {$\tick$} (x6)
(x6) edge [left]   node [align=center]  {$b_o$} (x7)
(x6) edge [above]   node [align=center]  {$a_o$} (x9)
(x7) edge [left]   node [align=center]  {$a_o$} (x8)
(x8) edge [below]   node [align=center]  {$b_o$} (x9)
(x9) edge [loop below,dashed]      node [align=center]  {$\tick$} (x9)

(x0) edge [right]   node [align=center]  {$b_e$} (x10)
(x10) edge [above,dashed]   node [align=center]  {$\tick$} (x11)
(x11) edge [above]   node [align=center]  {$b$} (x12)
(x12) edge [above,dashed]   node [align=center]  {$\tick$} (x13)
(x13) edge [above,dashed]   node [align=center]  {$b_o$} (x14)
(x14) edge [loop below,dashed]      node [align=center]  {$\tick$} (x14);
\end{tikzpicture}
\caption{Networked plant from Example \ref{ex:non-FIFO-CCH}.}
\label{fig:non-FIFO-CCH-NP}
\end{figure}
\end{example}

\section{Requirement Automata}
\label{section:requirements}
%\AR{Since the control commands are sent in advance,we need to first complete the requirements not only for uncontrollable events but also for controllable one (to see if in-advance command will cause blocking or not)}
%\AR{We may need to skip all the preliminaries and use an example to show that we need to first complete requirement automata regardless of being controllable or not, and then apply the method on $P||R^{tot}$}
To generalize the method to a wider group of applications, we solve the basic NSC problem for a given set of control requirements. It is assumed that the desired behavior of $G$, denoted by the TDES $R$, is represented by the automaton  $(Q,\Sigma_R,\delta_R,q_0,Q_M)$ where $\Sigma_R\subseteq \Sigma_G$.
Since most control requirements are defined to provide safety of a plant, we call a supervised plant \emph{safe} if it satisfies the control requirements. 
%Definition \ref{dfn:safety} defines the safety of a networked supervised plant.

\begin{defn}[Safety]
\label{dfn:safety}
Given a plant $G$ and requirement $R$, a TDES $\mathit{NSP}$ with event set $\SigmaNSP$ is safe w.r.t.\ $G$ and $R$ if its behavior stays within the legal/safe behavior as specified by $R$; $P_{\SigmaNSP\cap\Sigma_R}(L({\mathit{NSP}}))\subseteq P_{\SigmaNSP\cap\Sigma_R}(L(R))$.
\hfill $\blacksquare$
\end{defn}

%\setlength{\fboxrule}{0.9pt}
%\begin{center}
%\framebox{
%\centering
%\begin{minipage}{0.9\linewidth}
\textbf{Problem Statement:}
Given a plant model $G$ as a TDES, control requirement $R$ for $G$ (also a TDES),
observation (control) channel with delay $N_o$ ($N_c$) and maximum capacity $\Mmax$ ($\Lmax$),
provide a networked supervisor \NS such that
\begin{itemize}
    \item \NSP is nonblocking,
    \item \NSP is time-lock free,
    \item \NS is timed networked controllable w.r.t.\ $G$,
    \item \NS is timed networked maximally permissive, and
    \item \NSP is safe for $G$ w.r.t.\ $R$.
\end{itemize}

In the conventional non-networked supervisory control setting, if $R$ is controllable w.r.t.\ $G$ (as defined in Definition \ref{dfn:cont.TDES}), then an optimal nonblocking supervisor can be synthesized for $G$ satisfying $R$~\cite{Wonham:19}.
%\AR{We do not need the following definition if we generalize the previous definition of controllability of TDES. \MF{No, I do not think so. It is much better to be slightly more general in Def 6, than to again give almost the same def here. But in any case, please use \eqref{eq:P} instead of writing ``Equation~\ref{eq:P}''; these are not really equations, despite what \LaTeX~calls them. } }
%\begin{defn}[Controllable Requirement]
%\label{dfn:ctrl.R}
%Consider the plant $G$ and requirement $R$ represented in Equation \ref{eq:P} and Equation \ref{eq:R}, respectively. $R$ is said to be controllable w.r.t. $G$ if whenever $w\in L(P||R)$ and $wu\in L(G)$ for some 
%\begin{center}
%$u\in \Sigmauc\cup \begin{cases}
%       \text{$\varnothing$} &\quad\text{if $\exists\sigma\in\SigmaFor~w\sigma\in (L(P||R))$}  \\
%       \text{$\{\tick\}$} &\quad\text{otherwise,}
%\end{cases}$
%\end{center} 
%then $wu\in L(R)$.
%\hfill
%$\blacksquare$
%\end{defn}
%\AR{Check ctrl.R defenition in Wonhams book}
%\subsection{Uncontrollable Requirement}
If $R$ is not controllable w.r.t.\ $G$, then the supremal controllable sublanguage of $G||R$, indicated by $sup \mathcal{C}(G||R)$, should be calculated. Then, the synthesis is applied on $sup \mathcal{C}(G||R)$~\cite{Cassandras:99,Wonham:19}.

In a networked supervisory control setting, synthesizing a networked supervisor for $sup \mathcal{C}(G||R)$ does not always result in a safe networked supervised plant. This issue occurs due to the fact that in $sup \mathcal{C}(G||R)$, some events are already supposed to be disabled, to deal with controllability problems introduced by requirement $R$. In a conventional non-networked setting, this does not cause a problem because events are observed immediately when executed. However, when observations are delayed, there could be a set of states reached by the same observation. 
%\MF{So... the problem is about observability and in essence non-determinism? Is not this solved already, for instance by Karen Rudie long ago?}
%\AR{There exists language-based synthesis approaches for supervisory control under partial observations in which the observability condition is defined and if it does not hold, the supremal observable sublanguage should be computed! Since the observability condition is not closed under union, it is not possible to compute the supremal observable language, and so stronger conditions such as relative observability and normality are defined~\cite{Wonham:19-review}. Also, for NSC, as stated in~\cite{Lin20}, they defined network observality condition, but how to find the network observable sublanguage is still an open problem.}
%\AR{But, anyhow we do not deal with observability problem here as we assumed that all events occurring in the plant are observable. For this reason, we do not even have the definition of networked observaility (like the one we have for controlability). We just need to deal with non-determinism in synthesizing $\NS$ for $\NP$.}
Hence, if an event is disabled at a state, it should be disabled at all observationally equivalent ones.
Even for a controllable requirement, any disablement of events should be considered at all observationally equivalent states. 

To take care of this issue, any requirement automaton $R$ (whether controllable or uncontrollable) is made complete as $R^{\bot}$ in terms of both uncontrollable and controllable events.
\emph{Completion} was first introduced in~\cite{Flordal:07} where the requirement automaton $R$ is made complete in terms of only uncontrollable events. 
%as $R^\bot$ by adding a new blocking state called $q_d$ to the set of states of $R$. Then, whenever $R$ disables an uncontrollable event $\sigma_u$ in a state $q$, a transition $\langle q, \sigma_u, q_d\rangle$ is added in $R^\bot$.
By applying the synthesis on $G||R^\bot$, all original %\MR{Do not use initial in cases where you mean original.}
controllability problems in $G||R$ are translated to blocking issues.
%\MF{All \emph{initial} controllability problems are converted to blocking. But it is important that when synthesis then starts, controllability still has to be considered. I say this, because there has been some confusion about this. ``plantification'' as we call it, only points out the initially bad states (``bad'' in the sense of breaking controllability), it does not make \emph{all} controllability issues into blocking issues. The controllability fix-point calculation must still be done. Maybe this warrants a slight change, saying ``all initial controllability problems in $P||R$''. Or will that confuse more than help?}
Note that this translation is necessary to let the supervisor know about the uncontrollable events that are disabled by a given requirement. To solve the blocking issues, synthesis still takes the controllability definition into account. %\MF{Yes, that sounds good.}

\begin{defn}[Automata Completion]
\label{dfn:Rtot}
For a TDES $R=(Q,\Sigma_R,\delta_R,q_0,Q_M)$, the complete automaton $R^{\bot}$ is defined as
$R^{\bot}=(Q\cup\{q_d\},\Sigma_R,\delta^{\bot}_R,q_0,Q_M)$ with $q_d\notin Q$, where for every $q\in Q$ and $\sigma\in\Sigma_R$,
\begin{equation*}
\delta^{\bot}_R(q,\sigma)=\begin{cases}\text{$\delta_R(q,\sigma)$} &\quad\text{if $\delta_R(q,\sigma)!$}\\
\text{$q_d$} &\quad\text{otherwise}.
\end{cases}    
\end{equation*}
%\MF{So $q_d$ allows arbitrary continuations (the language from $q_d$ is $\Sigma_R^*$), is this what you want? I'm just asking, because that is not the way ``completion'' (what we call ``plantification'') is done in Supremica. It is unnecessary to allow the plant to continue once the dump state is reached, as we know that no marking can ever be reached. So it is a kind of optimization to not have self-loops in $q_d$. BTW\dots no example?}
%\AR{The self-loops were there due to obsevationally equivalent disablements that we consider in the algorithm ... but thinking about it again, it does not seem to hurt not to have those ... so I removed them ... but then I should not refer to \cite{Cassandras:99} anymore and maybe better not to use the term totalization? shall we then say we use completion from \cite{Flordal:07} but in terms of both controllable and uncontrollable events?}
%\AR{I removed the example to save space ... it is there now and we can decide  later to have it or not}
\hfill
$\blacksquare$
\end{defn}

To find a networked supervisor, Algorithm~\ref{algo} is applied on $\Pi(G||R^{\bot},N_c,N_o,\Lmax,\Mmax)$.
%To simplify, $G||R^{\bot}$ is denoted by $G^{t}$, the networked plant $\Pi(G^{t},N_c,N_o,\Lmax,\Mmax)$ by $NP^{t}$ and the networked supervised plant $\NS_{N_c}\|_{N_o}\,G^{t}$ by $\NSP^{t}$.
The obtained networked supervisor is already guaranteed to be timed networked controllable, timed networked maximally permissive, and it results in a nonblocking and time-lock free networked supervised plant.
Theorem \ref{theorem:safety} shows that the networked supervised plant is safe as well.

\begin{theorem}[Safe NSP]
\label{theorem:safety}
Given a plant $G$, requirement $R$, and the networked supervisor $\NS$ computed by Algorithm~\ref{algo} for $\Pi(G||R^{\bot},N_c,N_o,\Lmax,\Mmax)$: $\NS_{N_c}\|_{N_o}\,(G||R^{\bot})$ is safe for $G$ w.r.t.\ $R$.
%\hfill $\blacksquare$
\end{theorem}

\begin{proof}
See Appendix \ref{proof:safety}.
\hfill $\blacksquare$
\end{proof}

\section{Conclusions and Future Work}
\label{section:conclusions}
In this paper, we study the networked supervisory control synthesis problem.
%The objective is to provide a networked supervisor satisfying controllability, nonblockingness, and control requirements under communication delays.
We first introduce a networked supervisory control framework in which both control and observation channels introduce delays, the control channel is FIFO, and the observation channel is non-FIFO.
Moreover, we assume that a global clock exists in the system such that the passage of a unit of time is considered as an event \emph{tick} in the plant model. Also, communication delays are measured as a number of occurrences of the \emph{tick} event. In our framework, uncontrollable events occur in the plant spontaneously. However, controllable events can be executed only if they have been enabled by the networked supervisor. On the other hand, the plant can either accept a control command (enabled by the networked supervisor) and execute it or ignore the control command and execute some other event. For the proposed framework, we also provide an asynchronous composition operator to obtain the networked supervised plant. Furthermore, we adapt the definition of conventional controllability for our framework and introduce timed networked controllability. Then, we present a method of achieving the networked plant automaton representing the behavior of the plant in the networked supervisory control framework. For the networked plant, we provide an algorithm synthesizing a networked supervisor which is timed networked controllable, nonblocking, time-lock free, and maximally permissive. Finally, to generalize, we solve the problem for a given set of control (safety) requirements modeled as automata. We guarantee that the proposed technique achieves a networked supervisor that is timed networked controllable, nonblocking, time-lock free, maximally permissive, and safe.  

Our proposed approach can be adjusted to a setting with observation delay and control delay specified to each event, a setting with bounded control and observation delays, or to a setting with lossy communication channels.
In each case, only the timed asynchronous composition and networked plant operators need to be updated, the synthesis algorithm stays the same.
%\AR{I think that the following sentence is better to be given as a remark than a future work .... I already added a Remark}
%In this paper, delays are assumed to be constant. In future, we aim to generalize the approach for bounded amount of delays.
%\AR{What about different fixed delays for events? I think our setting is easily adjustable to this situation. but, I am not sure if how things change with bounded delays and if it is better to give this as remark because we already know how to deal with non-FIFO channels!}\AR{The main concern for any of these changes is how we should change Assumption 1 and the way we look ahead to determine controllable events ...  }
For  cases  with  large  state  spaces, we  must deal  with  the scalability problem of the networked plant. For such cases, it is suggested to switch to timed automata. A supervisory control synthesis method for timed automata has been recently proposed by the authors~\cite{Rashidinejad:20}. Networked supervisory control of timed automata will be investigated in future research.

%This issue will be solved by replacing discrete time by dense time and modeling the plant as timed automata.

%\AR{future work: dealing with plant with unobservable events??
%if the plant has unobservable events ... then, we need to define timed networked observability, consider it in the algo and prove it ...
%} \MR{I am not a big fan of future work parts. They typically are either research that is already solved, but not published yet, or plans that the author will never undertake. What is Martin's opinion? I am flexible here.}

\appendices
\section{Technical Lemmas}
Here, the notation $.$ is used to refer to an element of a tuple. For instance, $z.a$ refers to the (first) element $a$ of $z=(a,y,m,l)$.

\begin{lemma}[Nonblockingness over Projection\cite{Rashidinejad18}]
\label{lemma:projectionNB}
For any TDES $G$ with event set $\Sigma$ and any event set $\Sigma'\subseteq \Sigma$: if $G$ is nonblocking, then $P_{\Sigma'}(G)$ is nonblocking.
%\hfill$\blacksquare$
\end{lemma}
%\MF{$P$ for plant, and $P_\Sigma$ for projection. Can this not be avoided? Using $G$ for plant?}

\begin{proof}
Consider an arbitrary TDES $G = (A,\Sigma,\delta,a_0,A_m)$ and arbitrary $\Sigma'\subseteq \Sigma$. Suppose that $G$ is nonblocking. Consider an arbitrary reachable state $A_r \subseteq A$ in $P_{\Sigma'}(G)$. By construction $A_r$ is nonempty. Assume that this state is reached through the word $w \in \Sigma'$. Then, for each state $a \in A_r$, again by construction, $\delta(a_0,w') = a$ for some $w'\in \Sigma^*$ with $P_{\Sigma'}(w')=w$. Because $G$ is nonblocking, there exists a $v'\in \Sigma^*$ such that $\delta(a,v') =a_m$ for some $a_m \in A_m$. Consequently, from state $A_r$, it is possible to have a transition labelled with $P_{\Sigma'}(v')$ to a state $A'_r$ containing $a_m$. By construction, this state $A'_r$ is a marked state in $P_{\Sigma'}(G)$. Hence, the projection automaton is nonblocking as well.
\hfill $\blacksquare$
\end{proof}

\begin{lemma}[Time-lock Freeness over Projection\cite{Rashidinejad18}]
\label{lemma:projectionTLF}
For any TDES $G$ with event set $\Sigma$ and any event set $\Sigma'\subseteq \Sigma$, $\tick\in \Sigma'$: if $G$ is TLF
, then $P_{\Sigma'}(G)$ is TLF.
%\hfill $\blacksquare$
\end{lemma}

\begin{proof}
The proof is similar to the proof of Lemma \ref{lemma:projectionNB}. 
\hfill$\blacksquare$
\end{proof}

\begin{lemma}[\NSP Transitions]
\label{lemma:NSP}
Given a plant $G$, networked supervisor $\NS$ (for that plant) and networked supervised plant $\NSP$ (for those): $\deltaNSP(z_0,w).a=\delta_G(a_0,P_{\Sigma_G}(w))$ and
$\deltaNSP(z_0,w).y=\deltaNS(y_0,P_{\SigmaNS}(w))$, for any $w\in L(\NSP)$.
%\MR{Is it necessary to require that $w \in L(\NSP)$?}\AR{I am confused ... Do you mean we should not say anything about $w$ or saying $w\in \Sigma^*_{NSP}$?}
%\hfill$\blacksquare$ 
\end{lemma}

\begin{proof}
Take $w\in L(\NSP)$, we show that $\deltaNSP(z_0,w).a=\delta_G(a_0,P_{\Sigma_G}(w))$ and
$\deltaNSP(z_0,w).y=\deltaNS(y_0,P_{\SigmaNS}(w))$. This is proved by induction on the structure of $w$.
\textbf{Base case:} Assume $w=\epsilon$. Then, %$\delta_G(a_0,P_{\Sigma_G}(\epsilon))!\And \deltaNS(v_0,P_{\SigmaNS}(\epsilon))!$ and so $\deltaNSP(z_0,\epsilon)!$. Also, 
%$\deltaNSP(z_0,w)=(a_0,y_0,[],\epsilon)=z$ where 
$\deltaNSP(z_0,w).a=a_0=\delta_G(a_0,P_{\Sigma_G}(\epsilon))$ and $\deltaNSP(z_0,w).y=y_0=\deltaNS(y_0,P_{\SigmaNS}(\epsilon))$. 
\textbf{Induction step:} Assume that $w=v\sigma$ where the statement holds for $v$, i.e., $\deltaNSP(z_0,v).a=\delta_G(a_0,P_{\Sigma_G}(v))$ and $\deltaNSP(z_0,v).y=\deltaNS(y_0,P_{\SigmaNS}(v))$.
It suffices to prove that the statement holds for $v\sigma$, i.e., $\deltaNSP(z_0,v\sigma).a=\delta_G(a_0,P_{\Sigma_G}(v\sigma))$ and $\deltaNSP(z_0,v\sigma).y=\deltaNS(y_0,P_{\SigmaNS}(v\sigma))$.
%with $\deltaNSP(z_0,v)!$ which means that $\delta_G(a_0,P_{\Sigma_G}(v))!\And \deltaNS(v_0,P_{\SigmaNS}(v))!$. Also by 
%By induction we have \MF{"by induction"? Is this not "by definition"?} \AR{this is what we suppose that holds for $v$ and we prove for $v\sigma$, so it is by induction ...}
%%MF \MF{Sorry, but it holds by definition; Def 6. You also give (almost) the same expression in the body of Property/Lemma 2. What holds by induction, is what you use this starting point to prove. It goes like: 1. Base case. 2. Assume it holds for case n. 3. Then by induction it holds for case n+1. And I was taught by my math professor that an induction proof \emph{must} always end by something like "by the principle of induction this proves the claim", but I can live without that part.}\AR{edited ... please check}
%%MF Here are some nice examples of proof by induction https://faculty.math.illinois.edu/~hildebr/213/inductionsampler.pdf
%$\deltaNSP(z_0,v)=(\delta_G(a_0,P_{\Sigma_G}(v)),\deltaNS(y_0,P_{\SigmaNS}(v)),m',l')$ for some $m'\in M$ and $l'\in L$.
%It suffices to prove that $\deltaNSP(z_0,v\sigma)=(\delta_G(a_0,P_{\Sigma_G}(v\sigma)),\deltaNS(y_0,P_{\SigmaNS}(v\sigma)),m,l)$ for some $m\in M$ and $l\in L$.
%It suffices to prove that $z.a=\delta(a',P_{\Sigma_G}(\sigma))$ and $z.y=\deltaNS(y',P_{\SigmaNS}(\sigma))$.
Considering Definition \ref{dfn:operator}, for $\sigma$ enabled at $\deltaNSP(z_0,v)$ the following cases may occur:

$\sigma\in\SigmaNS\setminus\{\tick\}$, which refers to item 1) and item 5). Then, $\deltaNSP(z_0,v).a$ remains unchanged; $\deltaNSP(z_0,v\sigma).a=$ $\deltaNSP(z_0,v).a=$ $\delta_G(a_0,P_{\Sigma_G}(v))=$ $\delta_G(a_0,P_{\Sigma_G}(v)\epsilon)=$ $\delta_G(a_0,P_{\Sigma_G}(v\sigma))$, and
$\deltaNSP(z_0,v\sigma).y=\deltaNS(\deltaNS(y_0,P_{\SigmaNS}(v),\sigma)=\deltaNS(y_0,P_{\SigmaNS}(v
)\sigma)=\deltaNS(y_0,P_{\SigmaNS}(v\sigma))$.

$\sigma\in \Sigma_G\setminus\{tick\}$, which refers to item 2) and item 3). Then, $\deltaNSP(z_0,v\sigma).a=\delta_G(\delta_G(a_0,P_{\Sigma_G}(v)),\sigma)=\delta_G(a_0,P_{\Sigma_G}(v)\sigma)=\delta_G(a_0,P_{\Sigma_G}(v\sigma))$, and $\deltaNSP(z_0,v\sigma).y$ remains unchanged; $\deltaNSP(z_0,v\sigma).y=\deltaNSP(z_0,v).y=\deltaNS(y_0,P_{\SigmaNS}(v)\epsilon)=\deltaNS(y_0,P_{\SigmaNS}(v\sigma))$.

$\sigma=\tick$, which refers to item 4). Then, $\deltaNSP(z_0,v\sigma).a=\delta_G(\delta_G(a_0,P_{\Sigma_G}(v)),\sigma)=\delta_G(a_0,P_{\Sigma_G}(v)\sigma)=\delta_G(a_0,P_{\Sigma_G}(v\sigma))$, and $\deltaNSP(z_0,v\sigma).y=\deltaNS(\deltaNS(y_0,P_{\SigmaNS}(v)),\sigma)=\deltaNS(y_0,P_{\SigmaNS}(v)\sigma)=\deltaNS(y_0,P_{\SigmaNS}(v\sigma))$.
\textbf{Conclusion:} By the principle of induction, the statement ($\deltaNSP(z_0,w).a=\delta_G(a_0,P_{\Sigma_G}(w))$ and $\deltaNSP(z_0,w).y=\deltaNS(v_0,P_{\SigmaNS}(w))$) holds for all $w\in L(\NSP)$.
\hfill$\blacksquare$
\end{proof}

\begin{lemma}[NP Transitions]
\label{lemma:NP}
Given a plant $G$ with $x_0.a=a_0$, for any $w\in L(\NP)$:  $\deltaNP(x_0,w).a=\delta_G(a_0,P_{\Sigma_G}(w))$.
%\hfill$\blacksquare$
\end{lemma}

\begin{proof}
The proof is similar to the proof of Lemma \ref{lemma:NSP}.
\hfill $\blacksquare$
\end{proof}

\begin{lemma}[NP Enabling Commands]
\label{lemma:ontime}
Given a plant $G$, events from $\Sigma_e$ are enabled on time in the networked plant $\NP$ (for that plant); for any $w\sigma\in L(G)$, $\sigma\in \Sigma_c$: there exists $w_0\sigma_e w_1 \sigma\in L(\NP)$ where $\sigma_e\in\Sigma_e$ is the enabling event of $\sigma$, $P_{\Sigma_G}(w_0w_1)=w$ and $|P_{\{\tick\}}(w_1)|=N_c$.
%\hfill$\blacksquare$
\end{lemma}

\begin{proof}
Assume that $G'=P_{\Sigma_G\setminus\Sigmauc}(G)$ is represented by $(A', \Sigma_G, \delta'_G, a'_{0}, A'_{m})$, and let us do the proof by induction on
%the ordinal number of controllable events enabled in $G$.\MR{We need to do induction 
the number of controllable events in $w\in L(G)$. \textbf{Base case:} Assume that $\sigma$ is the $1^{th}$ controllable event enabled in $G$. Then, $w\in (\Sigmauc\cup\{\tick\})^*$.
According to Assumption 1, $|P_{\{\tick\}}(w)|\geq N_c$.
Let say $|P_{\{\tick\}}(w)|=N_c+i$ for some $i\in\mathbb{N}_0$. 
Then, $\tick^{N_c+i}\sigma\in L(G')$.
Also, assume that $w=w_iw_{N_c}$ for some $w_i,w_{N_c}$ where $|P_{\{\tick\}}(w_i)|=i$, and $|P_{\{\tick\}}(w_{N_c})|=N_c$.
%and $w_{N_c}$ starts with the event $\tick$ (so all uncontrollable events between $i^{th}$ \tick and the $(i+1)^{th}$ one are considered in $w_i$).
Considering Definition \ref{dfn:NP}, $\NP$ starts from $x_0=(a_0,\delta'_G(a'_0,\tick^{N_c}),[],\epsilon)$. Then, based on item 4), \tick occurs in $\NP$ when it is enabled in both $G$ and $G'$. For the first $i$~\ticks, whenever \tick is enabled in $G$, it is also enabled in $G'$ (there are $i$ \ticks enabled in $G'$ before $\sigma$ occurs).  
Meanwhile, if there is an event ready to be observed, then based on item 5), the corresponding observed event occurs in $\NP$ which does not change the current state of $G$ and $G'$.
%, and so \tick occurs after the execution of all those (ready) observed events.
Also, based on item 3), if an uncontrollable event is enabled in $G$, it occurs in $\NP$ without changing the state of $G'$.
%Whenever there is no more event ready to be observed or an uncontrollable event enabled in $G$, then
Otherwise, \tick occurs in $\NP$ by being executed in both $G$ and $G'$.
We call this situation as $G$ and $G'$ are synchronized on \tick.
Therefore, it is feasible that some $w_0$ is executed in $\NP$ based on the execution of $\tick^{i}$ in $G'$ and $w_i$ in $G$.  Then, $\delta_{NP}(x_0,w_0).a=\delta_G(a_0,w_i)$ and $\delta_{NP}(x_0,w_0).a'=\delta'_G(a_0,\tick^{N_c+i})$, and so $P_{\Sigma_G}(w_0)=w_i$.
After that, since  $(\delta_{NP}(x_0,w_0).a',\sigma)!$, based on item 1), $\sigma_e$ occurs in $\NP$, and $(\sigma,N_c)$ is added to $\delta_{NP}(x_0,w_0).l$. Note that based on item 3) (item 5)), uncontrollable events enabled in $G$ (events ready to be observed) can occur in between, but without loss of generality, let us assume that $\sigma_e$ is enabled first, and then uncontrollable (observed) events are executed.
So, $w_1$ will be executed in $NP$ based on the execution of $w_{N_c}$ in $G$. Therefore, $P_{\Sigma_G}(w_1)=w_{N_c}$, and $|P_{\{\tick\}}(w_{1})|=|P_{\{\tick\}}(w_{N_c})|=N_c$. Based on item 4), by the execution of each \tick, $\delta_{NP}(x_0,w_0\sigma_e).l$ is decreased by one. Also, $\sigma$ is the only controllable event enabled in $G$ so that $head(\delta_{NP}(x_0,w_0\sigma_ew_1).l)=(\sigma,0)$. Then, based on item 2), $\sigma$ will be executed in $\NP$. 
%and so there exists some $v_0,v_1\in (\Sigmauc\cup\{\tick\})^*$  with $|P_{\{\tick\}}(v_0)|=N_c$ such that $w=v_0v_1$. 
%Let say $\delta_G(a_0,v_0)=a_1$, $\delta(a_1,v_1)=a_2$ for some $a_1,a_2\in A$. Then, $\tick^{N_c}P_{\Sigma\setminus\Sigmauc}(v_1)\sigma\in L(G')$. Let say, $\delta'_G(a'_0,\tick^{N_c})=a'_1$, $\delta'_G(a'_1,P_{\Sigma\setminus\Sigmauc}(v_1))=a'_2$ for some $a'_1,a'_2\in A'$. 
%Also assume that $|P_{\{\tick\}}(v_1)|=i$
\textbf{Induction step:} Assume that $\sigma$ is the $n^{th}$ controllable event enabled in $G$ where the statement holds for all previous controllable events.
Let us indicate the $(n-1)^{th}$ controllable event by $\sigma^{n-1}$ such that $w_{n-1}\,\sigma^{n-1}\in L(G)$. As the statement holds for $\sigma^{n-1}$, there exists some $w^{n-1}_0\,\sigma^{n-1}_e\,w^{n-1}_1\,\sigma^{n-1}\in L(NP)$ such that $P_{\Sigma_G}(w^{n-1}_0\,w^{n-1}_1)=w_{n-1}$ and $|P_{\{\tick\}}(w^{n-1}_1)|=N_c$. It suffices to prove that for the next controllable event $\sigma^n$, $w_n\,\sigma^n\in L(G)$, there exists some $w^n_0\,\sigma_e\,w^n_1\,\sigma^n\in L(NP)$ with $P_{\Sigma_G}(w^n_0\,w^n_1)=w_n$ and $|P_{\{\tick\}}(w^n_1)|=N_c$.
Let us say $w_n=w_{n-1}\,\sigma^{n-1}w$ where $w\in (\Sigmauc\cup\{\tick\})^*$. Assume that $|P_{\{\tick\}}(w)|=j$, and
%\AR{we have 2 cases making the proof complex ... 1. $w^{n-1}_0\sigma^{n-1}_ew^{n-1}_1\sigma^{n-1}w'\sigma_ew_1\sigma$ if $j\geq N_c$ 2. $w^{n-1}_0\sigma^{n-1}_ew'_1\sigma_ew'_2\sigma^{n-1}w'_3\sigma$ if $j\leq N_c$ (the only thing we know by induction assumption now is the execution of the words including controllable events as well ...)}
$w_{n-1}=w^{n-1}_i\,w^{n-1}_{N_c}$ where $|P_{\{\tick\}}(w^{n-1}_i)|=i$, and $|P_{\{\tick\}}(w^{n-1}_{N_c})|=N_c$.
Moreover, let us say for $w_{n-1}\,\sigma^{n-1}\,w\,\sigma\in L(G)$, there exists $\tick^{N_c}\,w'_{n-1}\,\sigma^{n-1}\,\tick^j\sigma^n\in L(G')$ where $|P_{\{\tick\}}(w'_{n-1})|=i$.
Considering Definition \ref{dfn:NP}, $G'$ synchronizes with $G$ on executing \tick since whenever \tick is enabled in $G'$, it occurs in $\NP$ only if $G$ enables it as well. Also, uncontrollable events (observed events) occur as they are enabled in $G$ (as the corresponding event is ready to be observed), and due to the induction assumption, all controllable events occurring in $G$ before $\sigma^n$ are enabled on time, and so they will be executed in $\NP$.
By the execution of $w^{n-1}_0\,\sigma^{n-1}_e$ in $\NP$, $\deltaNP(x_0,w^{n-1}_0\,\sigma^{n-1}_e).a'=\delta'_G(a'_0,\tick^{N_c}\,w'_{n-1}\,\sigma^{n-1})$, and so $\deltaNP(x_0,w^{n-1}_0\,\sigma^{n-1}_e).a=\delta_G(a_0,w^{n-1}_i)$ ($G$ and $G'$ synchronize on \tick).
%Considering Definition \ref{dfn:NP}, $\sigma^{n-1}_e$ occurs in $\NP$ as $\sigma^{n-1}$ is enabled in $G'$. Staring from $x_0$, $\sigma^{n-1}$ is enabled in $G'$ after the execution of $w'_{n-1}$ which includes $i$ \ticks pass \AR{no! here we need to make sure that $w^{n-1}_i$ is executed in $G$ which holds based the induction assumption},
At this point, (before reaching $\sigma$) in $G'$, $\tick^j$ is enabled, and $w^{n-1}_{N_C}$ is enabled in $G$ (before reaching $\sigma^{n-1}$).
Then, one of the following cases may occur:

$j< N_c$. Then,
assume $w^{n-1}_{N_c}=w^{n-1}_jw^{n-1}_{N_c-j}$ for some $w^{n-1}_j,w^{n-1}_{N_c-j}$ where $|P_{\{\tick\}}w^{n-1}_j|=j$ and $|P_{\{\tick\}}w^{n-1}_{N_c-j}|=N_c-j$.
Then, the execution of $\tick^j$ in $G'$ is synchronized with the execution of $w^{n-1}_j$ in $G$ resulting in $w^{n-1}_0\sigma^{n-1}_ev_1\in L(NP)$ with $|P_{\{\tick\}}(v_1)|=j$ and $P_{\Sigma_G}(v_1)=w^{n-1}_{j}$. After that $\sigma_e$ occurs in $\NP$ (as it is enabled in $G'$) adding $(\sigma,N_c)$ to $l$. This follows by the execution of $w^{n-1}_{N_c-j}$ in $G$ and results in $w^{n-1}_0\sigma^{n-1}_ev_1\sigma^n_ev_2\in L(\NP)$ where $|P_{\{\tick\}}(v_2)|=N_c-j$ and $P_{\Sigma_G}(v_2)=w^{n-1}_{N_c-j}$.
At this point, $\sigma^{n-1}$ is executed in $\NP$ following by the execution of $v_3$ where $P_{\Sigma_G}(v_3)=w$, and so $|P_{\{\tick\}}(v_3)|=j$.
This results in $w^{n-1}_0\,\sigma^{n-1}_e\,v_1\,\sigma^n_e\,v_2\,\sigma^{n-1}\,v_3\in L(\NP)$ where
$P_{\{\tick\}}(v_2\sigma^{n-1}v_3)=N_c-j+j=N_c$, and $head(l)=(\sigma^n,0)$ (after the execution of $\sigma^{n-1}$, this is only $(\sigma^n,N_c)$ in $l$, and $l$ is decreased by one by the execution of each \tick), and so $\sigma^n$ occurs in $\NP$.
Hence, $w^n_0\,\sigma^n_e\,w^n_1\,\sigma^n\in L(\NP)$ for $w^n_0=w^{n-1}_0\,\sigma^{n-1}_e\,v_1$ and $w^n_1=v_2\,\sigma^{n-1}\,v_3$ where $P_{\Sigma_G}(w^n_0\,w^n_1)=P_{\Sigma_G}(w^{n-1}_0\,\sigma^{n-1}_e\,v_1\,v_2\,\sigma^{n-1}\,v_3)=w^n$ and $|P_{\{\tick\}}(w^n_1)|=N_c-j+j=N_c$.

$j> N_c$. Then, assume $w=w_{j-N_c}w_{N_c}$ for some $w_{j-N_c},w_{N_c}$ where $|P_{\{\tick\}}w_{j-N_c}|=j-N_c$ and $|P_{\{\tick\}}w_{N_c}|=N_c$.
The execution of $\tick^{N_c}$ in $G'$ is synchronized with the execution of $w^{n-1}_{N_c}$ in $G$ resulting in $w^{n-1}_0\,\sigma^{n-1}_e\,w^{n-1}_1\,\sigma_{n-1}\in L(NP)$. After that, the execution of the remaining $\tick^{j-N_c}$ in $G'$ will be synchronized with execution of $w_{j-N_c}$ in $G$ resulting in $w^{n-1}_0\,\sigma^{n-1}_e\,w^{n-1}_1\,\sigma^{n-1}\,v_1\in L(\NP)$ where $P_{\Sigma_G}(v_1)=w_{j-N_c}$. Then, $\sigma^n_e$ occurs in $\NP$ as it is enabled in $G'$ adding $(\sigma^n,N_c)$ to $l$. 
Finally, the execution of $w_{N_c}$ in $G$ results in $w^{n-1}_0\sigma^{n-1}_ew^{n-1}_1\sigma^{n-1}v_1\sigma^n_ev_2\in L(\NP)$ with $P_{\Sigma_G}(v_2)=w_{N_c}$.
As $N_c$ \ticks have passed, $head(l)=(\sigma^n,0)$, and $\sigma^n$ occurs in $\NP$. Hence, $w^n_0\,\sigma^n_e\,w^n_1\,\sigma^n\in L(\NP)$ for  $w^n_0=w^{n-1}_0\,\sigma^{n-1}_e\,w^{n-1}_1\,\sigma^{n-1}\,v_1$ and $w^n_1=v_2$ where $P_{\Sigma_G}(w^n_0\,w^n_1)=w_n$ and $|P_{\{\tick\}}(w^n_1)|=N_c$.

$j=N_c$. Then, after the execution of $w^{n-1}_0\sigma^{n-1}_e$ in $\NP$, $v_1$ occurs in $\NP$ related to the execution of $w^{n-1}_{N_c}$ in $G$ and $w'_n$ in $G'$. 
At this point, $\sigma^{n-1}$ is enabled in $G$ and $head(l)=(\sigma^{n-1})$. Also, $\sigma^n$ is enabled in $G'$. Therefore, either $\sigma^n_e\sigma^{n-1}$ or $\sigma^{n-1}\sigma^n_e$ occurs in $\NP$ both followed by the execution of some $v_2$ in $\NP$ such that $P_{\Sigma_G}(v_2)=w$. As $N_c$ \ticks have passed ($head(l)=(\sigma^n,0)$), and $\sigma^n$ is enabled in $G$, $\sigma^n$ occurs in $\NP$. 
This results in one of the following words; $w^{n-1}_0\sigma^{n-1}_ev_1\sigma^{n-1}\sigma^n_ev_2\sigma^n\in L(\NP)$ or $w^{n-1}_0\sigma^{n-1}_ev_1\sigma^n_e\sigma^{n-1}v_2\sigma^n\in L(\NP)$ where the statement holds in both cases as already discussed in the previous items.
\textbf{Conclusion:} By the principle of induction, the statement holds for all $\sigma\in\Sigma_c$ and $w\in\Sigma^*_G$ with $w\sigma\in L(G)$.
\hfill $\blacksquare$
\end{proof}

\begin{lemma}[NSP and NP]
\label{lemma:m,l}
Consider a plant $G$, a networked supervisor $\NS$ (for that plant), the observation channel $M$, and the control channel $L$.
The networked plant $\NP$ has the set of states $X$ and the networked supervised plant $\NSP$ has the set of states $Z$. Then, for any pair of $x\in X$ and $z\in Z$ reachable through the same $w\in \Sigma^*_{\mathit{NSP}}$: $x.m=z.m$ and $x.l=z.l$.
%\hfill$\blacksquare$
\end{lemma}

\begin{proof}
%For $\NP=(X,\SigmaNSP,\deltaNP,x_0,X_m)$  and $\NSP=(Z,\SigmaNSP,\deltaNSP,z_0,Z_m)$, we have with $x_0.m=[]$ and $x_0.l=\epsilon$, $z_0.m=[]$ and $z_0.l=\epsilon$.
Take $x\in X$, $z\in Z$, and $w\in \Sigma^*_{\mathit{NSP}}$ such that $x=\deltaNP(x_0,w)$ and $z=\deltaNSP(z_0,w)$.
By induction on the structure of $w$, it is proved that $x.m=z.m$ and $x.l=z.l$.
\textbf{Base case:} Assume $w=\epsilon$. Then, $x.m=x_0.m=[]$ ($x.l=x_0.l=\epsilon$), and $z.m=z_0.m=[]$ ($z.l=z_0.l=\epsilon$). Thereto, $x.m=z.m$ and $x.l=z.l$. \textbf{Induction step:} Assume $w=v\sigma$ where the statement holds for $v\in \Sigma^*_{\mathit{NSP}}$ and the intermediate states reached by $v$ so that $\deltaNP(x_0,v).m=\deltaNSP(z_0,v).m$ and $\deltaNP(x_0,v).l=\deltaNSP(z_0,v).l$. It suffices to prove that the statement holds for $v\sigma$, i.e., $\deltaNP(x_0,v\sigma).m=\deltaNSP(z_0,v\sigma).m$ and $\deltaNP(x_0,v\sigma).l=\deltaNSP(z_0,v\sigma).l$.
Considering Definition \ref{dfn:NP} and Definition \ref{dfn:operator}, in both operators, $\deltaNP(x_0,v).m$ ($\deltaNSP(z_0,v).m$) changes by the execution of $\sigma\in\Sigma_c\cup\Sigmauc\cup\tick\cup\Sigma_o$ (item 2), item 3), item 4), and item 5)), and $\deltaNP(x_0,v).l$ ($\deltaNSP(z_0,v).l$) changes by the execution of $\sigma\in\Sigma_e\cup \Sigma_c\cup\tick$ (item 1), item 2), and item 4)) in a similar way.
Therefore, starting from $\deltaNP(x_0,v)$ and $\deltaNSP(z_0,v)$ with $\deltaNP(x_0,v).m=\deltaNSP(z_0,v).m$ ($\deltaNP(x_0,v).l=\deltaNSP(z_0,v).l$), the execution of the same event $\sigma$ results in $\deltaNP(x_0,v\sigma).m=\deltaNSP(z_0,v).m$ ($\deltaNP(x_0,v\sigma).l=\deltaNSP(z_0,v).l$).
\textbf{Conclusion:} By the principle of induction, the statement ($x.m=z.m$ and $x.l=z.l$) holds for all $w\in \Sigma^*_{\mathit{NSP}}$, $x=\deltaNP(x_0,w)$ and $z=\deltaNSP(z_0,w)$.
\hfill$\blacksquare$
\end{proof}

\begin{lemma}[\NSP and Product]
\label{lemma:product}
Given a plant $G$ and a networked supervisor $\NS$ with event set $\SigmaNS$. If $L(\NS)\subseteq P_{\SigmaNS}(L(\NP))$, then $L(\NSP)=L(\NS||\NP)$.
%\hfill$\blacksquare$
\end{lemma}

%\AR{can we generalize this by saying for any $\NS$ in NSC framework? (Given a plant $G$ and a networked supervisor $\NS$ in the NSC framework: $L(\NSP)=L(\NS||\NP)$) and then in the proof  say that such a NS has event set $\SigmaNS$ and $L(\NS)\subseteq P_{\SigmaNS}(L(\NP))$ ? This helps with timed networked MPness proof!}

\begin{proof}
This is proved in two steps; 1. for any $w\in L(\NSP)$: $w\in L(\NS||\NP)$, and 2. for any $w\in L(\NS||\NP)$: $w\in L(\NSP)$.
%Let say $G=(A,\Sigma,\delta,a_0,A_m)$, $\NS=(Y,\SigmaNS,\deltaNS,y_0,Y_m)$, $\NP=(X,\SigmaNSP,\deltaNP,x_0,X_m)$, and $\NSP=(Z,\SigmaNSP,\deltaNSP,z_0,Z_m)$.
%where $\NSP$ and $\NP$ stand for the networked supervised plant and networked plant, respectively.

%\begin{enumerate}
1) Take $w\in L(\NSP)$. By induction on the structure of $w$, it is proved that $w\in L(\NS||\NP)$.
\textbf{Base case:} Assume that $w=\epsilon$. Then, $w\in L(\NS||\NP)$ by definition.
\textbf{Induction step:}
Let $w=v\sigma$ for some $v\in \Sigma^*_{NSP}$ and $\sigma\in \SigmaNSP$ where the statement holds for $v$, i.e., $v\in L(\NS||\NP)$.
It suffices to prove that the statement holds for $v\sigma$, i.e., $v\sigma\in L(\NS||\NP)$.
%Then, $\deltaNSP(z_0,v)=z_1$ and $\deltaNSP(z_1,\sigma)=z_2$ for some $z_1,z_2\in Z$.
%\MR{You (too) often use the word assumption for something you can derive / know. An assumption is a statement that you want to use, but do not know for a fact.}
Due to Lemma \ref{lemma:NSP}, $\deltaNSP(z_0,v).y=\deltaNS(y_0,P_{\SigmaNS}(v))$, $\deltaNSP(z_0,v\sigma).y=\deltaNS(\deltaNSP(z_0,v).y,P_{\SigmaNS}(\sigma))$, $\deltaNSP(z_0,v).a=\delta_G(a_0,P_{\Sigma_G}(v))$, and $\deltaNSP(z_0,v\sigma).a=\delta_G(\deltaNSP(z_0,v).a,P_{\Sigma_G}(\sigma))$. Due to the definition of synchronous product (in \cite{Cassandras:99}), since $\SigmaNS\subseteq\SigmaNSP$, one can say any $w\in L(\NS||\NP)$ if $w\in L(\NP)$ and $P_{\SigmaNS}(w)\in L(\NS)$. For $v\sigma\in L(\NSP)$, it is already showed that $P_{\SigmaNS}(v\sigma)\in L(\NS)$, and so it suffices to prove $v\sigma\in L(\NP)$.
%\MR{In the previous sentence you introduce $y_1$ and $y_2$ implicitly. Never do that. Always introduce new variables / notations explicitly. Also always wonder if the introduction is needed. Is $y_1$ so much more easy than $z_1.y$, etc.?}\AR{To me, introducing these variables makes the proof more understandable ... but, I will see how this work if I remove them ...}
For $v\in L(\NS||\NP)$: $v\in L(\NP)$ (since $\SigmaNS\subseteq\SigmaNSP$).
%For $v\in L(\NP)$, $\deltaNP(x_0,v)=x_1$ for some $x_1\in X$.
Then, due to Lemma \ref{lemma:NP}, $\deltaNSP(z_0,v).a=\delta_G(a_0,P_{\Sigma_G}(v))=\deltaNP(x_0,v).a$.
Moreover, both $\deltaNSP(z_0,v)$ and $\deltaNP(x_0,v)$ are reachable through $v$, and so due to Lemma \ref{lemma:m,l}, $\deltaNSP(z_0,v).m=\deltaNP(x_0,v).m$ and $\deltaNSP(z_0,v).l=\deltaNP(x_0,v).l$.
%\MR{Is the previous statement a fact? Do we have a lemma that states that this is true?}\AR{I agree, it may not be clear to everyone ... I will add a Lemma for this ... }
%Due to line \ref{line:reachable} of Algorithm \ref{algo}, $\ns(i)\subseteq NS(i-1)$, and $\ns(0)=NP$. Hence, one can say $\NS\subseteq P_{\SigmaNS}(NP)$ (line \ref{line:project} of Algorithm \ref{algo}).
Since $P_{\SigmaNS}(v\sigma)\in L(\NS)$ (for $v\sigma\in L(\NSP)$, $\deltaNS(y_0,v\sigma)!$ due to Lemma \ref{lemma:NSP}), and $L(\NS)\subseteq P_{\SigmaNS}(L(\NP))$, then one can say there exists $w'\in L(\NP)$, $P_{\SigmaNS}(w')=P_{\SigmaNS}(v\sigma)$. Without loss of generality, assume $w'=v'P_{\SigmaNS}(\sigma)$ where $P_{\SigmaNS}(v')=P_{\SigmaNS}(v)$.
%such that $\deltaNP(x_0,v')=x'_1$ and $\deltaNP(x'_1,P_{\SigmaNS}(\sigma))=x'_2$.
Let us complete the proof for different cases of $\sigma\in\SigmaNSP$.
%Note that for $\sigma\in\SigmaNS$, we already know $P_{\SigmaNS}(v)\sigma\in L(\NS)$. Also, $\Sigma_a\notin\SigmaNS$. Due to the definition of synchronous product \cite{Cassandras:99}, it suffices to prove that $\deltaNP(x_1,\sigma)!$.

$\sigma\in\Sigma_e$. Then, $\delta'_G(\deltaNP(x_0,v).a',\sigma)!$ since $\deltaNP(x_0,v).a'=$ $\deltaNP(x_0,v').a'$ and $\delta'_G(\deltaNP(x_0,v').a',\sigma)!$ ($P_{\Sigma_e\cup\{tick\}}(v)=P_{\Sigma_e\cup\{tick\}}(v')$, and due to Definition \ref{dfn:NP}, $x.a'$ changes by $w\in (\Sigma_e\cup\{tick\})^*$). So, due to item 1), $\deltaNP(\deltaNP(x_0,v),\sigma)!$.
%Moreover, we are sure that $|l.x_1|<\Lmax$ since $l.x_1=l.z_1$ and $|l.z_1|<\Lmax$.
%\AR{currently, we do not have the upper limit restriction for $\NSP$ ... shall we have it there as well? I think it would be more meaningful to have the same conditions for $\NSP$ as well ...}

$\sigma\in\Sigma_{c}$. Then,  $\delta_G(\deltaNP(x_0,v).a,\sigma)!$ since $\deltaNP(x_0,v).a=\deltaNSP(z_0,v).a$ and $\delta_G(\deltaNSP(z_0,v).a,\sigma)!$.
Also, the condition $\head(\deltaNP(x_0,v).l)=(\sigma,0)$ is satisfied since $\deltaNP(x_0,v).l=\deltaNSP(z_0,v).l$ and $\head(\deltaNSP(z_0,v).l)=(\sigma,0)$ (considering Definition \ref{dfn:operator}-item 2)), $\sigma$ can occur only if $\head(\deltaNSP(z_0,v).l)=(\sigma,0)$). So, due to Definition \ref{dfn:NP}-item 2), $\deltaNP(\deltaNSP(z_0,v),\sigma)!$.

$\sigma\in\Sigmauc$. Then, $\delta_G(\deltaNP(x_0,v).a,\sigma)!$  since $\deltaNP(x_0,v).a=\deltaNSP(z_0,v).a$ and $\delta_G(\deltaNSP(z_0,v).a,\sigma)!$. So, based on Definition \ref{dfn:NP}-item 3), $\deltaNP(\deltaNP(x_0,v),\sigma)!$.
%\AR{condition: $|m|<\Mmax$}

$\sigma=tick$. Then, $\delta_G(\deltaNP(x_0,v).a,\sigma)!$ since $\deltaNP(x_0,v).a=\deltaNSP(z_0,v).a$ and $\delta_G(\deltaNSP(z_0,v).a,\sigma)!$. 
In addition, $\delta'_G(\deltaNP(x_0,v).a',\sigma)!$ since $\deltaNP(x_0,v).a'=\deltaNP(x_0,v').a'$ and $\delta'_G(\deltaNP(x_0,v').a',\sigma)!$.
Also, $(\sigma,0)\notin \deltaNP(x_0,v).m$ for all $\sigma\in \Sigma_a$ since $\deltaNSP(z_0,v).m=\deltaNP(x_0,v).m$ and $(\sigma,0)\notin \deltaNSP(z_0,v).m$ (considering Definition \ref{dfn:operator}-item 4), \tick can occur if $(\sigma,0)\notin \deltaNSP(z_0,v).m$). Therefore, based on Definition \ref{dfn:NP}-item 4), $\deltaNP(\deltaNP(x_0,v),\sigma)!$.

$\sigma\in \Sigma_o$. Then, $(\sigma,0)\in \deltaNP(x_0,v).m$ because $\deltaNP(x_0,v).m=\deltaNSP(z_0,v).m$ and $(\sigma,0)\in \deltaNSP(z_0,v).m$ (due to Definition \ref{dfn:operator}-item 5)). So, due to Definition \ref{dfn:NP}-item 5), $\deltaNP(\deltaNP(x_0,v),\sigma)!$.
\textbf{Conclusion:} By the principle of induction, $w\in L(\NS||\NP)$ is true for any $w\in L(\NSP)$.
    
2) Take $w\in L(\NS||\NP)$, by induction, it is proved that $w\in L(\NSP)$ is true.
\textbf{Base case:} Assume that $w=\epsilon \in L(\NS||\NP)$. Then, $w\in L(\NSP)$ by definition.
\textbf{Induction step:} Let $w=v\sigma\in L(\NS||\NP)$ where the statement is true for $v$, i.e., $v\in L(\NSP)$. It suffices to prove that the statement holds for $v\sigma$, i.e., $v\sigma\in L(\NSP)$. Due to the definition of synchronous product (in \cite{Cassandras:99}), since $\SigmaNS\subseteq\SigmaNSP$, one can say any $w\in L(\NS||\NP)$ if $w\in L(\NP)$ and $P_{\SigmaNS}(w)\in L(\NS)$. %Then, for $v\sigma\in L(\NP)$, $\deltaNP(x_0,v)=x_1$ and $\deltaNP(x_1,\sigma)=x_2$ for some $x_1, x_2\in X$.
Due to Lemma \ref{lemma:NP}, $\deltaNP(x_0,v).a=\delta_G(a_0,P_{\Sigma_G}(v))$ and $\deltaNP(x_0,v\sigma).a=\delta_G(\deltaNP(x_0,v).a,\sigma)$.

%Also, for $P_{\SigmaNS}(v\sigma)\in L(\NS)$, $\deltaNS(y_0,P_{\SigmaNS}(v))=y_1$, $\deltaNS(\deltaNP(x_0,v\deltaNS(y_0,P_{\SigmaNS}(v)),P_{\SigmaNS}(\sigma))=y_2$ for some $y_1,y_2\in Y$. For, $v\in L(\NSP)$,  $\deltaNSP(z_0,v)=z_1$ for some $z_1\in Z$.

Also, due to Lemma \ref{lemma:NSP}, for $v\in L(\NSP)$, $\deltaNSP(z_0,v).y=\deltaNS(y_0,P_{\SigmaNS}(v))$, and $\deltaNSP(z_0,v).a=\delta_G(a_0,P_{\Sigma_G}(v))$.

Moreover, since both $\deltaNSP(z_0,v)$ and $\deltaNP(x_0,v)$ are reachable through $v$, based on Lemma \ref{lemma:m,l}, $\deltaNSP(z_0,v).m=\deltaNP(x_0,v).m$ and $\deltaNSP(z_0,v).l=\deltaNP(x_0,v).l$. Now, for different cases of $\sigma\in\SigmaNSP$, we prove that $\deltaNSP(\deltaNSP(z_0,v),\sigma)!$. 

$\sigma\in\Sigma_e$. Then, based on the assumption, $\deltaNS(y_0,P_{\SigmaNS}(v)\sigma)!$, and so considering Definition \ref{dfn:operator}-item 1), $\deltaNSP(\deltaNSP(z_0,v),\sigma)!$.

$\sigma\in\Sigma_{c}$. Then, $\delta_G(\deltaNSP(z_0,v).a,\sigma)!$ since $\deltaNSP(z_0,v).a=\deltaNP(x_0,v).a$ and $\delta_G(\deltaNP(x_0,v).a,\sigma)!$.
Also, the condition $\head(\deltaNSP(z_0,v).l)=(\sigma,0)$ is satisfied since $\deltaNSP(z_0,v).l=\deltaNP(x_0,v).l$ and $\head(\deltaNP(x_0,v).l)=(\sigma,0)$ (based on Definition \ref{dfn:NP}-item 2)). Hence, considering Definition \ref{dfn:operator}-item 2), $\deltaNSP(\deltaNSP(z_0,v),\sigma)!$.

$\sigma\in\Sigmauc$. Then, $\delta_G(\deltaNSP(z_0,v).a,\sigma)!$ since $\deltaNSP(z_0,v).a=\deltaNP(x_0,v).a$ and $\delta_G(\deltaNP(x_0,v).a,\sigma)!$, and so considering Definition \ref{dfn:operator}-item 3), $\deltaNSP(\deltaNSP(z_0,v),\sigma)!$.
%\AR{condition: $|m|<\Mmax$}

$\sigma=tick$. Then, $\delta_G(\deltaNSP(z_0,v).a,\sigma)!$ since $\deltaNSP(z_0,v).a=$ $\deltaNP(x_0,v).a$ and $\delta_G(\deltaNP(x_0,v).a,\sigma)!$. Also, $\deltaNS(\deltaNS(y_0,P_{\SigmaNS}(v)),\sigma)!$ due to the assumption. Moreover, $(\sigma,0)\notin \deltaNSP(z_0,v).m$ for all $\sigma\in \Sigma_a$ since $(\sigma,0)\notin \deltaNP(x_0,v).m$ (based on Definition \ref{dfn:NP}-item 4)) and $\deltaNSP(z_0,v).m=\deltaNP(x_0,v).m$. Therefore, based on Definition \ref{dfn:operator}-item 4), $\deltaNSP(\deltaNSP(z_0,v),\sigma)!$.

$\sigma\in \Sigma_o$. Then, $(\sigma,0)\in \deltaNSP(z_0,v).m$ because $\deltaNSP(z_0,v).m=\deltaNP(x_0,v).m$ and $(\sigma,0)\in \deltaNP(x_0,v).m$ (due to Definition  \ref{dfn:NP}-item 5)). Moreover, $\deltaNS(\deltaNS(y_0,P_{\SigmaNS}(v)),\sigma)!$ based on the assumption. So, due to Definition \ref{dfn:operator}-item 5), $\deltaNSP(\deltaNSP(z_0,v),\sigma)!$.
\textbf{Conclusion:} By the principle of induction $w\in L(\NSP)$ is true for any $w\in L(\NS||\NP)$.
\hfill$\blacksquare$
%\end{enumerate}
\end{proof}

\begin{corollary}[Lemma \ref{lemma:product}]
\label{corollary:product}
Given a plant $G$ and a networked supervisor $\NS$ with event set $\SigmaNS$ such that $L(\NS)\subseteq P_{\SigmaNS}(L(\NP))$:
\begin{enumerate}
    \item $L(\NSP)\subseteq L(\NP)$, and
    \item  $L_m(\NSP)\subseteq L_m(NP)$.
%\hfill $\blacksquare$
\end{enumerate}
\end{corollary}

\begin{proof}
This clearly holds since due to Lemma \ref{lemma:product}, $\NSP=\NS||\NP$ and $\SigmaNS\subseteq\SigmaNSP$ 
\hfill$\blacksquare$
\end{proof}

\begin{lemma}[Finite NP]
\label{lemma:finiteNP}
Given a plant $G$ with a set of states $A$ and a set of events $\Sigma_G$: $\NP$ is a finite automaton.
%\hfill $\blacksquare$
\end{lemma}

\begin{proof}
We need to prove that $\NP$ has a finite set of states and a finite set of events. Considering Definition \ref{dfn:NP}, $\NP$ has a set of states $X=A\times Q'\times A'\times M\times L$. To prove that $X$ is finite, it is sufficient to guarantee that $A, Q', A', M$ and $L$ are finite sets because as proved in~\cite{SetTheory} the Cartesian product of finite sets is finite. $A$ is finite as the plant is assumed to be given as a finite automaton. $A'$ is finite since for each $a'\in A'$, $a'\subseteq A$, and $A$ is finite.
%For each $q'\in Q'$, we know that $q'\subseteq Q$. So, we should prove that $Q$ is a finite set. $Q=A\times M$ is finite since $A$ and $M$ are finite as the maximum size of every element of $M$ is limited to a finite number $\Mmax$.
$M$($L$) is finite as the maximum size of every element of $M$ is limited to a finite number $\Mmax$($\Lmax$).
%Finally, $L$ is finite since the size of every element of $L$ is limited to $\Lmax$.
Moreover,
$\SigmaNSP=\Sigma_e\cup\Sigma_o\cup\Sigma_G$ is finite since $G$ is a finite automaton, and so $\Sigma_G$ is finite. $\Sigma_e$ and $\Sigma_o$ are finite since due to Definition \ref{dfn:Sigmao&Sigmae}, the size of $\Sigma_e$ is equal to the size of $\Sigma_c$, and the size of $\Sigma_o$ is equal to the size of $\Sigma_a$. 
\hfill $\blacksquare$
\end{proof}

\begin{lemma}[Nonblocking NS]
\label{lemma:NBNS}
The networked supervisor $\NS$ synthesized from Algorithm \ref{algo} is nonblocking.
%\hfill$\blacksquare$
\end{lemma}

\begin{proof}
Based on Property \ref{property:termination}, Algorithm \ref{algo} terminates, let say after $n$ iterations. Then, either $x_0\in \Uncon(\BS(n))$ or $\BS(n)=\varnothing$ where $\BS(n)=\BPre(\NS(n)\cup\BLock(\NS(n))\cup \TLock(\NS(n)))$. In case that $x_0\in \Uncon(\BS(n))$, the algorithm gives no result. Otherwise, the algorithm gives $\NS=P_{\SigmaNS}(\NS(n))$ where $\NS(n)$ is nonblocking since $\BLock(\NS(n))=\varnothing$. Moreover, due to Lemma \ref{lemma:projectionNB}, the projection preserves nonblockingness, and so $\NS$ is nonblocking.
\hfill$\blacksquare$
\end{proof}

\begin{lemma}[TLF NS]
\label{lemma:TLFNS}
The networked supervisor $\NS$ synthesized for a plant $G$ using Algorithm \ref{algo}  is TLF.
%\hfill$\blacksquare$
\end{lemma}

\begin{proof}
The proof is similar to the proof of Lemma \ref{lemma:NBNS}.
\hfill$\blacksquare$
\end{proof}

\section{Proofs of Properties and Theorems}

\subsection{Proof of Property \ref{property:NSP&P}}\label{proof:NSP&P}
%\begin{proof}
It suffices to prove that $w\in L(G)$ for any $w\in P_{\Sigma_G}(L(\NSP))$.
Take arbitrary $w\in P_{\Sigma_G}(L(\NSP))$. Then, according to Definition \ref{dfn:proj}, $P_{\Sigma_G}(w')=w$ for some $w'\in L(\NSP)$.
Then, due to Lemma \ref{lemma:NSP},  $\deltaNSP(z_0,w').a=\delta_G(a_0,P_{\Sigma_G}(w'))$ meaning that $w \in L(G)$.
%\AR{AND $\deltaNS(y_0,P_{\SigmaNS}(w')!$ so is this valid only if $\NS$ sends the right commands or not related to $\NS$?} \MR{I think yellow part is not needed for this property.}
%\hfill$\blacksquare$
%\end{proof}

\subsection{Proof of Property \ref{property:NPLE}}\label{proof:NPLE}
%\begin{proof}
%Let say $G=(A,\Sigma,\delta,a_0,A_m)$ and $\NP=(X,\SigmaNSP,\deltaNSP,x_0,X_m)$.
The proof consists of two cases:

%\begin{enumerate}
1) for any $w\in P_{\Sigma_G}(L(\NP))$: $w\in L(G)$. This is proved by induction on the structure of $w$. \textbf{Base case:} Assume $w=\epsilon$. Then, $w\in L(G)$ by definition. \textbf{Induction step:} Assume that $w=v\sigma$ for some $v\in \Sigma^*_G$ and $\sigma\in\Sigma_G$ where the statement holds for $v$, i.e., $v\in L(G)$.
%In other words, $\delta_G(a_0,v)=a_1$ for some $a_1\in A$.
It suffices to prove that the statement holds for $v\sigma$, i.e., $v\sigma \in L(G)$.
%or equivalently $\delta_G(\delta_G(a_0,v),\sigma)!$.
Due to the projection properties, for $v\sigma\in P_{\Sigma_G}(L(\NP))$, one can say there exists $v'\in \Sigma^*_{NSP}$,
%$\deltaNP(x_0,v')=x$ for some $x\in X$
$P_{\Sigma_G}(v')=v\sigma$. Without loss of generality,
%we have $\deltaNP(x_0,v'')=x'$ for some $x'\in X$
let say $v'=v''\sigma$ where $P_{\Sigma_G}(v'')=v$.
%and $\deltaNP(x',\sigma)=x$.
Then, due to Lemma \ref{lemma:NP}, $\deltaNP(x_0,v'').a=\delta_G(a_0,P_{\Sigma_G}(v''))=\delta_G(a_0,v)$, and $\deltaNP(\deltaNP(x_0,v''),\sigma).a=\delta_G(a_0,P_{\Sigma_G}(v''\sigma))=\delta_G(\delta_G(a_0,v),\sigma)$. So, $\delta_G(\delta_G(a_0,v),\sigma)!$ and the statement holds for $v\sigma$. \textbf{Conclusion:} By the principle of induction, the statement $w\in L(G)$ holds for all $w\in P_{\Sigma_G}(L(\NP))$. 
%Also, $v'\sigma \in L(\NP)$ since $v\sigma\in P_{\Sigma_G}(L(\NP))$. Due to Definition \ref{dfn:NP}, $\sigma$ can be executed in $L(\NP)$ only if it is enabled in $G$ and so $P_{\Sigma_G}(v')\sigma\in L(G)$. \MR{Besides $\sigma$ being enabled in the plant, isn't it necessary to be in the right state? I am not convinced yet. Say something about it?}

2) If $max_c\leq \Lmax$, for any $w\in L(G)$: $w\in P_{\Sigma_G}(L(\NP))$. This is proved by using induction on the structure of $w$. 
\textbf{Base case:} assume $w=\epsilon$. Then $w\in P_{\Sigma_G}(L(\NP))$ by definition.
\textbf{Induction step:} assume that $w=v\sigma$ for some $v\in L(G)$ and $\sigma\in\Sigma_G$ where the statement holds for $v$, i.e., $v\in P_{\Sigma_G}(L(\NP))$. It suffices to prove that $v\sigma \in P_{\Sigma_G}(L(\NP))$.
For $v\in P_{\Sigma_G}(L(\NP))$, there exists $v'\in \Sigma^*_{NSP}$, $P_{\Sigma_G}(v')=v$ due to the projection properties.
%so that $\deltaNP(x_0,v')=x_1$ for some $x_1\in X$.
Considering Definition \ref{dfn:NP}, one of the following cases may occur at $\deltaNP(x_0,v')$. 

$\sigma\in \Sigmauc$, then due to item 3), $\deltaNP(\deltaNP(x_0,v'),\sigma)!$ because $\delta_G(\deltaNP(x_0,v').a,\sigma)!$. Applying the projection on $v'\sigma\in L(\NP)$ results in $v\sigma\in P_{\Sigma_G}(L(\NP))$.
%\AR{Is the following explanation understandable?}

$\sigma\in\Sigma_c$, then $(\sigma,0)\in \deltaNP(x_0,v').l$ since due to Lemma \ref{lemma:ontime}, $N_c$ \ticks earlier, $\sigma_e$ was enabled in $\NP$. When $\sigma_e$ occurred, based on item 1), $(\sigma,N_c)$ was certainly put in $l$ as Assumption 2 holds. The occurrence of each \tick (from $N_c$ \ticks) causes $l-1$ as item 4) says.
Also, the control channel is FIFO ($l$ is a list), so even if a sequence of events have been enabled simultaneously, the ordering is preserved in $l$. So far,
$\head(\deltaNP(x_0,v').l)=(\sigma,0)$ and $\delta_G(\deltaNP(x_0,v').a,\sigma)!$ as assumed. So, due to item 2), $v'\sigma\in L(\NP)$, and $v\sigma\in P_{\Sigma_G}(L(\NP))$.

$\sigma=\tick$, then let us first empty $\deltaNP(x_0,v').m$ from any $(\sigma',0)$ by executing $v_o\in \Sigma^*_o$.
%such that $\deltaNP(\deltaNP(x_0,v'),v_o)=x_2$ (based on item 5)).
Then, $(\sigma',0)\notin \deltaNP(x_0,v'\,v_o).m$.
Also, $\deltaNP(x_0,v'\,v_o).a=\deltaNP(x_0,v').a$
%and $x_2.a'=x_1.a'$
since the execution of observed events only changes $\deltaNP(x_0,v').m$. $\delta_G(\deltaNP(x_0,v').a,\tick)!$ due to the assumption, and so $\delta_G(\deltaNP(x_0,v'\,v_o).a,\tick)!$.
Now, as the worst case, assume that at $\deltaNP(x_0,v'\,v_o).a'$, only $v_c\in \Sigma^{*a}_c$ is enabled, and after that either \tick occurs or nothing. 
Based on item 4) and item 1), this is then only $v_{c_e}$ executed at $\deltaNP(x_0,v'\,v_o)$.
%leading to $\deltaNP(x_2,v_{c_e})=x_3$ such that $x_3.a=x_2.a$ and $x_3.m=x_2.m$.
$\delta_G(\deltaNP(x_0,v'\,v_o\,v_{c_e}).a,\tick))!$, $(\sigma,0)\notin \deltaNP(x_0,v'\,v_o\,v_{c_e}).m$, and $\neg\delta'_G(\deltaNP(x_0,v'\,v_o\,v_{c_e}).a',\sigma')!$ for all $\sigma,\sigma'\in\Sigma_a$. So, based on item 4),  
$v'\,v_o\,v_{c_e}\,\tick\in L(\NP)$, and so $v\,\tick\in P_{\Sigma_G}(L(\NP))$.
%Therefore, there exists a $w_c\in \Sigma^{*a}_c$ such that $\delta'_G(x_2.a',w_c\tick)!$. Let us first execute $w_{ce}$ at $x_2$ resulted in $\deltaNP(x_2,v_{ce})=x_3$. Then, $x_3.a=x_2.a$ and $x_3.m=x_2.m$ since the execution of enabling events only changes $x_2.l$ and $a'.l$. Then, at $x_3$, we have: $(\sigma',0)\notin x_2.m$, $\forall\sigma\in\Sigma_a$, $\delta_G(x_3.a,\tick)!$, and $\delta'_G(x_3.a',\tick)!$. Hence, $v'v_ov_{ce}\tick\in L(\NP)$ for which we get $v\,\tick\in P_{\Sigma_G}(L(\NP))$.
\textbf{Conclusion:} By the principle of induction, the statement ($w\in P_{\Sigma_G}(L(\NP))$) holds for all $w\in L(G)$.
%\hfill $\blacksquare$
%\end{enumerate}
%\end{proof}

\subsection{Proof of Property \ref{property:termination}}\label{proof:termination}
%\begin{proof}
Algorithm \ref{algo} terminates if at some iteration $i$, $y_0\in \Uncon(\bs(i))$ or $\bs(i)=\varnothing$. 
At each iteration $i$, $\bs(i)\subseteq Y$ since initially $\bs(0)=\BS(\ns(0))$ where $\BS(\ns(0))=\BLock(\ns(0))\cup\TLock(\ns(0)$, and so $\bs(0)\subseteq Y$ by definition. Also, $\bs(i)$ is updated at line \ref{line:BPre} to $\BPre(\ns(i))\cup\BS(\ns(i))$ where $\BPre(\ns(i))\subseteq Y$ and $\BS(\ns(i))\subseteq Y$ by definition, and so $\bs(i)\subseteq Y$.
Since $Y$ is a finite set, it suffices to prove that at each iteration, at least one state is removed from $Y$. Then, it is guaranteed that the algorithm loops finitely often.
%\MR{For example the size of the set $Y$, i.e., $|Y|$. What is this function here? Then for each execution of the body of the while you have to show that the value of that function actually decreases (assuming the while-condition of course).} 
%\MR{Assuming that we can prove that $|Y|$ decreases every execution of the body of the while. Assume that we can show that $\bs(i) \subset Y$, then the while-loop cannot execute infinitely, since $|Y|$ becomes 0 and this implies that $\bs(i) = \varnothing$.} \MR{So my suggestion is to prove that (i) $|Y|$ decreases with each body-execution, and (2) $\bs(i) \subseteq Y$ is an invariant of the while loop. The latter means that we need to show that this is true initially (just before the loop is entered and is true again at the end of the body of the loop.} 
%\AR{I've already tried this in the next yellow box :) just wanted to check if this can be right and then I improve the text}
%\MR{I believe that the status of $Y$ is unclear. All I can deduce is that it is the state set of the result, but is it also aprogram variable that is adapted? Then I would expect $Y(i)$ or an explicit update of $Y$.}
%\AR{Can we say $\bs(i)\subseteq Y$ and at least one state is removed from $Y$ at each iteration? So, although some state may be added to $\bs(i)$ at the end of an iteration because of line \ref{line:BPre}, but still the set that $\bs(i)$ can belong to shrinks until either $\bs(i)=\varnothing$ or $Y=\{y_0\}$ where we assume $y_0\notin BS(i)$ and so $\bs(i)=\varnothing$ ...}
So, let's say $y_0\notin \Uncon(\bs(i))$ and $\bs(i)\neq\varnothing$  (because otherwise the algorithm terminates immediately).
Then, there exists some state $y'\in \bs(i)$.
By definition this gives $y'\in \Uncon(\bs(i))$. Also, since at the end of each iteration, the automaton is made reachable (line \ref{line:reachable}), $y'$ is reachable from $y_0$ (possibly through some intermediate states). According to line \ref{line:removeUncon}, at least $y'$ is removed from $Y$, and so the algorithm terminates.

\subsection{Proof of Theorem \ref{theorem:nonblockingness}}\label{proof:NBness}
%\begin{proof}
We need to prove that for all $z\in Reach(z_0)$, there exists a $w\in \Sigma^*_{NSP}$ such that $\deltaNSP(z,w)\in Z_m$. Take $z\in Reach(z_0)$, then we need to find $w\in \Sigma^*_{NSP}$ for which  $\deltaNSP(z,w)\in Z_m$.
Let us assume that $z$ is reachable from $z_0$ via $w_0\in \Sigma^*_{NSP}$, i.e., $\deltaNSP(z_0,w_0)=z$.
Then, due to Lemma \ref{lemma:NSP}, $z.y=\deltaNS(y_0,P_{\SigmaNS}(w_0))$.
%Let us indicate $z.y$ and $P_{\SigmaNS}(w_0)$ as $y_1$ and $v_0$, respectively.
Due to Lemma \ref{lemma:NBNS}, for $z.y\in Reach(y_0)$, there exists some $v\in\Sigma^*_{NS}$ such that $\deltaNS(z.y,v)\in Y_m$.
Moreover, due to line \ref{line:reachable} of Algorithm \ref{algo}, $L(\ns(i))\subseteq L(\ns(i-1))$, and $\ns(0)=NP$. Hence, $L(\NS)\subseteq L(P_{\SigmaNS}(NP))$ (line \ref{line:project} of Algorithm \ref{algo}).
%\MF{The difference between ns(I) and ns(I-1) is the removal of the unreachable states of ns(I-1), not? If so, their languages must be equal!}
%\AR{Transitions are disabled at line 8 Algorithm 1}
Then, due to the projection properties, for $P_{\SigmaNS}(w_0)v\in L(\NS)$, one can say there exists some $w'\in L(\NP)$, $P_{\SigmaNS}(w')=P_{\SigmaNS}(w_0)v$ such that $\deltaNP(x_0,w')\in X_m$ (due to the projection properties, any state $y$ is marked only if $y \cap X_m \neq \varnothing$). %\MR{Previous explanation about marked state is very confusing as suddenly you move to subsets of $X$.}
Without loss of generality, assume that $w'=w'_0w'_1$ for some $w'_0, w'_1 \in \Sigma^*_{\mathit{NSP}}$ with
$P_{\SigmaNS}(w'_0)=P_{\SigmaNS}(w_0)$ and $P_{\SigmaNS}(w'_1)=v$. Let $x'_1\in X$ be such that $\deltaNP(x_0,w'_0)=x'_1$, and then $\deltaNP(x'_1,w'_1)\in X_m$.
Moreover, due to Corollary \ref{corollary:product}, $w_0\in L(\NP)$, and so $\deltaNP(x_0,w_0)=x_1$ for some $x_1\in X$. 
So far, we have $x_1,x'_1$ are reachable from $x_0$ via $w_0,w'_0$, respectively, where $P_{\SigmaNS}(w_0)=P_{\SigmaNS}(w'_0)$. Thereto, $x_1$ is observationally equivalent to $x'_1$.
Then, $x_1\notin\Uncon(\BS(\ns(i))$ at any iteration $i$ because otherwise $x'_1\in \OBS(\Uncon(\BS(\ns(i))))$, and $w'_0$ will be undefined ($y_0\in Y\setminus\Uncon(\BS(\ns(i))$, and so there exists at least a controllable event leading $x_0$ to $x'_1$ which is undefined). This is the case for all other states observationally equivalent to $x_1$ (because otherwise $P_{\SigmaNS}(w_0)\notin L(\NS)$ which contradicts the assumption). Therefore, $x_1\notin\Uncon(\BS(\ns(i))$ for any iteration $i$ of the algorithm. So, at each iteration $i$, there exists a $w\in\Sigma^*_{\NSP}$ leading $x_1$ to a marked state which does not become undefined because if it does, then $x_1\in\Uncon(\BS(\NS(i+1))$ which is a contradiction.
%meaning that $x_1\notin\Uncon(\BS(\NS))$ \AR{needs to change: because otherwise $x'_1$ should have been removed as well (due to line \ref{line:obs.eq} of Algorithm \ref{algo})}. \MR{How do you know that $x_1$ is not removed?}
%\MR{The basic idea seems to be that. We need to know that $x_1$ survives synthesis and is thus nonblocking in NS. This is the case since if it had been removed, then also the ``equivalent'' state $x'_1$ would have been removed and it isn't. Right?}
%\MR{This part of the reasoning is crucial and not clear enough at the moment. I guess that we need something like: if $x_1$ is not in $NS$ it must have been removed because it is bad or because it has become unreachable (without being bad). If it is bad then also the last transition to $x'_1$ would have been removed which contradicts our assumption. If it has become unreachable then the reason has to be that a transition is removed and then hopefully this implies that also a transition from$x_0$ to $x'_2$ has been removed?}
%Therefore, there exists $w\in \Sigma^*_{NSP}$, $\deltaNP(x_1,w)\in X_m$. Then, due to Lemma \ref{lemma:NP} and Definition \ref{dfn:NP}, $\delta_G(a_0,P_{\Sigma_G}(w_0.w))\in A_m$. Moreover, $\deltaNS(y_0,P_{\SigmaNS}(w_0.w))\in Y_m$ (due to the projection property). Considering Definition \ref{dfn:operator}, from $z_1$, $w$ is executed since $\delta_G(a_1,P_{\Sigma_G}(w))!$ and $\deltaNS(y_1,P_{\SigmaNS}(w))!$ and it leads to a state $z\in Z$ where due to Lemma \ref{lemma:NSP}, $z.a\in A_m$ and $z.y\in Y_m$, and so $z\in Z_m$.
%\hfill $\blacksquare$
%\end{proof}

\subsection{Proof of Theorem \ref{theorem:TLF}}\label{proof:TLF}
%\begin{proof}
%Let say $G=(A, \Sigma, \delta, a_{0}, A_{m})$, and assume that the algorithm terminates and results in $\NS=(Y,\SigmaNS,\deltaNS,y_0,Y_m)$. Then, considering $\NSP=(Z,\SigmaNSP,\deltaNSP,z_0,Z_m)$, 
We need to prove that for all $z\in Reach(z_0)$, there exists a $w\in \Sigma^*_{\NSP}$ such that $\deltaNSP(z,w\,\tick)!$.
Take $z\in Reach(z_0)$, and assume $z$ is reachable from $z_0$ via $w_0\in \Sigma^*_{NSP}$, i.e., $\deltaNSP(z_0,w_0)=z$. Then, due to Lemma \ref{lemma:NSP}, $z.a=\delta_G(a_0,P_{\Sigma_G}(w_0))$ and $z.y=\deltaNS(y_0,P_{\SigmaNS}(w_0))$.
%Let us indicate $z.a$ and $z.y$ as $a_1$ and $y_1$, respectively.
Based on Definition \ref{dfn:operator}, we need to find $w\in \Sigma^*_{NSP}$ such that
$\delta_G(z.a,P_{\Sigma_G}(w)\,\tick)!$, $\deltaNS(z.y,P_{\SigmaNS}(w)\,\tick)!$, and $(\sigma,0)\notin m$ for all $\sigma\in \Sigma_a$.
As guaranteed by Lemma \ref{lemma:TLFNS}, $\NS$ is TLF, and so for $z.y\in Reach (y_0)$, there exists $v\in\Sigma^*_{NS}$ such that $\deltaNS(z.y,v\,\tick)!$.
Also, $L(\NS)\subseteq P_{\SigmaNS}(L(\NP))$ (as stated before), and so from the projection properties, one can say there exists $v'\in \Sigma^*_{NSP}$, $P_{\SigmaNS}(v')=v$, $\deltaNP(x,v'\,\tick)!$. Let us take $w=v'$  for which we already know $\deltaNS(y,P_{\SigmaNS}(w)\,\tick)!$.
Also, $(\sigma,0)\notin m$ for all $\sigma\in \Sigma_a$ because otherwise Definition \ref{dfn:NP}-item 4) could not be satisfied.
It now suffices to prove $\delta_G(z.a,P_{\Sigma_G}(w)\,\tick)!$. As Property \ref{property:NPLE} says, $P_{\Sigma_G}(L(\NP))\subseteq L(G)$, and so $P_{\Sigma_G}(w)\,\tick\in L(G)$ for $w\,\tick\in L(\NP)$.
%\hfill $\blacksquare$
%\end{proof}

\subsection{Proof of Theorem \ref{theorem:controllability}}\label{proof:controllability}
%\begin{proof} 
We need to prove that if we take any $w\in L(\NSP)$ and $u\in \Sigmauc\cup\{tick\}$ such that $P_{\Sigma_G}(w)u\in L(G)$. Then, $wu\in L(\NSP)$ for $u\in\Sigmauc$, and for $u=\tick$ when there does not exist any $\sigma_f\in \hatSigmaFor\cup\Sigma_o$ such that $w\sigma_f\in L(\NSP)$.

%\begin{itemize}
Take $w\in L(\NSP)$ and $u\in\Sigmauc$.
%For $w\in L(\NSP)$, assume $\deltaNSP(z_0,w)=z$ for some $z\in Z$. 
From Lemma \ref{lemma:NSP}, $\deltaNSP(z_0,w).a=$ $\delta_G(a_0,P_{\Sigma_G}(w))$. Based on Definition \ref{dfn:operator}-item 2), $u$ occurs only if it is enabled by $G$. So, $\deltaNSP(\deltaNSP(z_0,w),u)!$ since $\delta_G(\deltaNSP(z_0,w).a,u)!$ due to the assumption.

Take $u=\tick$ where $\nexists_{\sigma\in\hatSigmaFor\cup\Sigma_o}~w\sigma\in L(\NSP)$. Considering  Definition \ref{dfn:operator}-item 4), $\tick$ occurs in $\NSP$ after $w$ if the following conditions hold; 1. $P_{\Sigma_G}(w)\,\tick \in L(G)$, 2. $P_{\SigmaNS}(w)\,\tick\in L(\NS)$, and 3. $\nexists \sigma\in\Sigma_o, \deltaNSP(z_0,w\sigma)!$. The first and the last conditions hold based on the assumption. So, we only need to prove $P_{\SigmaNS}(w)\,\tick\in L(\NS)$.  
Due to Corollary \ref{corollary:product}, for $w\in L(\NSP)$: $w\in L(\NP)$. 
%  such that $\deltaNP(x_0,w)=x_1$ for some $x_1\in X$.
Due to Property \ref{property:NPLE}, for $P_{\Sigma_G}(w)\,\tick \in L(G)$, there exists $w'\in L(\NP)$, $P_{\Sigma_G}(w')=P_{\Sigma_G}(w)$ such that $w'.\tick\in L(\NP)$.
%   Let say $\deltaNP(x_0,w')=x'_1$ and $\deltaNP(x'_1,\tick)=x'_2$.
Considering Definition \ref{dfn:NP}, $w\,\tick\in L(\NP)$ for the following reasons; 1. $\deltaNP(x_0,w).a=\deltaNP(x_0,w').a$ and $\deltaNP(x_0,w).a'=\deltaNP(x_0,w').a'$ since $P_{\Sigma_G}(w')=P_{\Sigma_G}(w)$. Hence, $\delta_G(\deltaNP(x_0,w).a,\tick)!$ and $\delta_G(\deltaNP(x_0,w).a',\tick)!$ (since $\deltaNP(x_0,w'\,\tick)!$).
2.  $m\in M$ changes only by the execution of $\sigma\in\Sigma_G$. So, $\deltaNP(x_0,w).m=\deltaNP(x_0,w').m$ since $P_{\Sigma_G}(w')=P_{\Sigma_G}(w)$. Also, $(\sigma,0)\notin \deltaNP(x_0,w').m$ for any $\sigma\in\Sigma_a$ since $\deltaNP(x_0,w'\,\tick)!$, and so $(\sigma,0)\notin \deltaNP(x_0,w).m$ for any $\sigma\in\Sigma_a$.
Due to the assumption, $w\sigma\notin L(\NSP)$ for $\sigma\in\SigmaFor\cup\Sigma_e\cup\Sigma_o$. Also, due to Theorem \ref{lemma:product}, $\NSP=\NS||\NP$.  In case that $w\sigma\in L(\NP)$ for some $\sigma\in\SigmaFor\cup\Sigma_o$, then, due to line \ref{line:ctrl} of Algorithm \ref{algo}, it could not be disabled by $\NS$. Also, if $w\sigma\in L(\NP)$ for some $\sigma\in\Sigma_e$ where both $\tick$ and $\sigma$ become disabled by $\NS$, then by definition, $\deltaNP(x_0,w)\in \BPre(\NS)$ and will be removed which violates the assumption ($w\in L(\NSP)$). Hence, $w\,\tick\in L(\NP)$ and \tick does not become disabled by Algorithm \ref{algo}, and so $P_{\SigmaNS}(w)\,\tick\in L(\NS)$.
%Moreover, from the assumption we have $\nexists_{\sigma\in\hat{\Sigma}_f}~w\sigma\in L(\NSP)$. This means that either $\nexists_{\sigma\in\hat{\Sigma}_f}~w\sigma\in L(\NP)$ at all which means that $F(\deltaNS(y_0,w))=\varnothing$ and so \emph{tick} cannot be disabled by $\NS$, or $\exists_{\sigma\in\hat{\Sigma}_f}~w\sigma\in L(\NP)$ but it is disabled by $\NS$. If both the forcible event and $\emph{tick}$ has been disabled by $\NS$, then $\deltaNS(y_0,w)\in \BPre(\NS)$ and will be removed considering line \ref{line:BPre} of Algorithm \ref{algo}. This results in $w\notin L(\NSP)$ which violates the assumption.
%\end{itemize}
%\hfill $\blacksquare$ 
%\end{proof}

\subsection{Proof of Theorem \ref{theorem:MPness}}\label{proof:MPness}
%\begin{proof}
To prove that $\NS$ is (timed networked) maximally permissive for $G$, we need to ensure that for any other proper networked supervisor (say $\NS'$) in the same NSC framework (with event set $\SigmaNS$): %communicating through the same observation and control channels for which $\NS'_{N_c}\|_{N_o}\,G$ is nonblocking, TLF, timed networked controllable:
$P_{\Sigma_G}(L(\NS'_{N_c}\|_{N_o}\,G))\subseteq P_{\Sigma_G}(L(\NSP))$.
First, assume that $L(\NS')\nsubseteq P_{\SigmaNS}(L(\NP))$.  Then, any extra transition of $\NS'$ that is not included in $P_{\SigmaNS}(L(\NP))$ does not add any new transition to $P_{\Sigma_G}(L(\NS'_{N_c}\|_{N_o}\,G))$.
Let say $v\sigma\in L(\NS')$ and $v\in P_{\SigmaNS}(L(\NP))$, but $v\sigma\notin P_{\SigmaNS}(L(\NP))$ for $\sigma\in\SigmaNS$. Also, %assume \MR{It is not an assumption, but a fact isn't it?}\AR{not necessarily. For instance, there could be some $\sigma_o$ enabled in $NS$ which is not executed in $\NSP$} that
there exists $w\in L(\NS'_{N_c}\|_{N_o}\,G)$ with $P_{\SigmaNS}(w)=v$.
If $\sigma=\tick$, then $\sigma$ cannot be executed in $\NS'_{N_c}\|_{N_o}\,G$ because based on Definition \ref{dfn:operator}-item 4), $\tick$ should be enabled by $G$ which is not the case; \tick is not enabled in $\NP$, and so due to Property \ref{property:NPLE}, it is not enabled in $G$.
If $\sigma\in\Sigma_o$, then it does not matter if $\sigma$ occurs in $\NS'_{N_c}\|_{N_o}\,G$ because it does not change  $P_{\Sigma_G}(L(\NS'_{N_c}\|_{N_o}\,G))$.
%$\sigma$ does not occur in $\NS'_{N_c}\|_{N_o}\,G$ because based on item 5) of Definition \ref{dfn:operator}, $(\sigma,0)$ should be in $m$ which is not the case because if it was then $\sigma$ occured in $\NP$ as well.
If $\sigma\in\Sigma_e$, then as Lemma \ref{lemma:ontime} says, $\NP$ enables all enabling events of $\Sigma_c$ that are executed in the plant on time ($N_c$ \ticks ahead). So, any extra enabling event by $\NS'$ will not be executed by the plant, and so it does not enlarge $P_{\Sigma}(L(\NS'_{N_c}\|_{N_o}\,G))$. Therefore, we continue the proof for the case that
$L(\NS')\subseteq P_{\SigmaNS}(L(\NP))$ (where Lemma \ref{lemma:product} and Corollary \ref{corollary:product} hold for $NS'$).
Take an arbitrary $w\in P_{\Sigma_G}(L(\NS'_{N_c}\|_{N_o}\,G))$, it suffices to prove that $w\in P_{\Sigma_G}(L(\NSP))$. Let say
%$G=(a_0,\Sigma,\delta,,A_m)$, $\NP=(x_0,\SigmaNSP,\deltaNP,X_m)$,  $\NSP=(z_0,\SigmaNSP,\deltaNSP,Z_m)$, and 
$\NS'_{N_c}\|_{N_o}\,G=(z'_0,\SigmaNSP,\delta_{NS'P},Z'_m)$. 
For $w\in P_{\Sigma_G}(L(\NS'_{N_c}\|_{N_o}\,G))$, due to the projection properties, there exists $v'\in L(\NS'_{N_c}\|_{N_o}\,G)$ such that $P_{\Sigma_G}(v')=w$ where $\delta_{NS'P}(z'_0,v')$ is a TLF and non-blocking state ($NS'$ is proper due to the assumption).
Also, any uncontrollable active event/non-preemptable \tick enabled at $\delta_G(a_0,w)$ is enabled at $\delta_{NS'P}(z_0,v')$, and it leads to a nonblobking and TLF state.
Based on Lemma \ref{lemma:NSP}, $P_{\SigmaNS}(v')\in L(\NS')$ for $v'\in L(\NS'_{N_c}\|_{N_o}\,G)$, and due to Corollary \ref{corollary:product}, $v'\in L(\NP)$.
Moreover, due to Lemma \ref{lemma:product}, $L(\NS'_{N_c}\|_{N_o}\,G)=L(\NS'||\NP)$, so regarding the definition of synchronous product, for any $w'\in L(\NP)$ and $P_{\SigmaNS}(w')=P_{\SigmaNS}(v')$: $w'\in L(\NS'_{N_c}\|_{N_o}\,G)$. 
$\delta_{NS'P}(z'_0,w')$ is a TLF and non-blocking state because $\NS'_{N_c}\|_{N_o}\,G$ is nonblocking and TLF due to the assumption.
Also, any uncontrollable active event or non-preemptable \tick enabled at $w'$ leads to a nonblocking and TLF state since $\NS'$ is  controllable for $G$ by the assumption.
Therefore, one can say $\deltaNP(x_0,v')\notin \OBS(\Uncon(\BS(NP))$.
Considering Algorithm \ref{algo}, initially, $\ns(0)=NP$ where $v'\in L(\NP)$ and $\deltaNS(y_0,v')\notin \OBS(\Uncon(\BS(\ns(0)))$.
The last statement holds for any iteration  of the algorithm until the last one (say $n$) so that $\deltaNS(y_0,v')\notin \OBS(\Uncon(\BS(\NS(n)))$ because otherwise all $y\in \OBS(\Uncon(\BS(\NS(n)))$ are removed (based on line \ref{line:obs.eq}),  and so $P_{\SigmaNS}(v')\notin L(\NS)$ because  it leads $\NS||\NP$ ($\NS||\NP=\NS(n)$) to a state in $\Uncon(\BS(\NS(n)))$.
Then, based on Lemma \ref{lemma:product}, $\NSP$ becomes blocking/time-lock/uncontrollable which violates the assumption.
Hence, (considering line \ref{line:obs.eq}) $v'$ is not undefined by Algorithm \ref{algo}, and so $P_{\SigmaNS}(v')\in L(\NS)$. Based on Lemma \ref{lemma:product}, $L(\NSP)=L(\NS||\NP)$. $P_{\SigmaNS}(v')\in L(\NS)$ and $v'\in L(\NP)$, so $v'\in L(\NSP)$ where applying the projection on $\Sigma_G$ gives $w\in P_{\Sigma_G}(L(\NSP))$.

\subsection{Proof of Theorem \ref{theorem:safety}}\label{proof:safety}
%\begin{proof}
To simplify, let us denote $G||R^{\bot}$ by $G^{t}$, the networked plant $\Pi(G^{t},N_c,N_o,\Lmax,\Mmax)$ by $NP^{t}$ and the networked supervised plant $\NS_{N_c}\|_{N_o}\,G^{t}$ by $\NSP^{t}$.
We need to prove that if we take any $w\in P_{\SigmaNSP\cap\Sigma_R}(L(\NSP^{t}))$, then $w\in P_{\SigmaNSP\cap\Sigma_R}(L(R))$. In our setting, $\SigmaNSP\cap\Sigma_R=\Sigma_R$ since $\SigmaNSP=\Sigma_e\cup\Sigma\cup\Sigma_o$ and $\Sigma_R\subseteq\Sigma_G$.
Hence, it suffices to prove that for any $w\in P_{\Sigma_R}(L(\NSP^{t}))$: $w\in L(R)$. 
Take $w\in P_{\Sigma_R}(L(\NSP^{t}))$, then due to Definition \ref{dfn:proj}, there exists $w'\in L(\NSP^{t})$ such that $P_{\Sigma_R}(w')=w$. Also, based on Property \ref{property:NSP&P}, $P_{\Sigma_G}(L(\NSP^t))\subseteq L(G^t)$, and so $P_{\Sigma_G}(w')\in L(G||R^{\bot})$.
Applying the projection on $\Sigma_R$ gives $P_{\Sigma_R}(w')\in L(R^{\bot})$.
For $w\in P_{\Sigma_R}(L(\NSP^{t}))\cap L(R^{\bot})$, $w\in L(R)$ since the blocking state $q_d$ added to $G||R$ to make $G||R^{\bot}$ is removed by $\NS$ as guaranteed by Theorem \ref{theorem:nonblockingness}.
%\hspace*{\fill} $\blacksquare$
%\end{proof}

%\section*{Acknowledgment}

%\ifCLASSOPTIONcaptionsoff\newpage\fi

\bibliographystyle{IEEEtran}
\bibliography{references.bib}

%\begin{comment}
\begin{IEEEbiography}[{\includegraphics[width=0.9in,height=1.25in,clip,keepaspectratio]{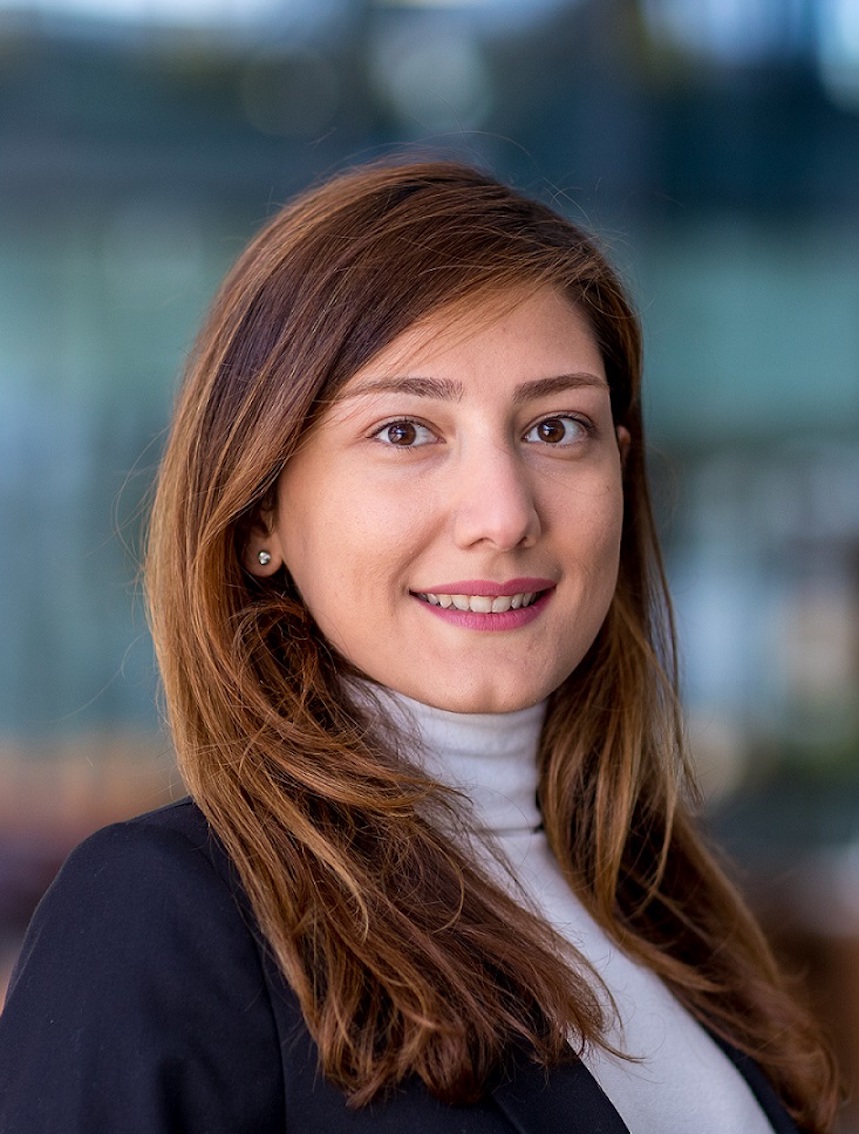}}]{Aida Rashidinejad}
received the M.Sc. degree in electrical-control engineering from Amirkabir University of Technology (Tehran Polytechnic), Tehran, Iran, in 2014. She is currently working towards PhD degree in mechanical engineering-control systems from Eindhoven University of Technology, Eindhoven, The Netherlands. Her current research interests include supervisory control synthesis, networked control, and cyber-physical systems. 
\end{IEEEbiography}

\begin{IEEEbiography}[{\includegraphics[width=1in,height=1.25in,clip,keepaspectratio]{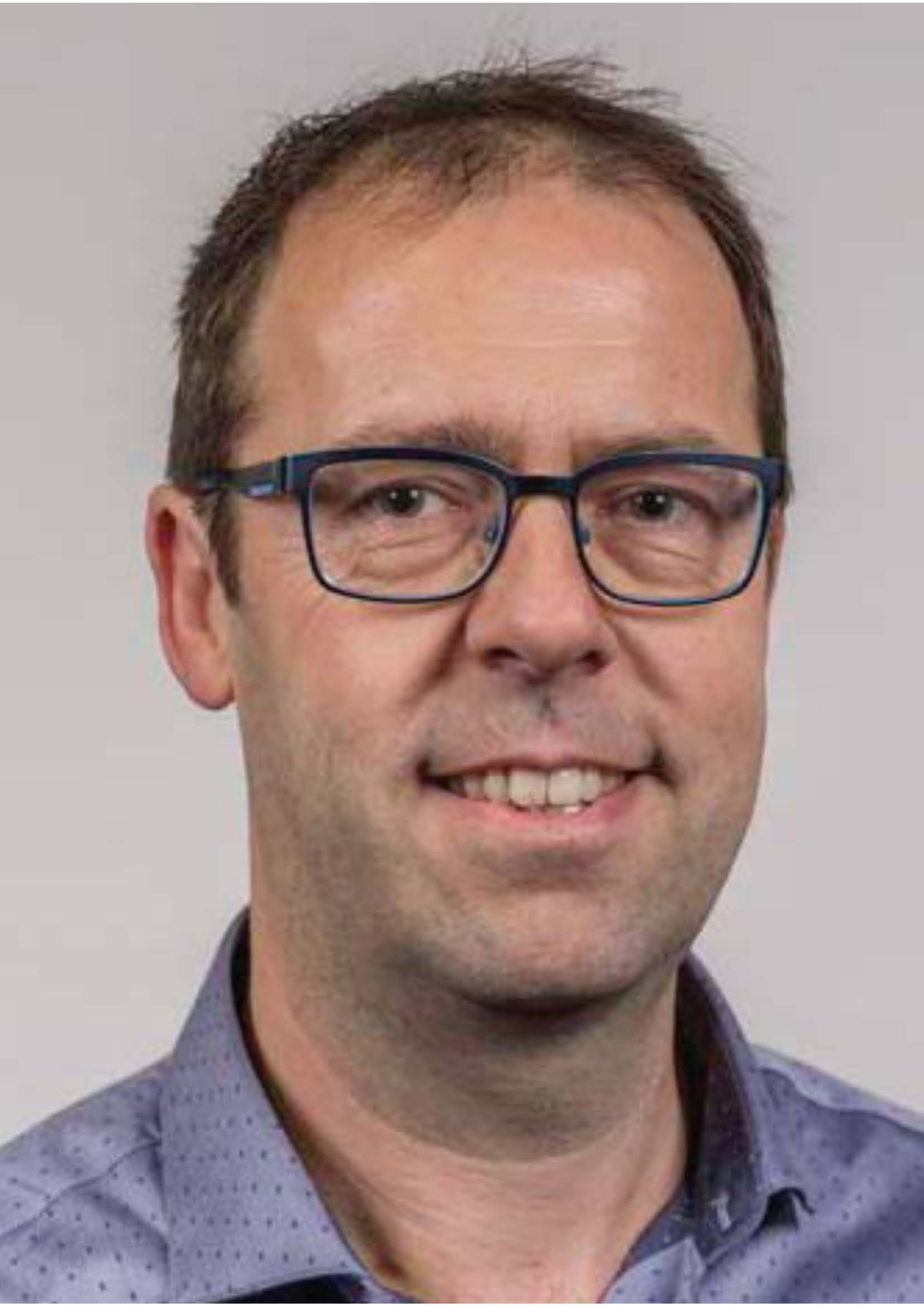}}]{Michel Reniers}
(S’17) is currently an Associate Professor in model-based engineering of
supervisory control at the Department of Mechanical Engineering at TU/e. He has authored
over 100 journal and conference papers. His
research portfolio ranges from model-based systems engineering and model-based validation
and testing to novel approaches for supervisory
control synthesis. Applications of this work are
mostly in the areas of cyber-physical systems.
\end{IEEEbiography}

%\newpage

\begin{IEEEbiography}[{\includegraphics[width=1in,height=1.25in,clip,keepaspectratio]{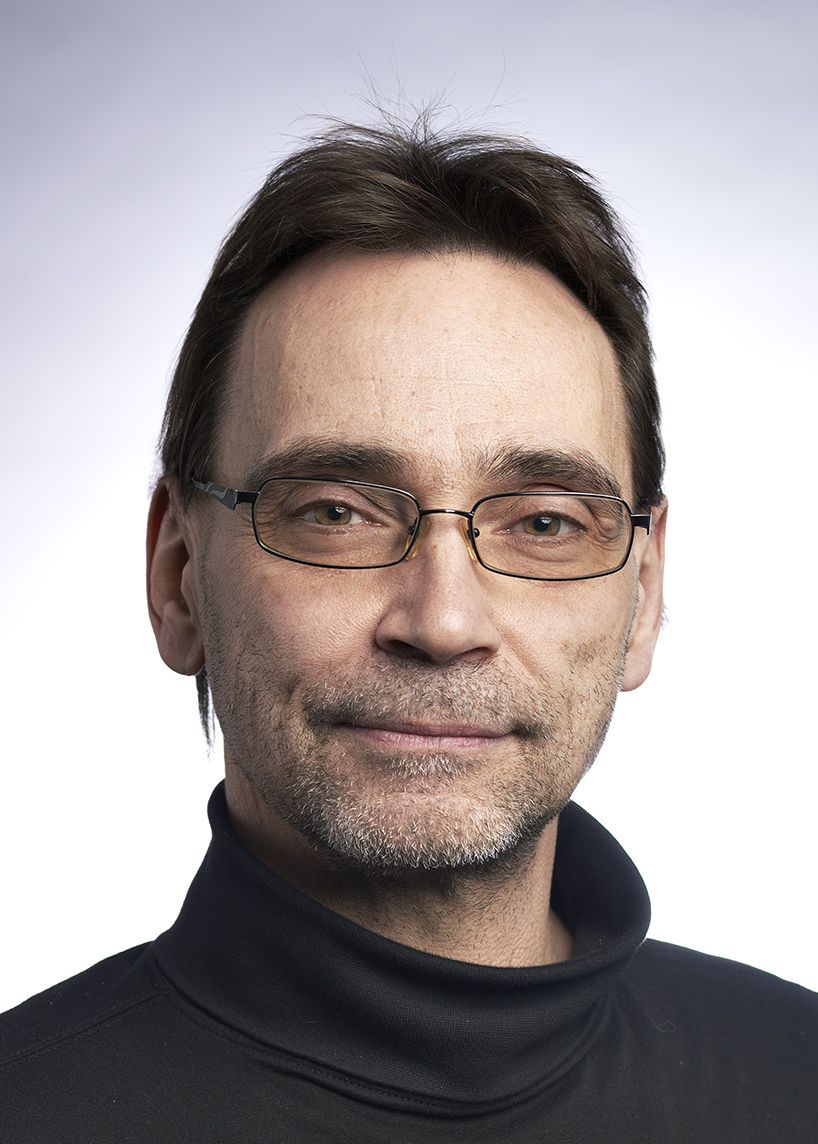}}]{Martin Fabian}
is Professor in Automation and Head of the Automation Research group at the Department of Electrical Engineering, Chalmers University of Technology. His research interests include formal methods for automation systems in a broad sense, merging the fields of Control Engineering and Computer Science. He has authored more than 200 publications, and is co-developer of the formal methods tool Supremica, which implements several state-of-the-art algorithms for supervisory control synthesis.
\end{IEEEbiography}


%\vfill


%\end{comment}

\end{document}